\documentclass[12pt]{iopart}

\expandafter\let\csname equation*\endcsname\relax
\expandafter\let\csname endequation*\endcsname\relax

\usepackage[utf8]{inputenc}
\usepackage{amsmath}
\usepackage{amsfonts}
\usepackage{amssymb}
\usepackage{soul}
\usepackage{amsthm}
\usepackage[colorlinks=true]{hyperref}
\usepackage[normalem]{ulem}
\usepackage{bm}
\usepackage[T1]{fontenc}
\usepackage{graphicx}
\usepackage{subcaption}
\usepackage{caption}
\usepackage{subcaption}
\usepackage{relsize}
\usepackage{mathrsfs}
\DeclareUnicodeCharacter{2212}{-}
\usepackage[usenames,dvipsnames]{xcolor}

\makeatletter
\renewcommand\footnoterule{%
  \kern-3\p@
  \hrule\@width2.5cm
  \kern2.6\p@}
\makeatother

 \def\be{\begin{equation}}
\def\ee{\end{equation}}
 \def\ba{\begin{align}}
\def\ea{\end{align}}
\def\bea{\begin{eqnarray}}
\def\eea{\end{eqnarray}}

\def\P{{\cal P}}

\newcommand{\bseq}{\begin{subequations}}
\newcommand{\eseq}{\end{subequations}}

\definecolor{darkred}{rgb}{0.5,0,0}
\definecolor{darkred2}{rgb}{0.7,0,0}
\definecolor{darkgreen}{rgb}{0,0.5,0}
\definecolor{darkblue}{rgb}{0,0,0.5}
\definecolor{prussian}{rgb}{0.0, 0.19, 0.33}
\definecolor{richelectricblue}{rgb}{0.03, 0.57, 0.82}
\definecolor{teal}{rgb}{0.0, 0.5, 0.5}
\definecolor{mediumseagreen}{rgb}{0.24, 0.7, 0.44}
\definecolor{lust}{rgb}{0.9, 0.13, 0.13}
\definecolor{ballblue}{rgb}{0.13, 0.67, 0.8}
\definecolor{darkcyan}{rgb}{0.0, 0.55, 0.55}
\definecolor{mountainmeadow}{rgb}{0.19, 0.73, 0.56}
\definecolor{palecarmine}{rgb}{0.69, 0.25, 0.21}
\definecolor{richcarmine}{rgb}{0.84, 0.0, 0.25}
\definecolor{tangelo}{rgb}{0.98, 0.3, 0.0}
\definecolor{venetian}{rgb}{0.784,0.031,0.082}
\definecolor{bdfrance}{rgb}{0.192,0.549,0.906}

\hypersetup{colorlinks=true, citecolor=venetian,
linkcolor=bdfrance, urlcolor=lust}
\usepackage{tensor}
\usepackage{mathtools}
\usepackage{amsbsy}
\usepackage{bm}
\usepackage{float}

\newcommand{\nn}{\nonumber}

\begin{document}

\title[Imaging and chaos in a Hartle-Thorne spacetime]{Chaotic photon orbits and shadows of a non-Kerr object described by the Hartle-Thorne spacetime}

\author{K Kostaros$^1$, and G Pappas$^1$}

\address{$^1$Department of Physics, Aristotle University of Thessaloniki, Thessaloniki 54124, Greece}
\ead{gpappas@auth.gr}

\date{today}


\begin{abstract}
The data from the event horizon telescope (EHT) have provided a novel view of the vicinity of the horizon of a black hole (BH), by imaging the region around the light-ring. They have also raised hopes for measuring in the near future, features of the image (or the shadow) related to higher order effects of photons traveling in these regions, such as the appearance of higher order bright rings produced by more than one windings of photons around the light-ring. While the prospect of measuring these fine features of Kerr BHs is very interesting in itself, there are some even more intriguing prospects for observing novel features of possible non-Kerr objects, in the case that the subjects of our images are not the BH solutions of general relativity. In the hope of sufficient resolution being available in the future, we explore in this work the structure and properties of null geodesics around a Hartle-Thorne spacetime that includes a deformation from the Kerr spacetime characterised by the quadrupole deformation $\delta q$. These spacetimes have been found to exhibit a bifurcation of the equatorial light-ring to two off-equatorial light-rings in a range of $\delta q$s and spin parameters. In addition to this, there is a range of parameters where both the equatorial and the off-equatorial light-rings are present. This results in the formation of a pocket that can trap photon orbits. We investigate the properties of these trapped orbits and find that chaotic behaviour emerges. Some of these chaotic orbits are additionally found to be ``sticky'' and get trapped close to periodic orbits for long times. We also explore how these novel features affect the shadow and find that the off-equatorial light-rings produce distinctive features that deform its circular shape, while the chaotic behaviour associated to the pocket creates features with fractal structure. These results are shown to be quite general, extending to higher order Hartle-Thorne spacetimes.
\end{abstract}

%
\vspace{2pc}

\submitto{\CQG}

\maketitle

%

\section{Introduction}
\label{sec:1}

In 2019, the Event Horizon Telescope (EHT) collaboration released the first image of a supermassive black hole (BH) \cite{EHT2019ApJ.I,EHT2019ApJ.V,EHT2019ApJ.VI}, providing a millimeter-band radio image of the shadow of the BH at the centre of the M87 galaxy, some $17.9 \textrm{Mpc}$ away \cite{Geb2011ApJ}. 100 years from the first observation of light deflection from the Sun by the Eddington-Dyson expeditions \cite{Dyson1920RSPTA}, and four decades since the pioneering work of the 70s on the topic \cite{1973blho.conf..215B,1973ApJ...183..237C,1979A&A....75..228L}, a BH shadow image constitutes the confirmation of strong light bending taking place near such an extremely compact object. 

More than that, the M87* image marks the beginning of an era of detailed observations from the very close vicinity of the horizon of a BH, which will provide us with the opportunity to further test the nature of these objects, as well as their properties, confirming or challenging our assumptions on them as well as the prevailing theory of gravity, i.e., general relativity (GR) \cite{Broderick2014ApJ}. 

The EHT prospects have sparked a lot of work in the literature that is exploring possible signatures of so called non-Kerr objects, either BHs different from Kerr or horizonless ultra compact objects (UCOs) \cite{Cardoso_Pani:2017,Cardoso:2019rvt}, that could lurk at the centres of galaxies. The approaches can be mostly (but not exclusively) divided into two categories. (i) On the one hand the literature has explored the impact of the change of geometry on the characteristics of the shadow of the compact object observed (such as shape or size) \cite{Johannsen2010ApJ,Johannsen2013ApJ,Medeiros:2019cde,Konoplya:2021slg}. This is done by assuming parametric deviations from the Kerr spacetime that are designed to be theory independent and can be used to put constraints on possible deviations in the style of the parameterised post-Newtonian approach \cite{Glampedakis:2006sb,Vigeland:2009pr,Vigeland:2011ji,Johannsen:2011dh,Johannsen2013PhRvD,Cardoso:2014rha,Konoplya:2016jvv,Papadopoulos:2018nvd,Carson_Yagi2020}. (ii) On the other hand, the subject of investigation 
is the appearance of the shadow of specific compact objects different from Kerr that are either extensions of Kerr BHs with additional scalar fields \cite{Cunha2015PhRvL,Cunha2016IJMPD,Cunha2016PhRvD,Cunha2017PhysRevD,Cunha2017PhysRevLett,Cunha2018GReGr} or horizonless compact objects such as boson stars \cite{Vincent:2015xta,Cunha2016PhRvD}, gravastars \cite{Sakai:2014pga}, wormholes \cite{Nedkova:2013msa,Ohgami:2015nra,Kasuya:2021cpk,Bugaev:2021dna}, and even more exotic alternatives \cite{Allahyari:2019jqz,Qin:2020xzu,Bacchini:2021fig,Lima:2021las,Khodadi:2020jij}.

It is a well established fact that rotation, in the case of the shadow of Kerr black holes, tends to have a very small effect on its circular shape, while it mostly shifts its position on the observer's plane \cite{Johannsen2010ApJ}. Work on parametric deviations from Kerr has produced an intriguing result. For the metrics used in these studies, the deformations from Kerr generally also produce small deviations from circularity, which introduce a general confusion problem with respect to Kerr and non-Kerr shadows \cite{Johannsen2010ApJ,Johannsen2013ApJ,Medeiros:2019cde,Psaltis_etal2020}. Significant deviations are generally observed only for very large values of the deformations. On the other hand, deformed BHs, scalarized BHs, boson stars or wormholes can have qualitatively and quantitatively significantly more different shadows. For the latter cases quite rich phenomenology can be observed, such as chaotic or fractal features, as in the case of scalarized BHs \cite{Cunha2016PhRvD} or deformed BHs \cite{WangPhysRevD2018} for example. Similarly rich phenomenology has also been observed in a slightly different class of systems, i.e., double black holes such as the Majumdar-Papapetrou  di-hole solution, where again chaotic and fractal features are present \cite{Shipley-Dolan2016}. Common thread between these cases is that the spacetimes involved are not integrable in the sense that they do not admit a full set of integrals of motion that would lead to the separation of variables for the geodesic motion, as is the case for the Kerr spacetime. 

In this work, we will assume a Hartle-Thorne (HT) at $\mathcal{O}(\Omega^2)$ rotating compact object \cite{Hartle1967,HT68} parameterised by the mass $M$, the spin $\chi =J/M^2$, and the quadrupole deformation $\delta q$, from the value of the corresponding Kerr quadrupole, 
as a model for non-Kerr compact objects, and explore the properties of their photon orbits and the shadows they produce. We are using a HT object since this is a general way of perturbatively constructing rotating compact objects from their static counterparts and therefore can be considered as a natural approximation for any such rotating object. For our analysis we use parameters, i.e., values for the quadrupole deformation $\delta q$, that correspond to prolate objects. Prolate deformations are what one expects from a compact object so as to be able to produce ultra-compact configurations that can serve as BH mimickers \cite{Glampedakis2018PhRvD}. Such HT objects with prolate $\delta q$s have been recently found to exhibit some very interesting features, such as off-equatorial planar light-rings and triple light-rings that under the right conditions can form pockets that can trap photon orbits \cite{Glampedakis_Pappas2019}. The pocket formation is a feature that is known to lead to chaotic behaviour for the orbits in other systems \cite{Cunha2016PhRvD,Shipley-Dolan2016} and this is exactly what we find in the HT case as well. Even though we mainly work with a second order HT spacetime, our findings hold in the case of the third order HT as well, as it is discussed in Section \ref{sec:App:2}, therefore making these results quite general, considering also the rest of the examples from the literature  \cite{Cunha2016PhRvD,Shipley-Dolan2016,Glampedakis_Pappas2019,WangPhysRevD2018}.

In section \ref{sec:2} we briefly present the HT metric and in section \ref{sec:3} briefly describe the equations governing the evolution of null orbits. Then, in section \ref{sec:4} we first study the chaotic behaviour of the photon orbits trapped in the pocket and find that apart from chaos, the orbits also exhibit a property that we will refer to as ``stickiness'', where chaotic orbits tend to spend long periods of time close to periodic orbits. We also study the evolution of the pocket from being fully detached from the exterior and small, up to the point when the throat that connects it to infinity appears, and see how chaos emerges. We then study the characteristics of the shadow of these HT objects in section \ref{sec:5}, where we find that the presence of the off-equatorial light-rings results in distinct features in the shadow that are clearly non-circular, while the existence of chaotic orbits when the pocket is present endow the shadow with fractal structures. Section \ref{sec:App:2} discusses the issue of how general these results on photon orbits and shadows are and their extension to the 3rd order HT. The final section are the conclusions. In what follows we will generally use geometric units with $G=c=1$, unless it is stated otherwise.

\section{The Hartle-Thorne spacetime}
\label{sec:2}

The HT spacetime is a stationary and axisymmetric spacetime that describes the exterior spacetime of a rotating compact object and is constructed perturbatively in terms of the rotation rate. The original scheme developed by Hartle and Thorne \cite{Hartle1967,HT68} aimed at describing both the interior structure and the exterior of a compact fluid configuration starting from a non-rotating static configuration and describing the rotating solution as a perturbation with the expansion parameter being the rotation rate. The spacetime is described by the line element,
\begin{equation}
    \begin{split}
        ds^2= &  -e^\nu \left(1+2h\right) dt^2 + e^\lambda \left(1+\frac{2\mu}{r-2m}\right)dr^2 \\&+r^2\left(1+2k\right)\{d\theta^2 +  \sin^2\theta \left[ d\phi -\left(\Omega - \omega\right)dt\right]^2 \} 
        + \mathcal{O}\left(\Omega^3 \right),
    \end{split}
    \label{eq:HT}
\end{equation}
where $\Omega$ is the angular velocity of the configuration, while the metric potentials $h\left(r,\theta\right)$, $\mu\left(r,\theta\right)$, $k\left(r,\theta\right)$ and $\omega \left( r, \theta\right)$ are expanded in terms of the Legendre polynomials $P_{\ell}(\cos\theta)$ as,
\begin{equation}
\begin{split}
&h\left(r,\theta\right)=h_0\left(r\right) +h_2\left(r\right) P_2, \quad \mu\left(r,\theta\right)=\mu_0\left(r\right) + \mu_2\left(r\right) P_2,\\
 &k\left(r,\theta\right)=k_2\left(r\right) P_2,  \quad \textrm{and} \quad \omega \left(r,\theta\right)=\omega_1 \left(r\right) P_1 ' .
\end{split}
\end{equation}
Given that $J$ is the angular momentum, $M$ is the non-rotating mass and $\delta q$ is the quadrupole deviation from the Kerr quadrupole, the above functions written in terms of the spin parameter $\chi = J/M^2$, the quadrupole deviation defined by $Q = -\chi^2M^3 (1 - \delta q)$, the second order correction to the mass $\delta m$, and the dimensionless distance $x = r/M$, take the form \cite{Benhar_etal2005,Yagi_2014,Yagi_2015,Glampedakis2018PhRvD,Glampedakis_Pappas2019} 
\begin{equation}
m=M, \quad e^\nu = e^{-\lambda} = 1- \frac{2}{x}, \quad \omega_1= \Omega - \frac{2\chi}{Mx^3}, \quad \frac{\mu_0}{M}= \chi^2 \left(\delta m - \frac{1}{x^3}\right),\nn
\end{equation}
\begin{equation}
h_0=\frac{\chi^2}{x-2}\left(\frac{1}{x^3}-\delta m \right),\nn
\end{equation}
\begin{equation}
\begin{split}
h_2=& \frac{5}{16}\chi^2 \delta q \left(1- \frac{2}{x}\right) \left[3x^2 \log\left(1-\frac{2}{x}\right) + \frac{2}{x}\frac{\left(1-1/x\right)}{\left(1-2/x\right)^2}\left(3x^2-6x-2\right)\right] \\ &+\frac{\chi^2}{x^3}\left(1+\frac{1}{x}\right),\nn
\end{split}
\end{equation}
\begin{equation}
k_2 = - \frac{\chi^2}{x^3}\left(1+\frac{2}{x}\right)- \frac{5}{8} \chi^2 \delta q\left[3\left(1+x-\frac{2}{x}-3\left(1-\frac{x^2}{2}\right)\right) \log\left(1-\frac{2}{x}\right)\right],\nn
\end{equation}
\begin{equation}
\begin{split}
\frac{\mu_2}{M} =  - \frac{5}{16}\chi^2 \delta q x \left(1-\frac{2}{x}\right)^2 &\left[3x^2 \log\left(1-\frac{2}{x}\right)  +\frac{2}{x}\frac{\left(1-1/x\right)}{\left(1-2/x\right)^2}\left(3x^2-6x-2\right)\right.\\ &\left.-\frac{\chi^2}{x^2}\left(1-\frac{7}{x}+\frac{10}{x^2}\right)\right].\nn
\end{split}
\end{equation}
For our purposes, we will use this spacetime expressed in terms of the total mass $M=m+\delta m$ that includes the correction due to rotation, spin $\chi=J/M^2$ (where $M$ is now the total mass), and $\delta q$, which amounts to just setting $\delta m=0$ in the above equations. We will also assume that with respect to calculating geodesics, the spacetime metric is ``as it is given'', i.e., the spacetime is described by the truncated to the given order metric and no further approximations are made. 

Lastly, we will assume that the HT spacetime ends at the ``surface'' of the compact object. Here we are using the term ``surface'' loosely, by which we mean the boundary through which photons will be lost in our integration of the geodesics. In principle one could try to find such a boundary in the form of a horizon for the HT spacetime, but even if some null surface can be found, it is not certain that it will be a horizon. Furthermore, there may be problematic regions such as the locations where the metric function $g_{rr}$ goes to zero. We avoid these difficulties by assuming that the regions of the spacetime that they may exist in, are hidden under the surface of the compact object which is as small as possible so as to cover the pathologies, while it leaves the allowed space for the photons (given by the zero-velocity separatrix that we will discuss in sections \ref{sec:3},\ref{sec:4}) undisturbed. The material that the compact object is made of, is assumed to be such that it evades all the usual issues with the emission of photons, so that the object can act as a BH mimicker \cite{Cardoso:2019rvt}. We further discuss these issues in the \ref{sec:App:0} where we further describe our algorithm for calculating the shadows.

\section{Null geodesics}
\label{sec:3}

A stationary and axisymmetric spacetime, admits two Killing vector fields. A timelike $\xi^\alpha$, associated to time translations, and a spacelike $\eta^\alpha$, associated to rotations with respect to an axis of symmetry.

The line element of such a spacetime can generally take the form \cite{wald1984}, 
\begin{equation}
ds^2=g_{tt} dt^2 +g_{rr}dr^2 + g_{\theta\theta}d\theta^2+ g_{t\phi}dt d\phi +g_{\phi\phi}d\phi^2.  
\end{equation}
The two aforementioned Killing fields are associated to two conserved quantities, the energy $E$ and the angular momentum $L$, both per unit mass,
\begin{equation}
E=-\xi^\alpha u_{\alpha}= -\left(g_{tt} \frac{dt}{d\lambda} + g_{t\phi}\frac{d\phi}{d\lambda}\right),
\label{eq:9}
\end{equation} 
\begin{equation}
L=\eta^\alpha u_{\alpha}=g_{t\phi}\frac{dt}{d\lambda}+ g_{\phi\phi}\frac{d\phi}{d\lambda},
\label{eq:10}
\end{equation}
where $\lambda$ is the affine parameter. The equations of motion for the geodesics in such a spacetime can be derived by the Lagrangian
\begin{equation}
\mathcal{L}=\frac{1}{2}g_{\alpha b} \dot{x}^\alpha \dot{x}^b,
\end{equation}
in which case the generalised momenta are given by 
\begin{equation}
p_{\alpha}=\frac{\partial \mathcal{L}}{\partial \dot{x}^\alpha},
\end{equation}
and the Hamiltonian is defined as
\begin{equation}
\mathcal{H}=\frac{1}{m} \sum p_{\alpha} \dot{x}^\alpha - \mathcal{L} = -E \dot{t}+L \dot{\phi} + g_{rr}\dot{r}^2 +g_{\theta\theta} \dot{\theta}^2 -\mathcal{L},
\end{equation}
where $p_r=g_{rr}\dot{r}$ and $p_{\theta}=g_{\theta\theta} \dot{\theta}$ are the radial and poloidal momenta respectively. Solving (\ref{eq:9}-\ref{eq:10}) for $\dot{t}$ and $\dot{\phi}$, 
we get the final expression for the Hamiltonian 
\begin{equation}
\begin{split}
\mathcal{H}&= \frac{1}{2}\left(g_{rr}\dot{r}^2 +g_{\theta\theta}\dot{\theta}^2 - \frac{L^2 g_{tt} + 2EL g_{t\phi} + E^2 g_{\phi\phi}}{\mathcal{D}}\right) \\
&=\frac{1}{2}\left(\frac{p_{r}^2}{g_{rr}}+\frac{p_{\theta}^2}{g_{\theta\theta}}-\frac{L^2 g_{tt} + 2EL g_{t\phi} + E^2 g_{\phi\phi}}{\mathcal{D}}\right),
\end{split}
\label{eq:14}
\end{equation}
where we have defined $\mathcal{D}=g_{t\phi}^2 - g_{tt}g_{\phi\phi}$. One can identify in this Hamiltonian the effective potential 
\begin{equation}
V_{eff}=-\frac{L^2 g_{tt} + 2EL g_{t\phi} + E^2 g_{\phi\phi}}{\mathcal{D}}.
\end{equation}
The contour of $V_{eff}=0$ marks the forbidden region for geodesic motion (where $V_{eff}>0$), acting as a zero-velocity separatrix. For the motion description of massive and massless particles, 
the Hamiltonian will be $\mathcal{H}=-1/2$ and $\mathcal{H}=0$ respectively. 
The equations of motion are then derived from Hamilton's canonical equations
\begin{equation}
\dot{x}^\alpha=\frac{\partial \mathcal{H}}{\partial p_\alpha},
\quad 
\dot{p}_\alpha=-\frac{\partial \mathcal{H}}{\partial x^\alpha}.
\end{equation}
In order to integrate these equations, initial positions and momenta need to be specified. The full system of equations stated explicitly is, \vspace{-30pt}

{\centering
\be
\begin{split}
\dot{r}= \frac{p_r}{g_{rr}} ,\quad &\quad\dot{p_r}=-\frac{\partial \mathcal{H}}{\partial r},
\\
\dot{\theta}=\frac{p_\theta}{g_{\theta\theta}} , \quad&\quad\dot{p_\theta}=-\frac{\partial \mathcal{H}}{\partial \theta},
\\
\dot{t}= \frac{E g_{\phi\phi}+L g_{t\phi}}{\mathcal{D}} , &\quad\dot{p_t}=0,
\\
\dot{\phi}=-\frac{L g_{tt} + E g_{t\phi}}{\mathcal{D}} ,  &\quad\dot{p_\phi}=0.
\end{split}
\ee
}
This system is accompanied by a set of initial positions $(t(0),r(0),\theta(0),\phi(0))$ and initial momenta $(p_t(0),p_r(0),p_{\theta}(0),p_{\phi}(0))$, chosen so as to satisfy the condition $\mathcal{H}=0$ for photons. In general, the choice of initial positions depends on the specifics of the problem, while for the momenta we have $p_t(0)=-E$ and $p_{\phi}(0)=L$. Alternatively to the momenta, one can use the velocities $(u^t, u^r, u^{\theta}, u^{\phi})$ where the connection between the two is straightforward (for example, $p_r=g_{rr}u^r$, and $p_{\theta}=g_{\theta\theta}u^{\theta})$. For the integration of photon geodesics, we also define the two impact parameters,    
\be 
  b\equiv- \frac{p_{\phi}}{p_t}=\frac{L}{E}, \quad \textrm{and} \quad \alpha=\frac{p_{\theta}}{p_t}.
\ee 
We note here that $b=L/E$ is the usual definition of the impact parameter for orbits, with positive $b$'s corresponding to co-rotating orbits. For the shadow applications, we will use a modification of this definition, i.e., $b=p_{\phi}/p_t$ which is minus the above definition, that will reflect the fact that we will be considering photons that arrive on the image plane of an observer and therefore the impact parameters will be expressed in that context. 

Throughout this paper, we use SageMath 9.2 \cite{sagemath}, an open source software, in order to numerically integrate the geodesic equations. We introduce the spacetime under study, i.e., the HT spacetime, as a 4-dimensional Lorentzian manifold $\mathcal{M}$ and a coordinate chart on it. We define the metric tensor $g_{\mu\nu}$ in the coordinate frame we set and create a map $\mathcal{M}\rightarrow \mathbb{E}^3$ for graphical purposes, when needed. We declare a starting point $\left(t_0,r_0,\theta_0,\phi_0\right)$ on $\mathcal{M}$ and an initial tangent vector at this point, where we express the four-velocity components in terms of the impact parameters $(b,\alpha)$, with $b$ essentially being the apparent displacement of the image perpendicular to the projected axis of symmetry while $\alpha$, the displacement parallel to the axis. The implemented geodesic integrator invokes the LSODA algorithm that automatically selects between the implicit Adam method and a method based on backwards differentiation formulas (for stiff problems).\footnote{The SageMath integrator maintains the conserved quantities to the order of a few $\times 10^{-8}$. The calculations have been also done using Mathematica and a completely independent implementation that uses a 4th order implicit Runge-Kutta, which is symplectic. The accuracy of the integration in this case, as indicated by the conservation of $\mathcal{H}=0$, is at the level of $10^{-8}$ or better.}

\begin{figure}[ht!]
    \centering
    \includegraphics[width=0.6\textwidth]{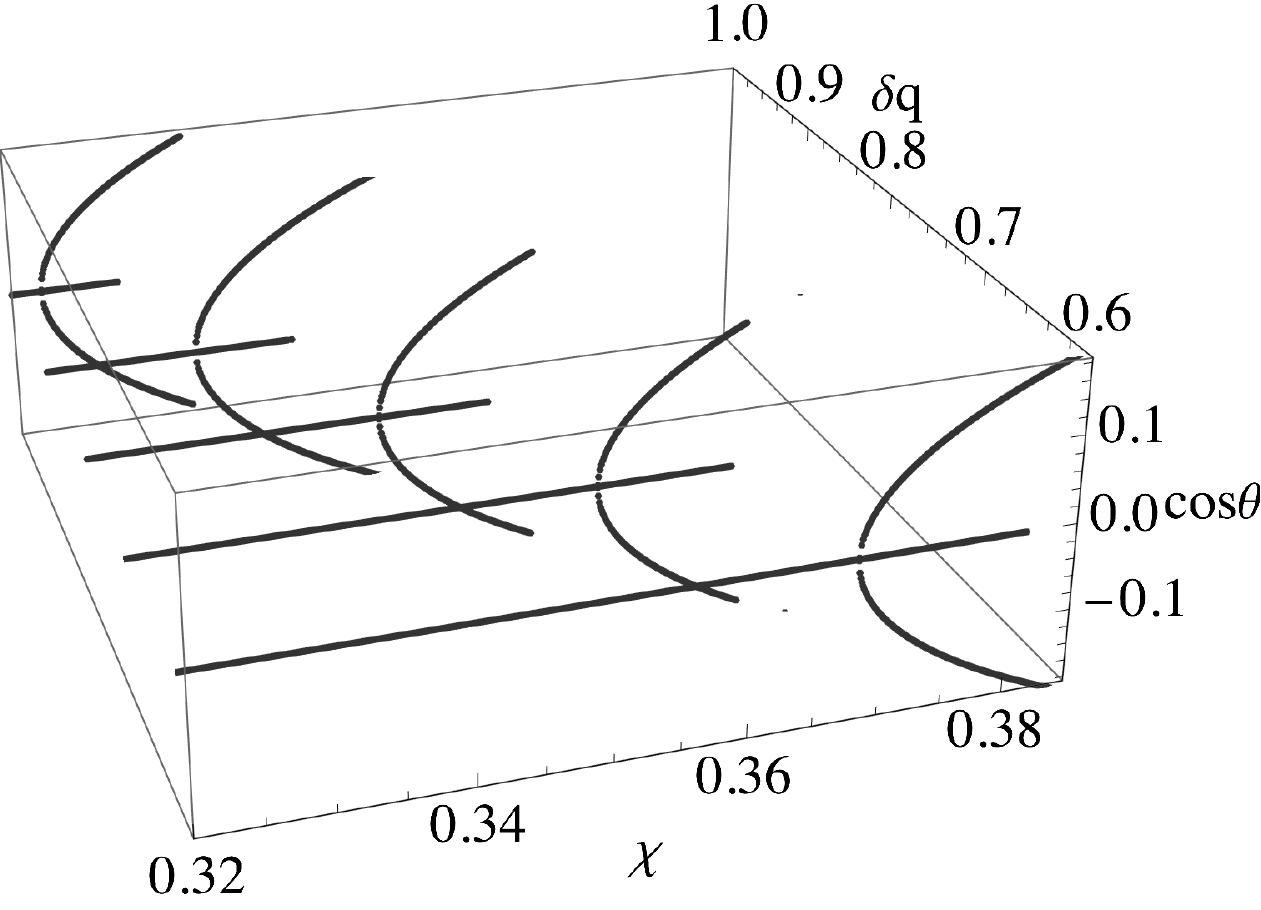}%
    \caption{The light-ring bifurcation that forms the pockets can be seen here for different values of $\delta q$. The plot shows $\mu=\cos\theta$ as a function of $\chi$. The triple light-ring cases can be clearly seen. Beyond some value for the spin, the equatorial ring is lost and only the two non-equatorial rings remain.}
    \label{fig:bifurc1}
\end{figure}
%
As it was shown in \cite{Glampedakis_Pappas2019}, the HT spacetime can have photon orbits and light-rings with interesting properties that can be characteristically different from those of a Kerr or Kerr-like BH. For negative values of the quadrupole deviation parameter $\delta q$, HT has Kerr-like equatorial light-rings. However, for appropriate positive values of $\delta q$ and above some critical spin parameter $\chi_c(\delta q)$, the co-rotating equatorial light-ring bifurcates to a symmetric pair of non-equatorial rings. The two off-equatorial light-rings together with the equatorial one, coexist for a short range of the spin parameter $\chi$ and can form pockets that trap photon orbits. This is a general feature of the HT metric that carries to higher order expansions, at least up to $\mathcal{O}(\Omega^3)$ (see \ref{sec:App:2}). Beyond the parameter range where all three rings are present, the equatorial co-rotating light-ring disappears and the two off-equatorial rings move to higher latitudes. Fig. \ref{fig:bifurc1} shows this behaviour of the light-rings as a function of spin $\chi$, for different values of the deformation $\delta q$. As the figure shows, the bifurcation and triple light-ring phenomenon is present in a relatively wide region of the parameter space. The range between $(0,1)$ for $\delta q$ is relevant for the description of black hole mimickers. Calculations for gravastars for example have shown that for a compactness between $0.4$ and the black hole limit of $0.5$, $\delta q$ is within the range of $(0,1)$, with the value $\delta q\rightarrow0$ corresponding to the black hole limit \cite{Glampedakis2018PhRvD}. In our analysis, we will use as an indicative deformation the value $\delta q=1$. The photons that propagate in the vicinity of both the pocket and the non-equatorial light-rings will be the focus of our interest.

\section{Chaotic orbits in the Hartle-Thorne spacetime}
\label{sec:4}

With general relativity being a strongly nonlinear theory, chaos is often more visible, either emerging even for systems that are integrable in Newtonian theory, such as the case of a fixed binary system \cite{Contopoulos1993NewtonianAR,Lynden-Bell:2002coq,Shipley-Dolan2016} (also known in Newtonian theory as the ``Euler'' problem \cite{euler1767probleme,euler1766motu,euler1767motu,LLbook,Apostolatos:2013wja,Eleni:2019wav}), or being amplified such as in the case of a Schwarzschild black hole endowed with a quadrupole \cite{Letelier2001}. 
Chaos in GR can be studied by investigating the geodesic dynamics of test particles in a given spacetime. This is done by exploring the stochasticity induced when perturbations are introduced in a given spacetime. A straight forward way of doing this is by endowing the background spacetime with additional higher order multipoles (i.e., higher than the angular momentum) and chaos has been found this way in several interesting cases \cite{Sota:1995ms,ChenWang2003,Gueron:2001ex,Zelenka:2017aqn,WenBiaoHan2010,Letelier2001}, where it is demonstrated that the relativistic dynamics are more chaotic than their respective Newtonian counterparts. Since the HT metric is a non-Kerr rotating spacetime with higher order moments deviating from those of a Kerr BH, its geodesics are non-integrable and therefore chaos may arise in the dynamics of test particles. This section explores the chaotic behaviour that can be found in the orbits of photons in the vicinity of such a central compact object. The goal is to eventually explore, if chaos exists, how it would affect the shadow of such an object, which we do in the next section.

As a final note we should point out that in order to detect chaos in GR, one has to take into account the fact that time is now tied to the coordinate system and no longer absolute, as in the Newtonian case, and therefore coordinate-independent methods are needed \cite{Lukes2014}. As such methods, we will mainly employ the use of Poincar\'e sections.

\subsection{Chaos Detection and Characterization}
A Poincar\'e section is a lower-dimensional subspace of the phase space of a dynamical system. In stationary and axisymmetric spacetimes, the energy $E=-p_t$ and the angular momentum $L_z = p_\phi$ are integrals of motion. This allows us to study a simpler system with two degrees of freedom, i.e., the meridian plane $(r,\theta)$ \cite{Zelenka:2017aqn}. The physical trajectories of a system with two degrees of freedom, lie on the three-dimensional energy surface $H(p_1,p_2,q_1,q_1)=H_0$ in phase space and, for bounded motion, after a long time interval, the solution curves will intersect any 2D-plane repeatedly \cite{CHEBTERRAB1996171}. When an orbit is stable, these intersection points lie on smooth curves that, as time flows, draw surfaces in the phase space. For perturbed systems, these surfaces made of solution curves, break their topology forming "island" chains near resonances. Within these ''islands'' the topology can be again broken to other chains and so on, forming what are called Birkhoff chains. These surfaces, called Kolmogorov-Arnold-Moser or KAM surfaces, isolate thin layers of chaos. As the perturbation increases, the transitions between the layers merge and this results in very complicated structures in the phase space \cite{Zelenka:2017aqn}. These complicated structures are the driving engines of deterministic chaos. We should mention that Poincar\'e sections can be drawn either using the momenta $p_a$ or equivalently the velocities $u^a$.

Once chaos is found in a system, it has to be characterised in order for its specific properties to be extracted. An often used method to measure chaos is through the power spectrum  $P_z(\omega)$ of the time series of the $z$ component of the particle's position, that is defined as
\begin{equation}
    P_z(\omega)=\left |  \int_{0}^{T} z(t)e^{i\omega t} \,dt \right |^2.
\end{equation}
The pattern of the power spectrum can either be that of white noise or obey some power law. Koyama et. al \cite{Koyama2007} found that the power spectrum obeys such a power law, called \textit{1/f fluctuations}  or \textit{pink noise}, when a chaotic orbit stagnates in the vicinity of periodic orbits for long time intervals. These orbits are called \textit{sticky}. The \textit{stickiness} phenomenon, first reported by Contopoulos \cite{Cont19701,Cont19702}, is a dynamical property of some Hamiltonian systems that emerges from the coexistence of regular and chaotic dynamics. This coexistence creates regions that act like fractal scattering zones near the boundaries of islands where chaotic trajectories are forced to behave regularly \cite{SantosMoises2019}.

Another useful tool that allows us to quantitatively study the characteristics of the system is the \textit{rotation number}. 
In order to calculate the rotation number we first have to identify a central invariant point of the section, usually on the $p_r=0$ axis and at the centre of a tori. We then successively measure the angle between two vectors that join the invariant point to the piercings that create the two dimensional tori. We sum up these angles for each orbit and the rotation number is
\begin{equation}
    \nu_\theta = \lim_{N\to\infty} \frac{1}{2\pi N}\sum^N_{j=1}\theta_j.
    \label{eq:rot}
\end{equation}
This number characterises the frequency structure of the phase space for each trajectory \cite{Cardenas2018}.
In this paper, we use a variation of the definition \eqref{eq:rot}, that will indicate to us if an orbit is indeed sticky. We pick the centre of our section in $(r,u^r)$ as the invariant point and we consider two vectors in the phase space joining it to two successive piercings. We proceed by calculating the average angle $\theta_{N\textrm{avg}}$ of $N$ consecutive piercings, for a single orbit, and we plot the final curve as a time series, expecting that when the motion is regular the value will be constant, indicating thus the trapping of an orbit. This value depends on the position of the island in phase space, around which the orbit is trapped.

We will employ these three tools in order to study the properties of photon orbits in the HT spacetime, that pass near the central compact object and interact with the feature of interest, i.e., the pocket formed by the three light-rings. In general, when we will refer to time in the evolution of these orbits we will mean the affine parameter of the geodesics or alternatively the ``affine time'', unless otherwise stated.

\subsection{Light Trapping Region}
\label{sec:ltr}
%
The shape of the zero-velocity separatrix (just {\it separatrix} from this point forward) depends on the impact parameter $b$ and the two free parameters of the HT metric, the spin parameter $\chi$ and the quadrupole deviation $\delta q$ (the mass $M$ is only a scale which we will set to $M=1$ length unit). For $\delta q=1$, $b\in [4.3M,4.31M]$ and $\chi \in [0.3225,0.328]$, all three photon ring solutions coexist \cite{Glampedakis_Pappas2019}. Our goal is to form and study a pocket disconnected from the compact object so that we can study photons that will be initially trapped for long periods of time in the pocket but eventually be able to escape to infinity. 

In order to determine at what stage of the pocket narrowing the photons can be trapped, we fix the spin parameter to $\chi=0.327352$ and integrate geodesics for some values of the impact parameter $b$ in the interval $b\in (4.3M,4.304332M)$, while measuring the average length of coordinate time $\Delta t$ a photon spends in the pocket. 
The first instance where we find orbits that can be characterised as temporarily trapped, occurs at $b=4.30378M$. 
As the impact parameter $b$ increases, so does the average coordinate time and the percentage of the orbits that get trapped, peaking at $\Delta t \approx 16000M$ for $b=4.304332M$, where the photons never leave the area for the given integration time. The exterior gets cut off at $b=4.304333M$.

\begin{figure}[h]
\centering
\begin{subfigure}[l]{.3\textwidth}
  \centering
  \includegraphics[width=0.99\textwidth]{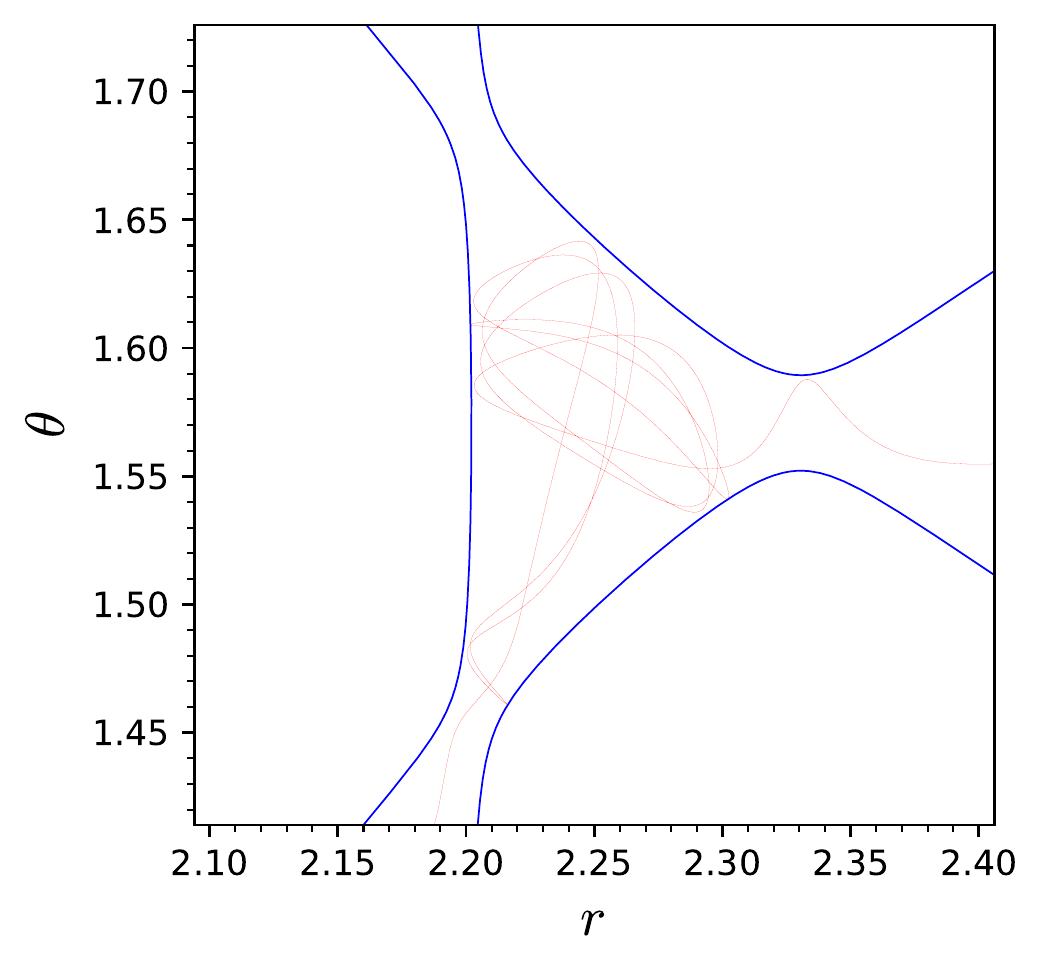}
  \caption*{ }
\end{subfigure}%
\begin{subfigure}[c]{.3\textwidth}
  \centering
  \includegraphics[width=0.99\textwidth]{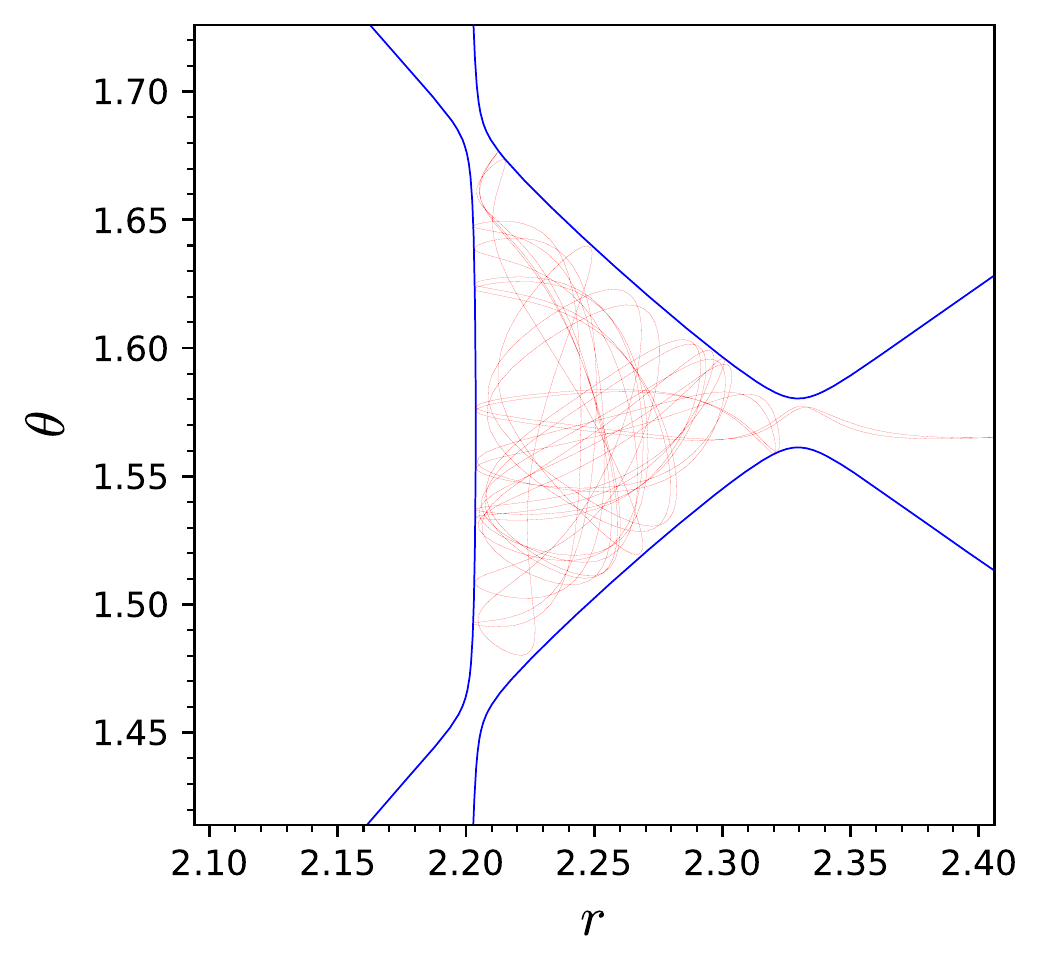}
  \caption*{}
\end{subfigure}
\begin{subfigure}[r]{.3\textwidth}
  \centering
  \includegraphics[width=0.99\textwidth]{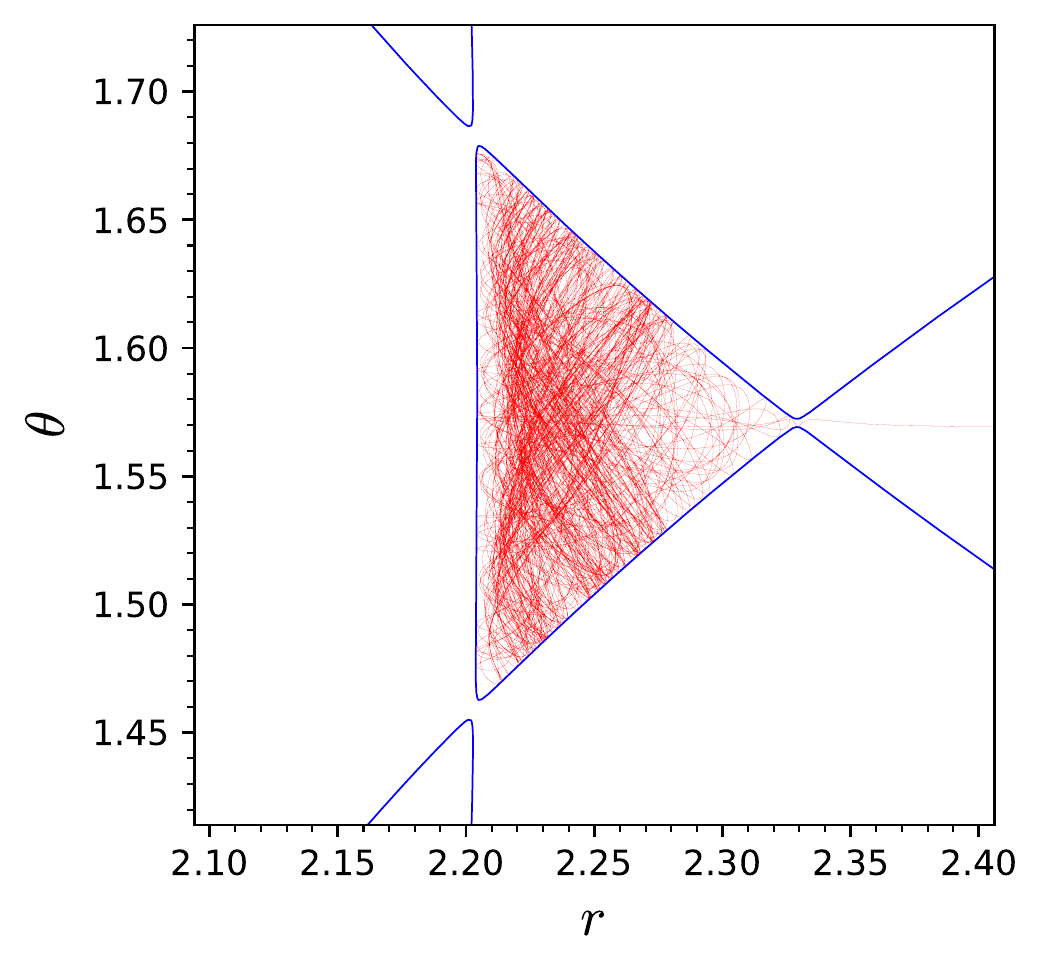}
  \caption*{}
\end{subfigure}
\caption{Left: Open pocket for $b\approx4.30405M$, the photon plunges onto the surface of the object through the lower throat after $\Delta t \approx 700M$. Middle: Open pocket for $b\approx 4.30426M$, the photon is temporarily trapped for $\Delta t \approx 4200M$, before escaping to spatial infinity. Right: Marginally Open pocket for $b=4.304332M$, the light ray enters the area and stays trapped for the remaining time of the integration, reaching $\Delta t \approx 16000M$. The pocket is now disconnected from the compact object but remains connected to infinity through a narrow throat.}
\label{fig:twosepswithrays}
\end{figure}

It is interesting to see how the orbit's characteristics change as one changes the pocket from open to almost closed. For this we present in Fig. \ref{fig:twosepswithrays} the orbits for three pocket ``sizes''. For $b \approx 4.30405M$ the pocket is relatively open and a photon with $\alpha=0.061M$ is trapped for a short period of time before falling into the object, reaching $\Delta t \approx 700M$. As $b$ increases to $b\approx 4.30426M$ the pocket is closer to being ``closed'' and for a photon with $\alpha=0.022M$, the period of time it remains in the pocket significantly increases, becoming $\Delta t \approx 4200M $, before escaping to infinity. 
Finally for $b=4.304332M$, the allowed region just disconnects from the object, i.e., the upper and lower throats close, and we find the longest coordinate time periods of trapping for photons 
which can be as long as the full integration interval ($\Delta t \approx 16000M$). %
In general, the range of spins for which we have 3 light-rings is relatively narrow. The same applies for the range in $b$ for which we have a trapping pocket (2 inner throats closed with the 3rd outer one narrowly open). Therefore one needs to choose such a spin so as to have the inner throats closing first with changing $b$ while the outer throat is still narrowly open so as to form a marginally open pocket. In the following subsections we will use a value for the spin of the central object equal to $\chi=0.327352$, as we did here, in order to have the 3 light-rings and the pocket.

\subsection{Open HT System}
\label{openht}

For our fixed value of the spin and for impact parameters up to $b\approx 4.30427M$, the system has two ``throats'' connecting the pocket to the compact object and one throat connecting it to infinity. It is in that sense that our system can be characterised as an open Hamiltonian system with three escapes. A similar, with respect to these features, system that has been thoroughly studied by Shipley and Dolan \cite{Shipley-Dolan2016} is that of a Majumdar-Papapetrou di-hole which was in turn found to strongly resemble that of H\'enon and Heiles \cite{Henon-Heiles-1964AJ,Zotos2014,Zotos2017}. 

\begin{figure}[H] 
 \centering
  \begin{subfigure}[b]{0.24\linewidth}
    \centering
         \includegraphics[width=0.96\textwidth]{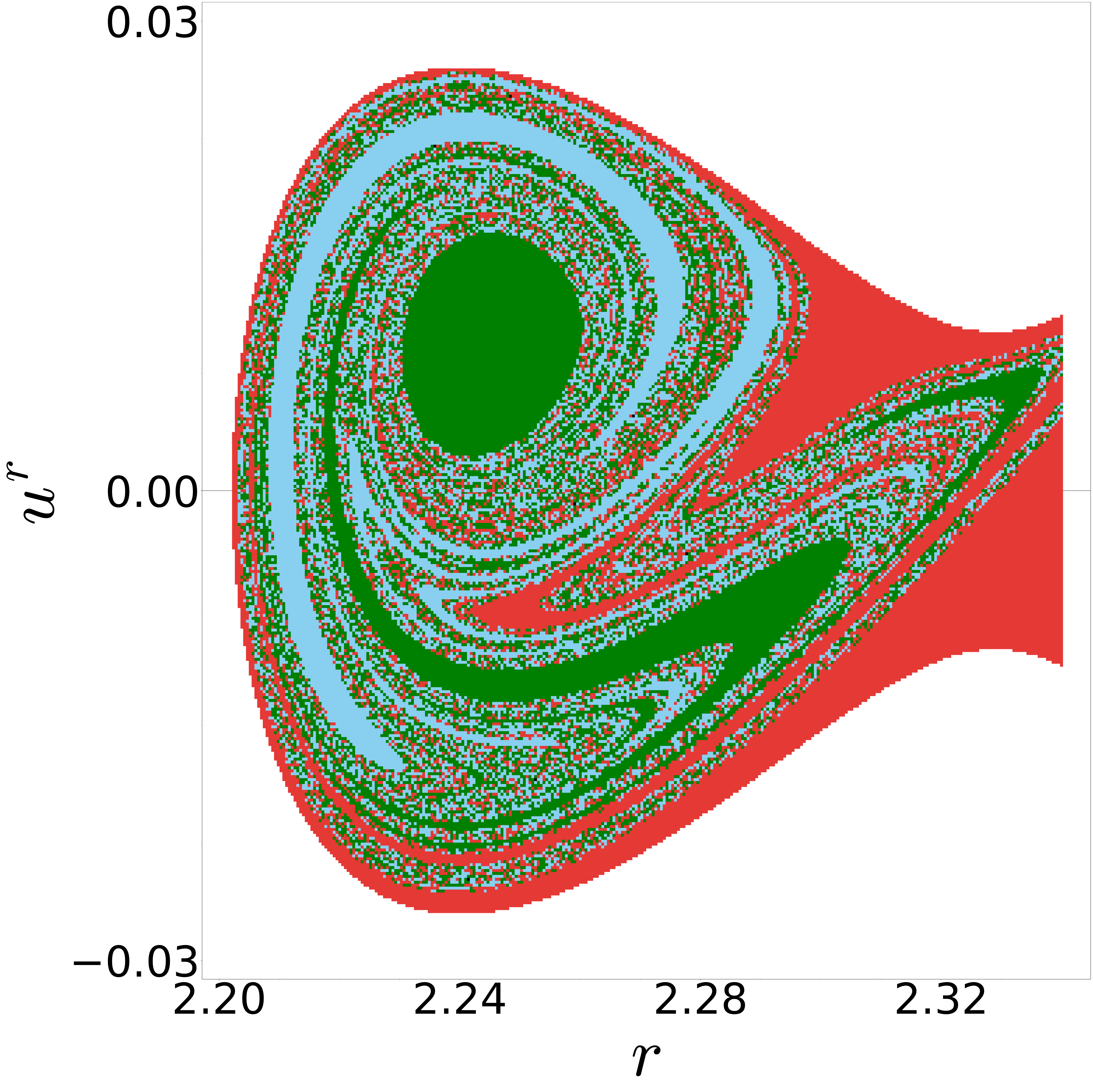} 
    \caption*{} 
    \label{} 
  \end{subfigure}
  \begin{subfigure}[b]{0.24\linewidth}
    \centering
      \includegraphics[width=1\textwidth]{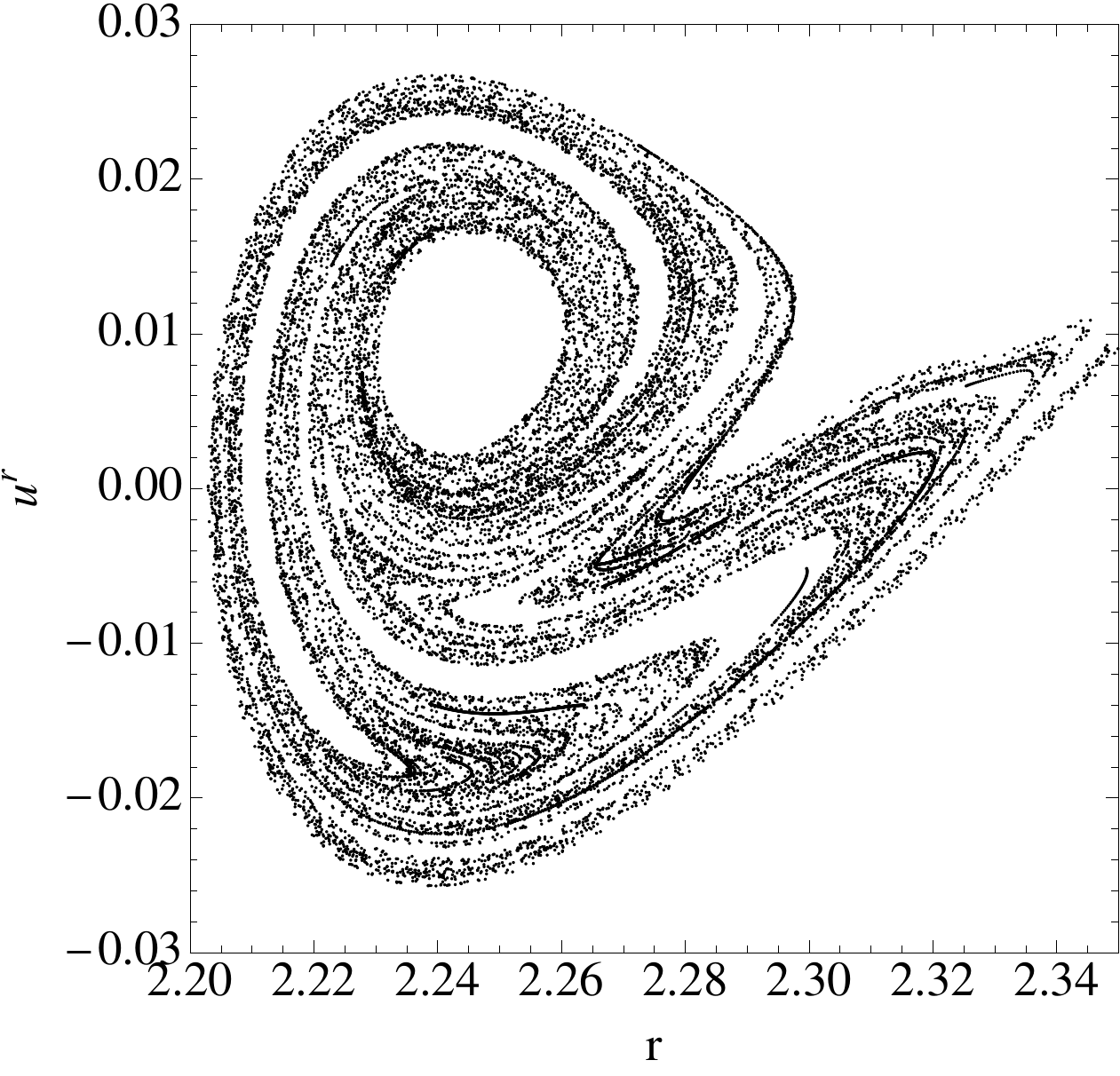} 
    \caption*{} 
    \label{} 
  \end{subfigure} 
    \begin{subfigure}[b]{0.24\linewidth}
    \centering
          \includegraphics[width=0.96\textwidth]{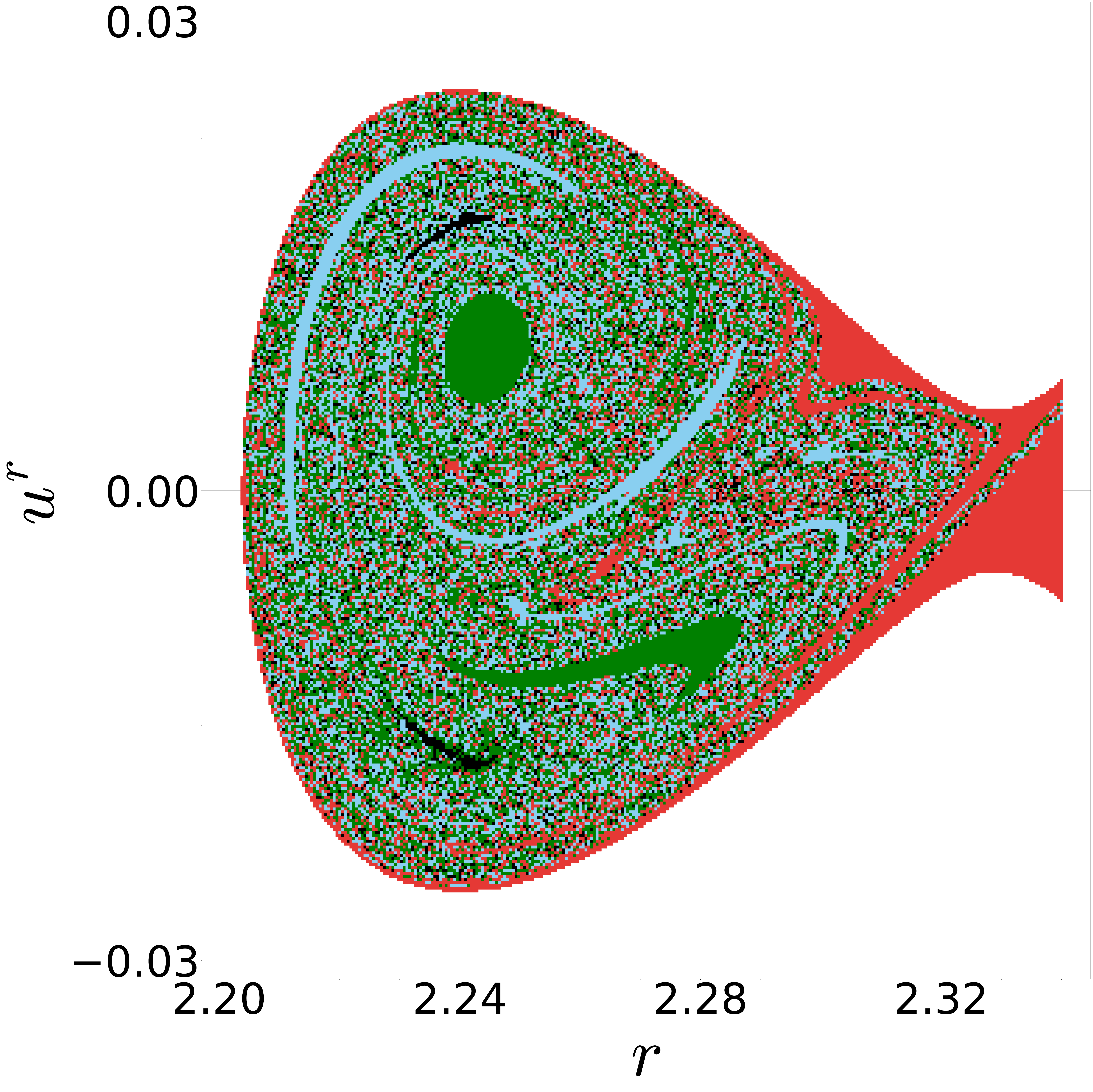}  
    \caption*{} 
    \label{} 
  \end{subfigure}
  \begin{subfigure}[b]{0.24\linewidth}
    \centering
    \includegraphics[width=1\textwidth]{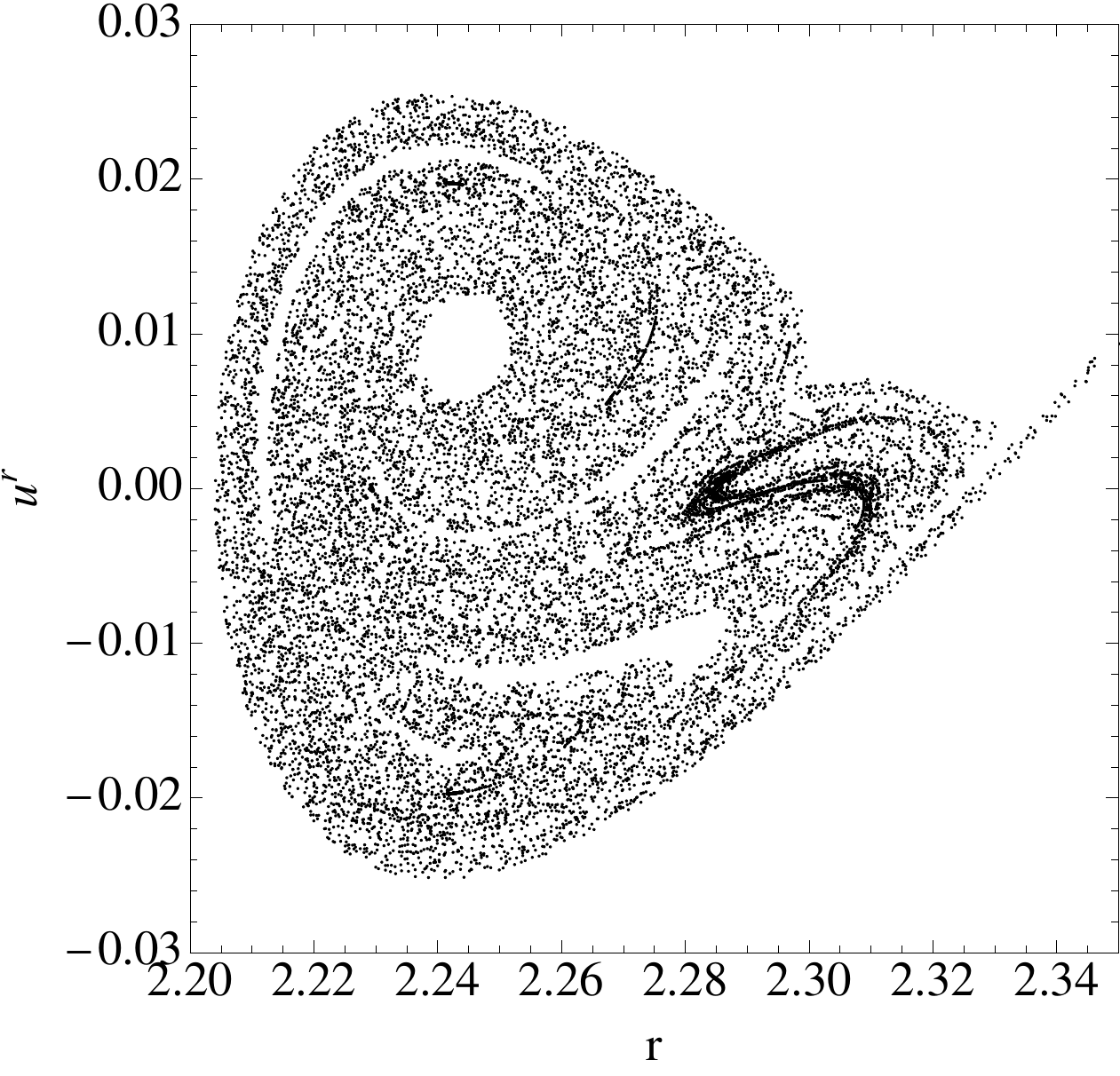} 
    \caption*{} 
    \label{} 
  \end{subfigure} 
    \caption{Left 2 plots ($b\approx4.30405M$): Exit basins for the open HT system that has three escapes and the corresponding Poincar\'e section.
    Right 2 plots ($b\approx 4.30426M$): The Exit basins have shrank and one can also see the trapping regions as black points.}
  \label{fig:basins1}
  \end{figure}

When multiple escapes exist, one can define the set of initial conditions in phase space that lead to a particular escape channel as \textit{exit basins} \cite{Zotos2017}. The details of what this is will become clear in the following examples. To help with the visualisation of the exit basins we also use Poincar\'e sections. 
The exit basins in the  HT spacetime (we remind that $\chi=0.327352$ and $\delta q =1$) will depend on the value of $b$, which determines the shape of the separatrix. %
In Fig. (\ref{fig:basins1}), 
we show two exit basins for two values of the impact parameter, $b\approx 4.30405M$ and $b\approx 4.30426M$, along with the corresponding Poincar\'e sections. 

We colour coded the initial conditions as red when the null rays escape to infinity, green when they plunge onto the compact object through the upper throat, light-blue for the lower one and black for those that are trapped for the entirety of the integration interval. 
The exit basins can generally be broad and well defined or extended with a complicated structure that contains fractal regions \cite{Zotos2017}. In these regions the escape throat can not be predicted. What we see in the two examples presented here is that in the first case we have well defined exit basins of blue, green, and red, which as $b$ increases (and the throats become narrower), shrink while the mixed/fractal regions expand. In the second case with higher $b$ we can see that a trapping region is formed along with the first islands of stability, represented by black pixels in the third plot of Fig.(\ref{fig:basins1}), %
where the orbits get trapped and do not escape the pocket through any of the exits. One can also notice in both cases that the fractal regions of the exit basins correspond to the regions of the Poincar\'e sections that are populated with points, while the well defined regions of the exit basins leave the corresponding areas of the Poincar\'e sections empty (since these orbits escape). 

\subsection{Marginally Open HT System}
\label{sec:marg}

We now turn our attention to the case where the pocket has been disconnected from the compact object leaving only one throat connected to infinity, that is admitted for $b=4.304332M$ (plot on the right of Fig. \ref{fig:twosepswithrays}). The time periods that the photons spend in the pocket in this case are long enough to get a good sense of their qualitative phenomenology. 
%
\begin{figure}[h]
\centering
\begin{subfigure}[h]{0.45\linewidth}
\includegraphics[width=0.9\linewidth]{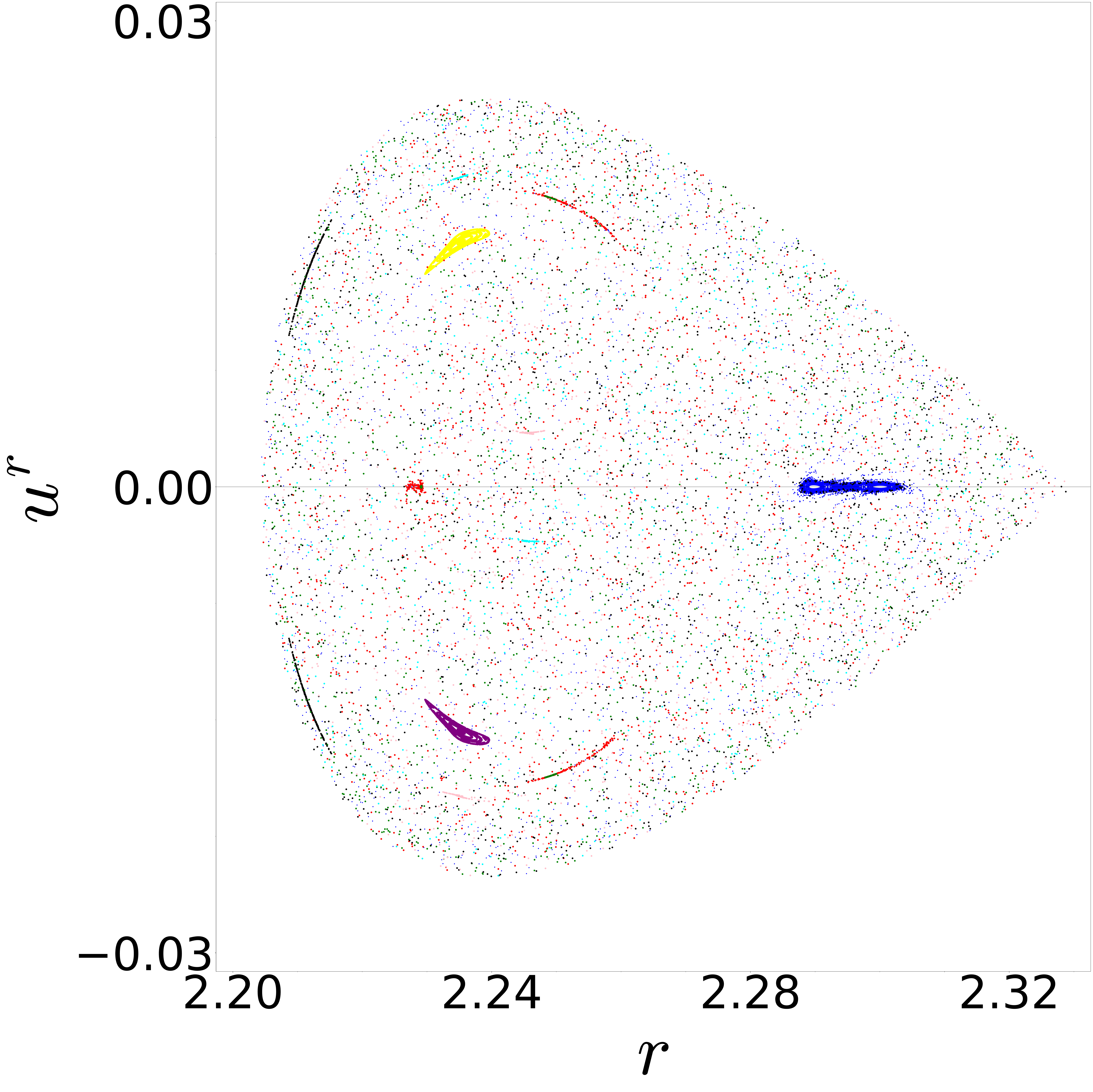}
\caption*{}
\end{subfigure}
\quad
\begin{subfigure}[h]{0.45\linewidth}
\includegraphics[width=0.9\linewidth]{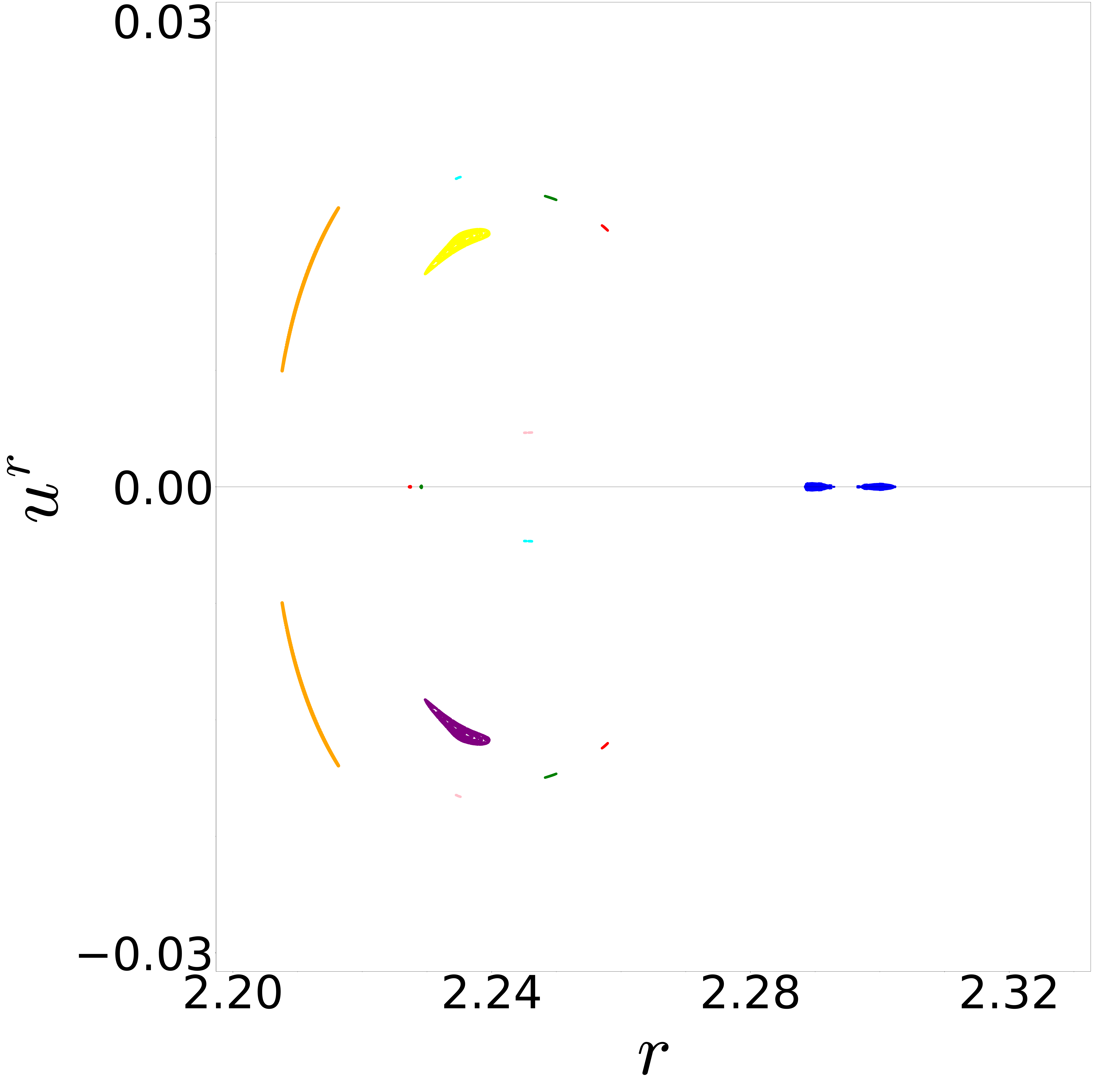}
\caption*{}
\end{subfigure}%
\caption{Poincar\'e section in the $(r,u^r)$-plane for the marginally open pocket.The impact parameter is $b=4.304332M$. Left: The domain is mostly chaotic for the whole available space, with small islands of stability surviving. Right: Just the islands of stability.}
\label{fig:marg1}
\end{figure}
%
We choose the equatorial plane ($\theta=\pi/2$) as the section where the orbits with $u^\theta > 0 $ are recorded. A few hundred light rays with different initial radii and with $u^r(0)=0$ were integrated for $30000M\textrm{-}50000M$ units of the affine parameter. Fig. \ref{fig:marg1} shows the Poincar\'e section for this configuration. As one can see on the plot on the left, the chaotic domain occupies almost the entire available space. However, there are islands of stability surviving in the chaotic sea, shown separately on the right plot of Fig. \ref{fig:marg1}, which we will examine further. We should point out at this instance that these islands of stability signify the existence of stable periodic photon orbits inside the pocket. Although not very well visible here, these stable periodic orbits are associated to accompanying unstable photon orbits. This issue will be discussed further when we will investigate the transition to the chaotic behaviour observed here.

Following \cite{Contop2003}, the different orbits that we will show can be categorised as follows. (i) $n_1 : n_2$ resonant tube orbits: These orbits surround stable resonant periodic orbits and form a structure that is essentially a low order Lissajous figure (low in the sense that $n_1$ and $n_2$ are small integers), and is called a ``tube''. That is, the $n_1 : n_2$ tube orbits exhibit a pattern of oscillations, where the orbit oscillates $n_1$ times in the direction of $r$ and $n_2$ in the direction of $\theta$.  (ii) Chaotic orbits: These orbits have no particular structure and tend to fill the available space.

\begin{figure}[H]
\centering
\begin{subfigure}[h]{0.22\linewidth}
\includegraphics[width=\linewidth]{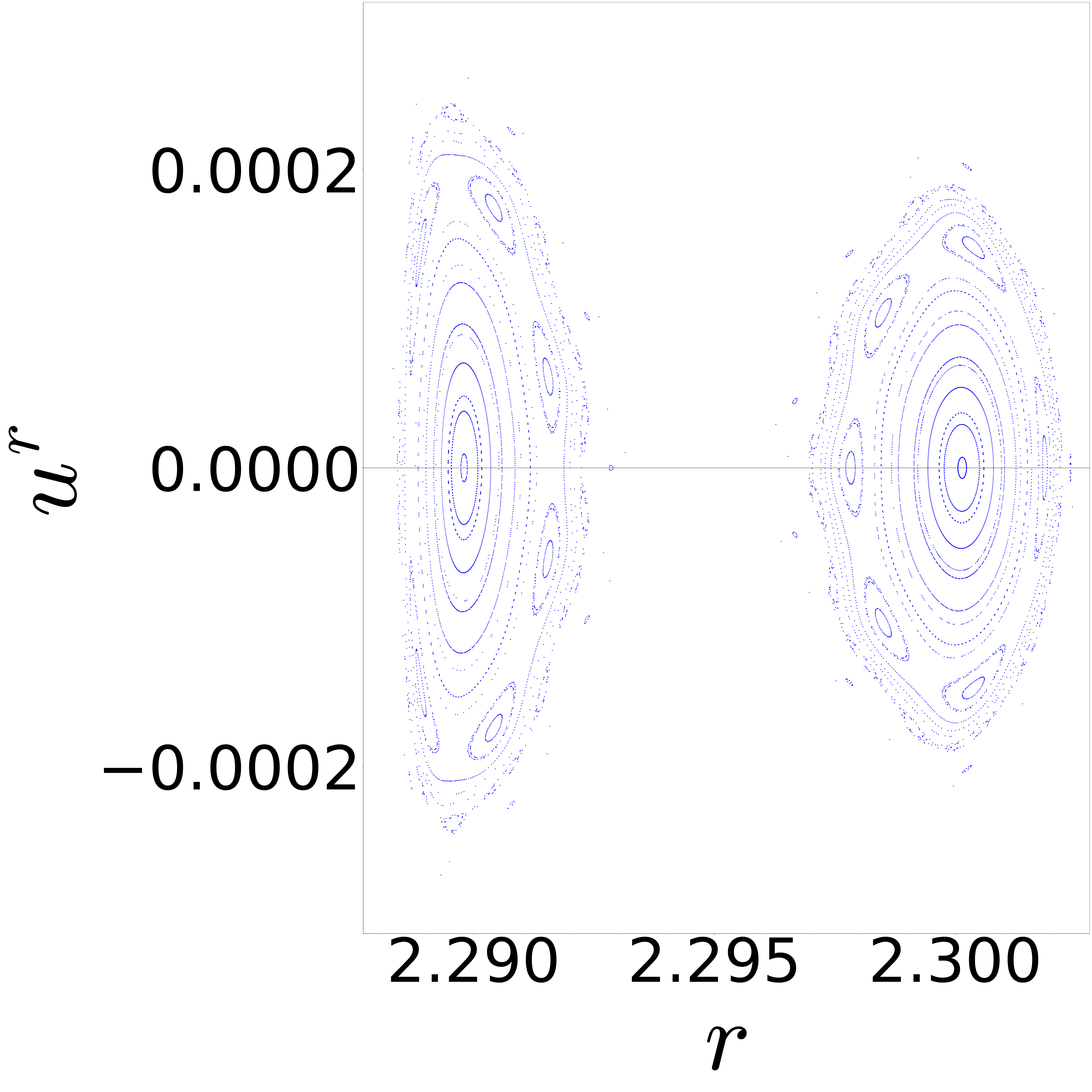}
\caption*{}
\end{subfigure}
\quad
\begin{subfigure}[h]{0.22\linewidth}
\includegraphics[width=\linewidth]{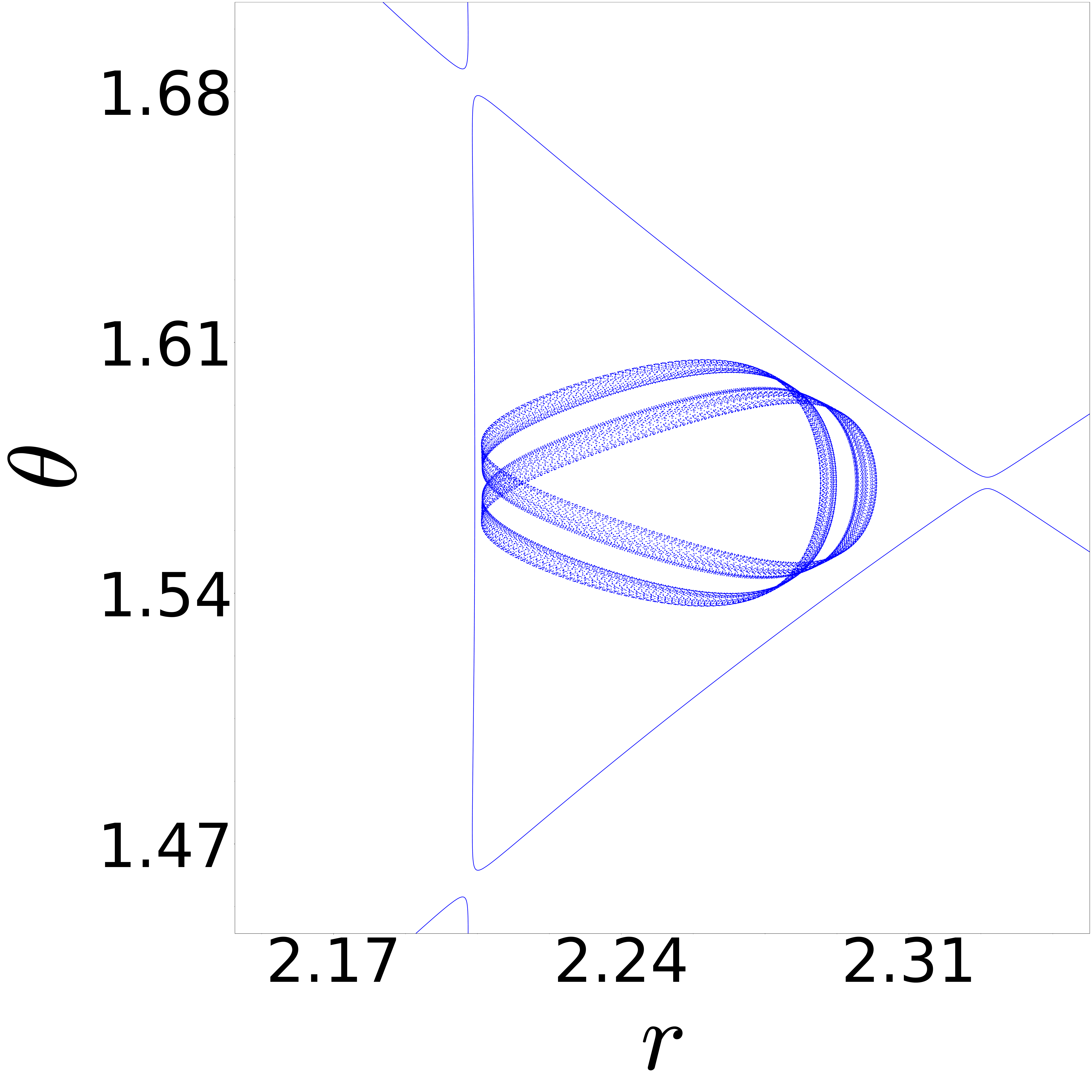}
\caption*{}
\end{subfigure}%
\quad
\begin{subfigure}[h]{0.22\linewidth}
\includegraphics[width=\linewidth]{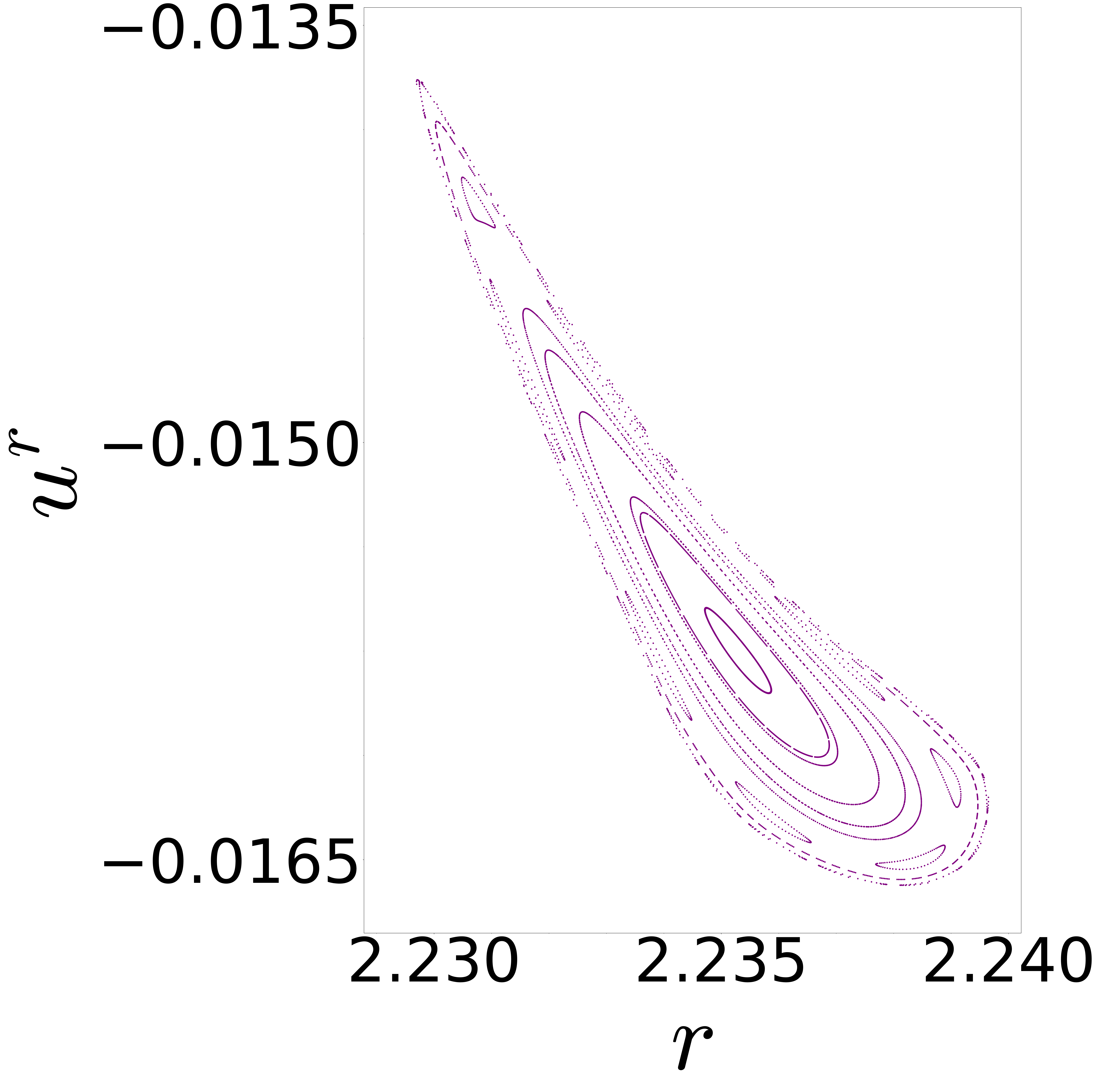}
\caption*{}
\end{subfigure}
\quad
\begin{subfigure}[h]{0.22\linewidth}
\includegraphics[width=\linewidth]{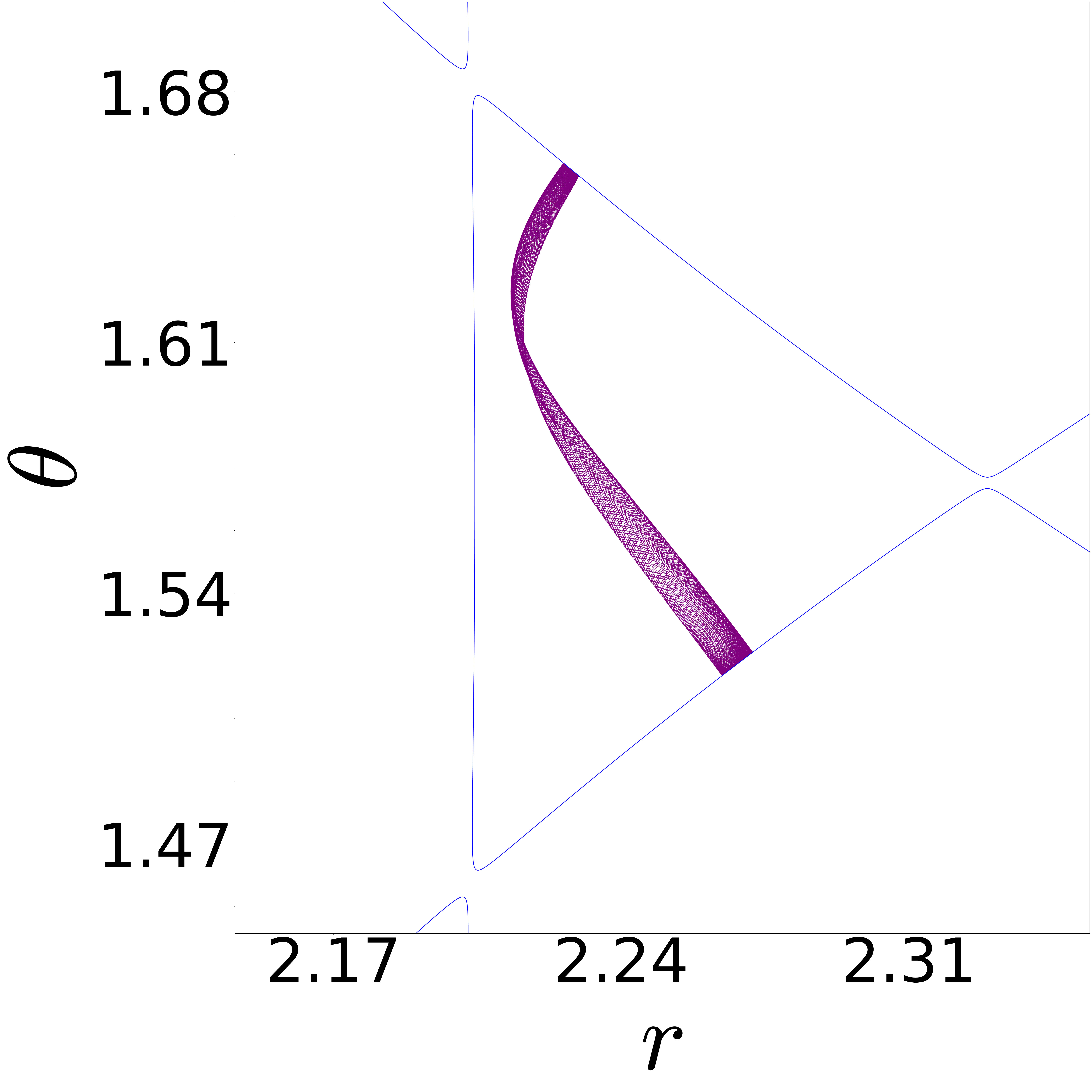}
\caption*{}
\end{subfigure}%
\caption{Two Left plots: A magnification of the islands of stability around two stable fixed points, where one can see the smaller islands around the stable points that form Birkhoff chains with the unstable points, and the corresponding 2:2 tube orbit. Two Right plots: Single island of stability around a stable fixed point with smaller islands of stability with higher periodicity inside the last KAM curve, and the corresponding 2:1 tube orbit.}
\label{fig:margblue}
\end{figure}
%
In Fig. \ref{fig:margblue} we show two islands of stability (left plot) that enclose stable fixed points in their centers. Smaller islands of higher periodic orbits appear to have arisen within the main islands that form Birkhoff chains with the unstable points and narrow stochastic layers seem to exist between the latter \cite{Zelenka:2017aqn}. The last KAM curve seems to have been destroyed and have become a cantorus with infinite gaps allowing the communication of the inner chaotic layers with the chaotic sea. The corresponding form of the orbit is a 2:2 resonant tube orbit. 
Fig.~\ref{fig:margblue} also shows a single island of stability (third plot from the left) with an outer layer of smaller islands of stability. Although the last KAM curve seems broken, something that should allow communication with the chaotic sea, there are no visible chaotic layers inside it which indicate that the breaks are probably due to insufficient integration time. The orbit is a 2:1 resonant tube. This island has a symmetrical counterpart with $u^r>0$, to which it is not connected though (shown in Fig.~\ref{fig:marg1} with yellow), i.e., different orbits populate that island albeit they are reflection symmetric with respect to the equatorial plane to the orbits shown on the right in Fig.~\ref{fig:margblue}.%

\begin{figure}[h]
\centering
\begin{subfigure}[h]{0.25\linewidth}
\includegraphics[width=\linewidth]{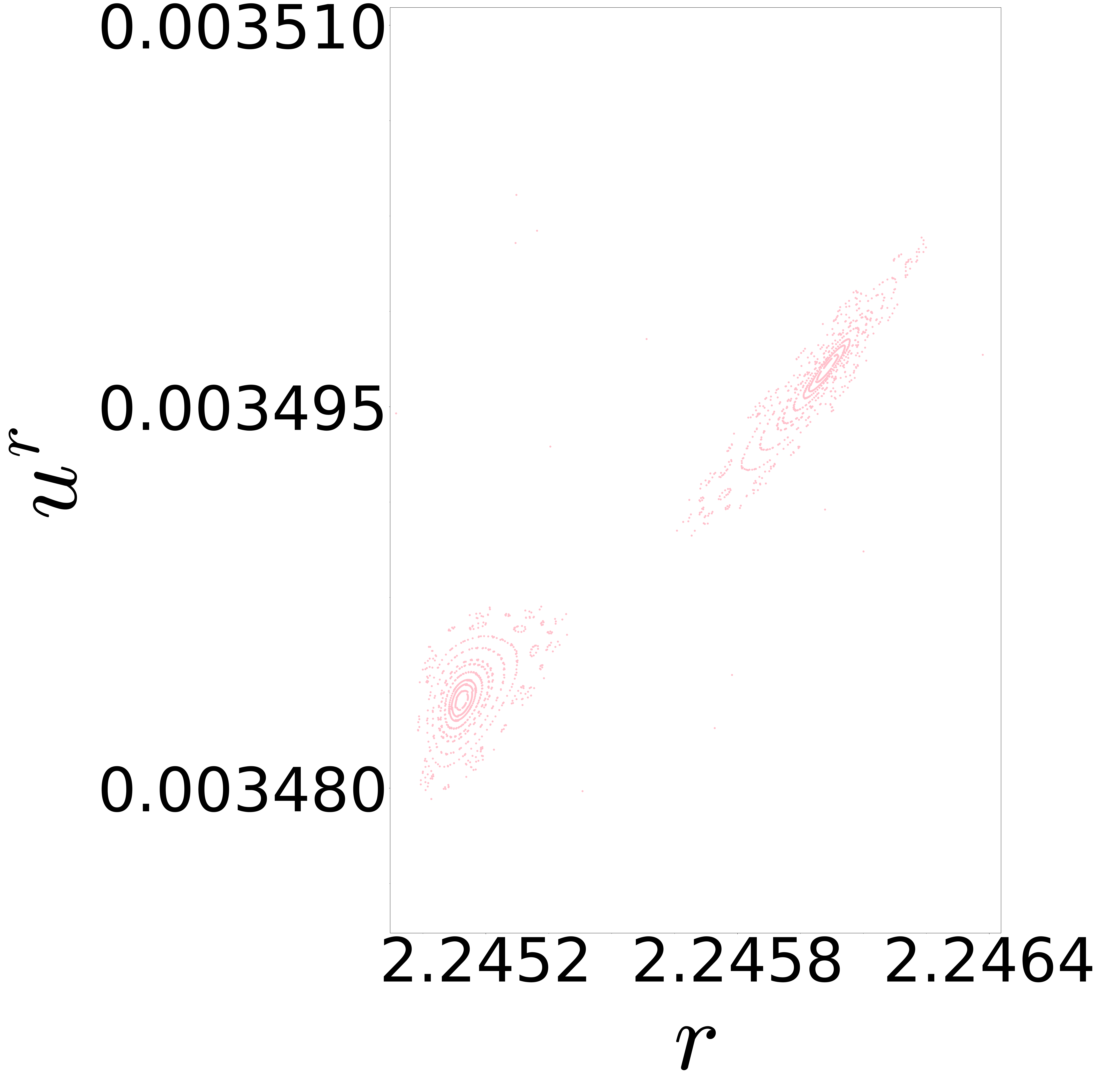}
\caption*{}
\end{subfigure}
\begin{subfigure}[h]{0.25\linewidth}
\includegraphics[width=\linewidth]{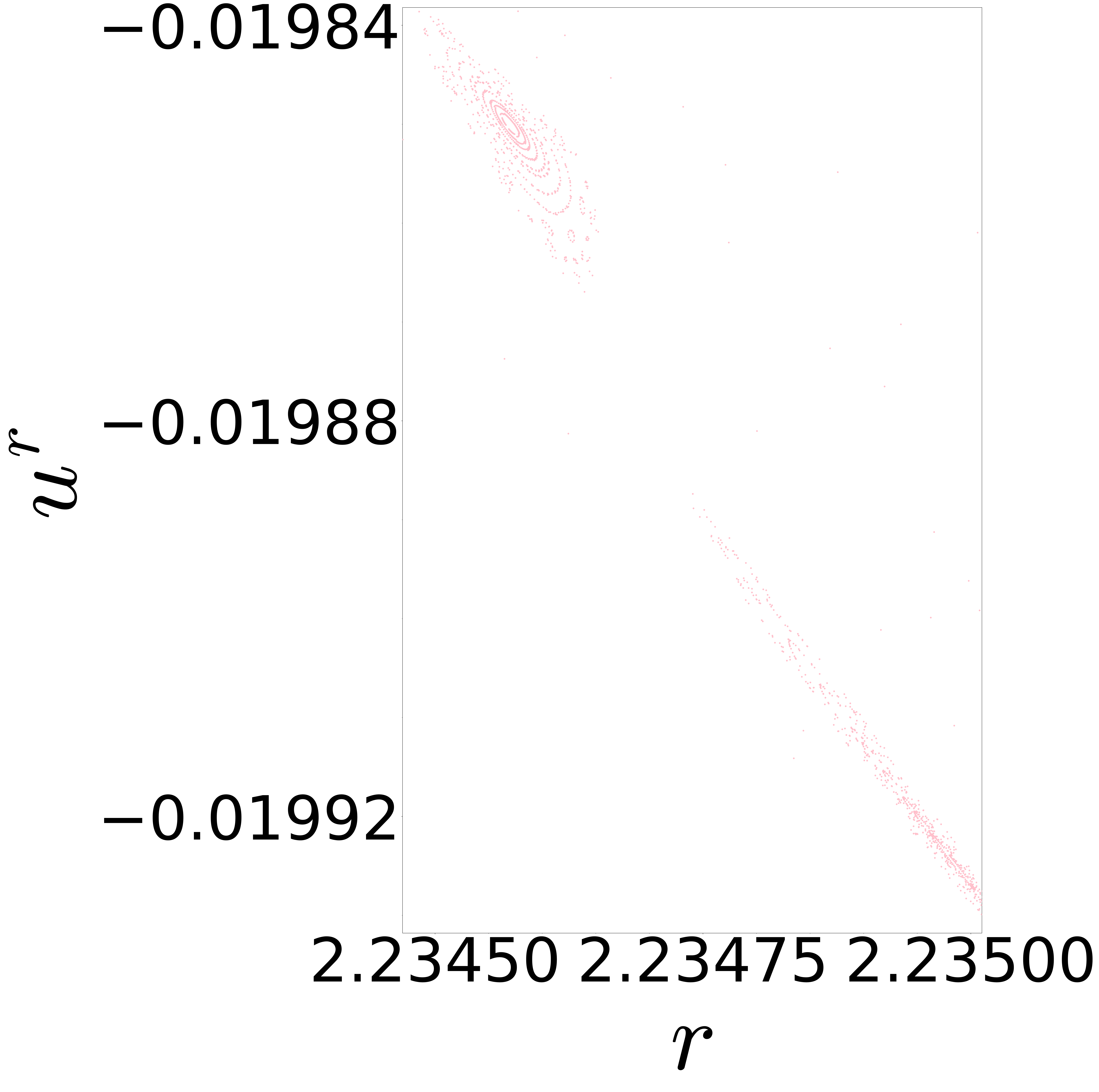}
\caption*{}
\end{subfigure}%
\begin{subfigure}[h]{0.25\linewidth}
\includegraphics[width=\linewidth]{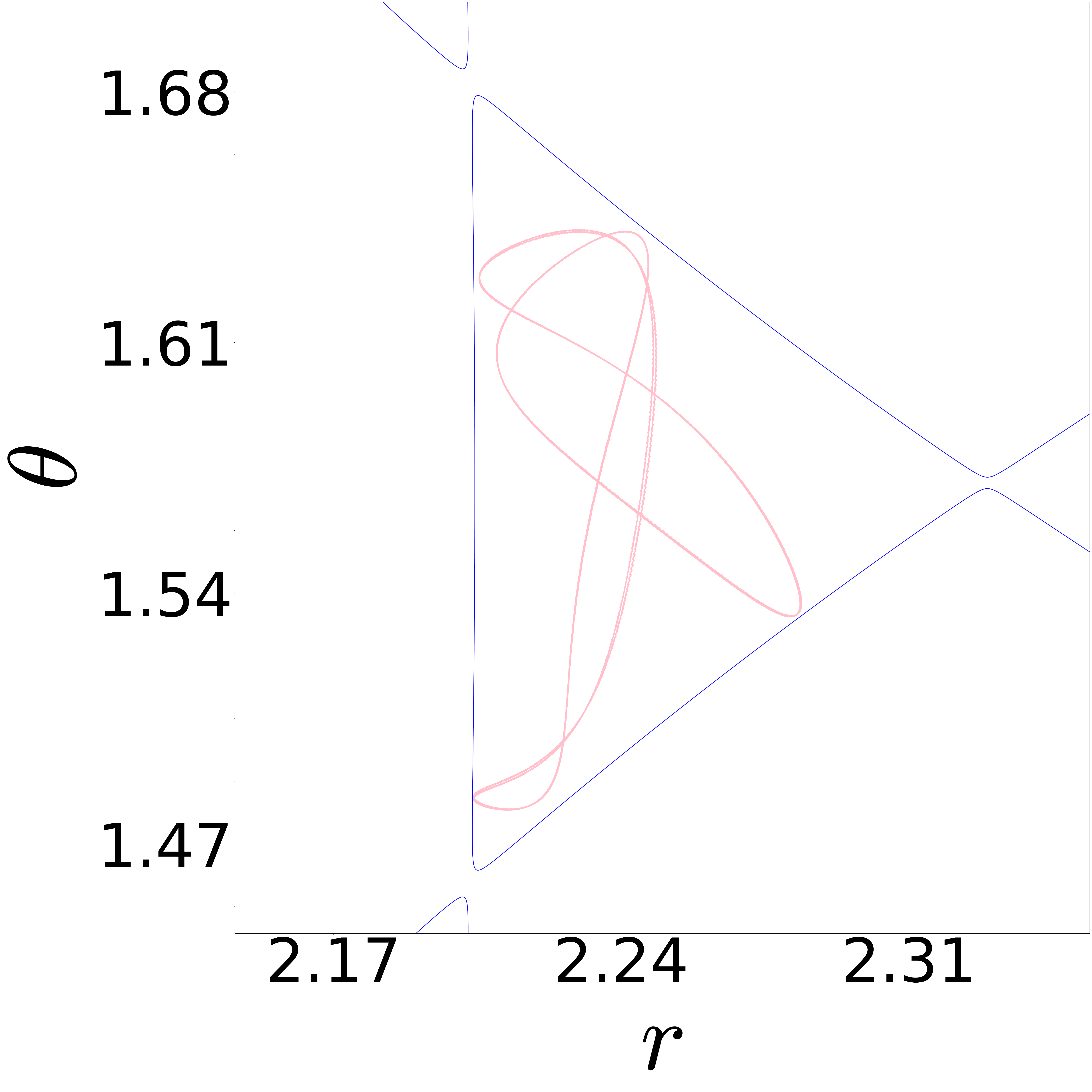}
\caption*{}
\end{subfigure}\\%
\begin{subfigure}[h]{0.25\linewidth}
\includegraphics[width=\linewidth]{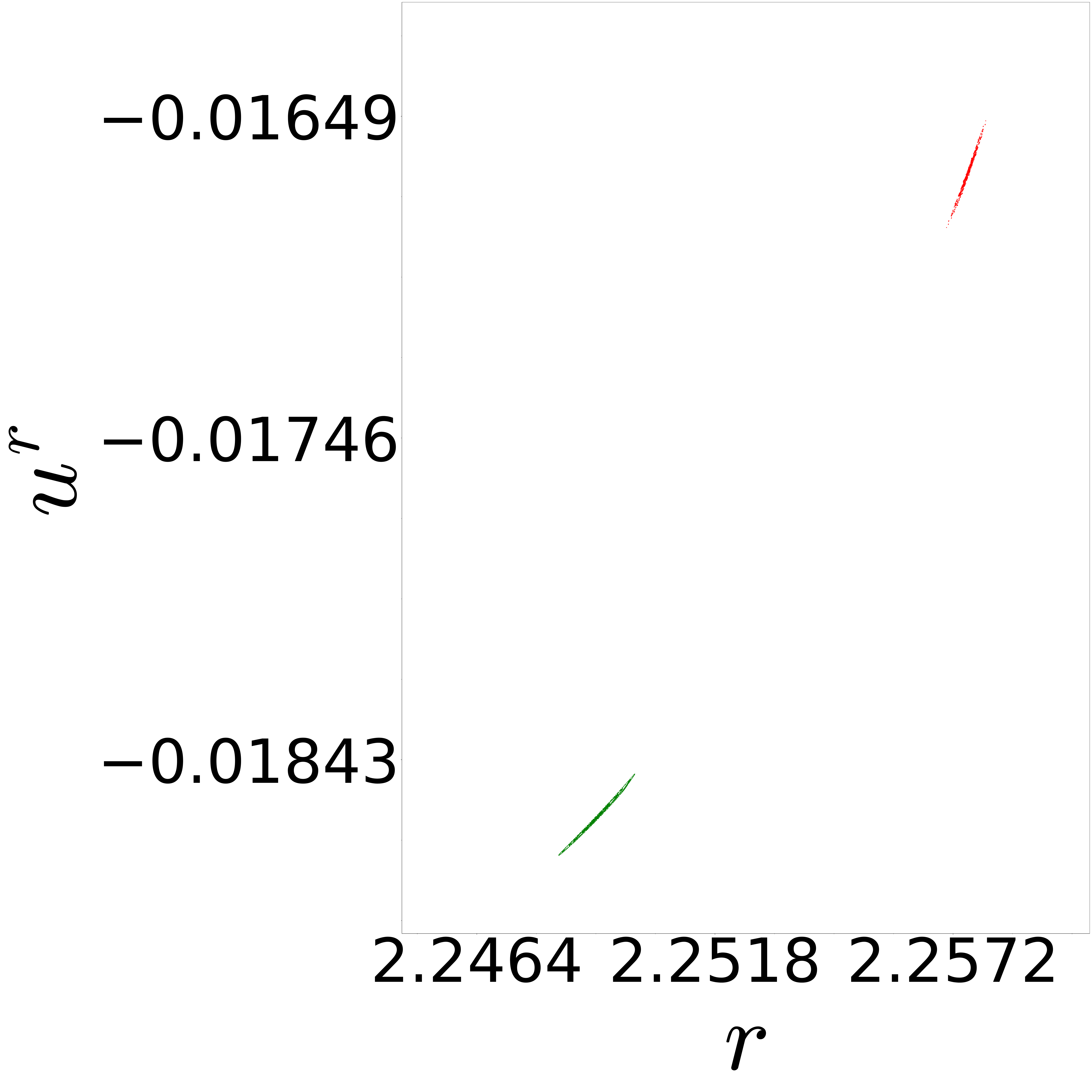}
\caption*{}
\end{subfigure}
\begin{subfigure}[h]{0.25\linewidth}
\includegraphics[width=\linewidth]{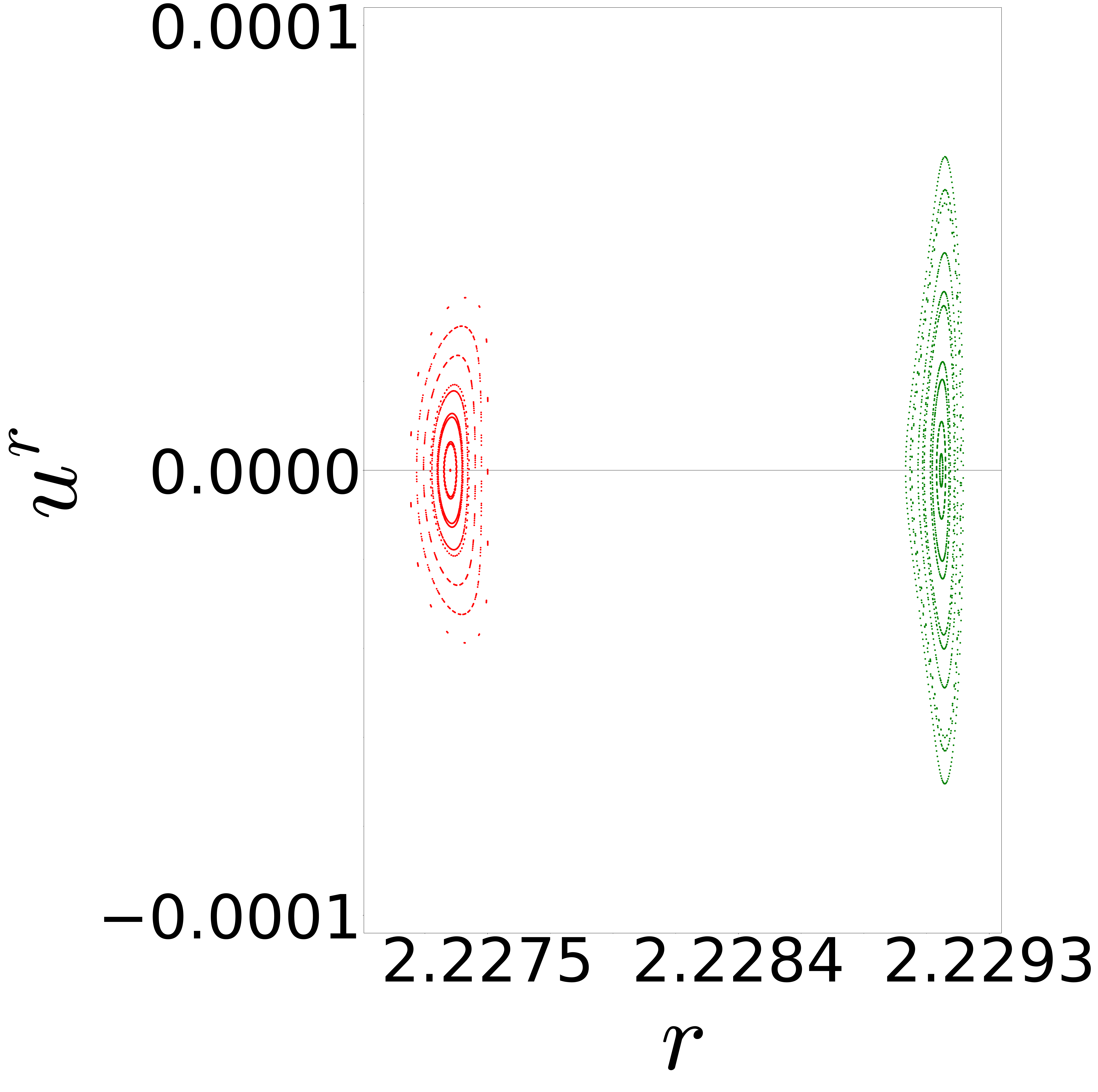}
\caption*{}
\end{subfigure}%
\quad
\begin{subfigure}[h]{0.25\linewidth}
\includegraphics[width=\linewidth]{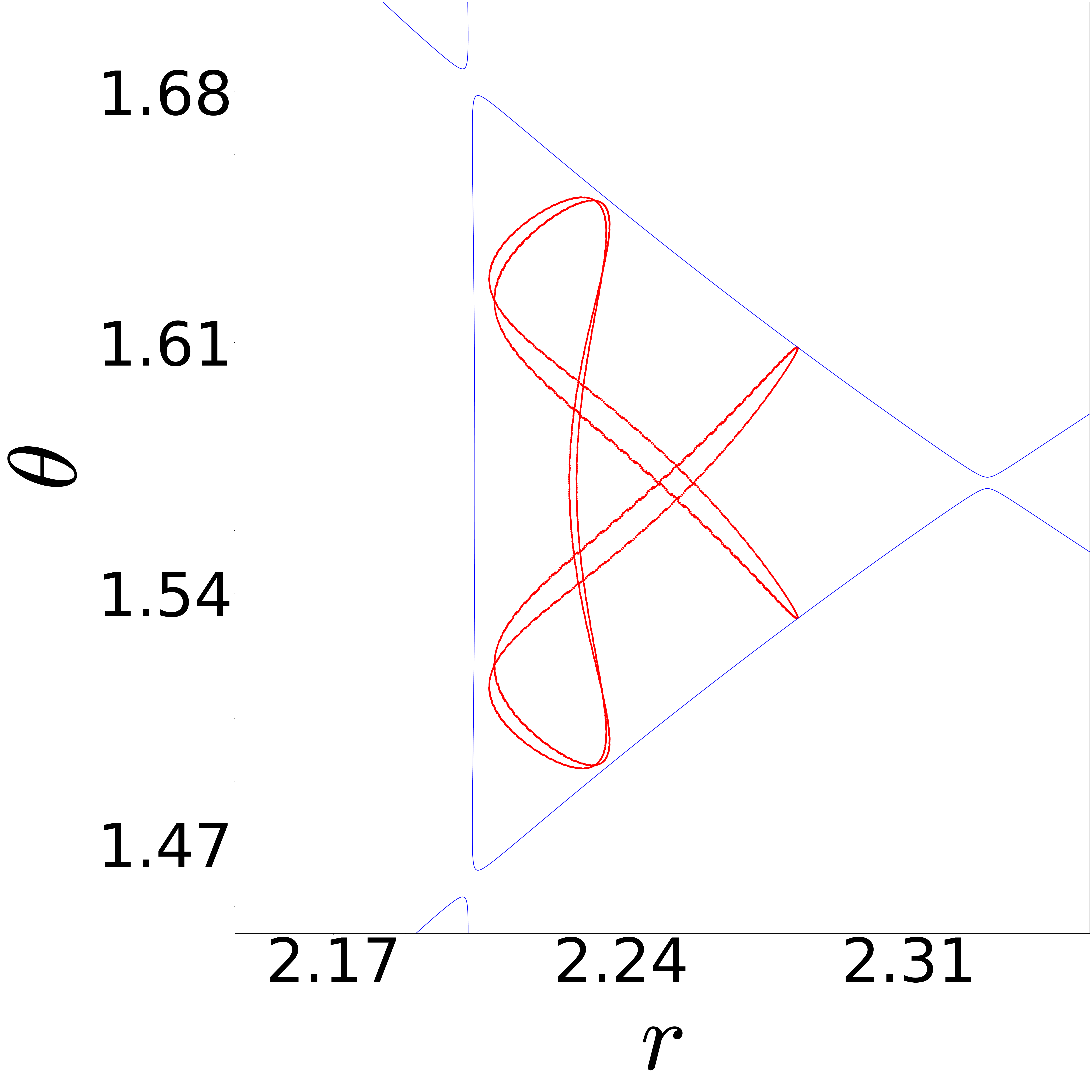}
\caption*{}
\end{subfigure}%
\caption{Top Left, Middle: Two sets of two islands that are formed by the same orbit. Top Right: Corresponding 4:2 resonant tube (Identical for the cyan orbit in Fig. \ref{fig:marg1}). Bottom Left, Middle: Two sets of islands for the red/green orbits. There is also a symmetric pair of islands for positive $u^r$. Bottom Right: The corresponding 4:2 form of the red (green) orbit.}
\label{fig:margpink}
\end{figure}
%
The pink quadruple islands shown in Fig. \ref{fig:margpink} correspond to tube orbits of 4:2 resonance and have a reflection symmetric with respect to the $u^r=0$ line (but separate) counterpart that is shown in cyan in Fig. \ref{fig:marg1}. 
Some more examples of orbits of the type of tube 4:2 resonance orbits are those forming the red and green triple islands shown at the bottom row of Fig.~\ref{fig:margpink}, %
which also have one additional island symmetric to the one shown in the leftmost plot but with $u^r>0$. The red and green islands are separate, created from different orbits, even though they look similar.   

The above have been a limited exploration of the existence of islands of stability and periodic orbits inside the pocket that is formed by the separatrix.  
Interestingly, one notices that to the extent that we have looked for islands of periodic orbits, we have managed to find islands with multiplicity of $n=1$ (Fig.~\ref{fig:margblue}), $n=2$ (Fig.~\ref{fig:margblue}), $n=3$ (Fig.~\ref{fig:margpink}), and $n=4$ (Fig.~\ref{fig:margpink}). This indicates that it is very likely that islands of multiplicity $n$ exist with $n>4$. This characteristic of the islands of stability that we observe is very likely related to the way that the system transitions to chaos, which is something that we will consider in a following section. Before that, we will turn our attention to the trapping of photons near regions of periodic orbits.    

\subsection{Sticky Light Rays}
\label{sec:sticky}

\begin{figure}[t]
\centering
\begin{subfigure}[h]{0.4\linewidth}
\includegraphics[width=0.8\linewidth]{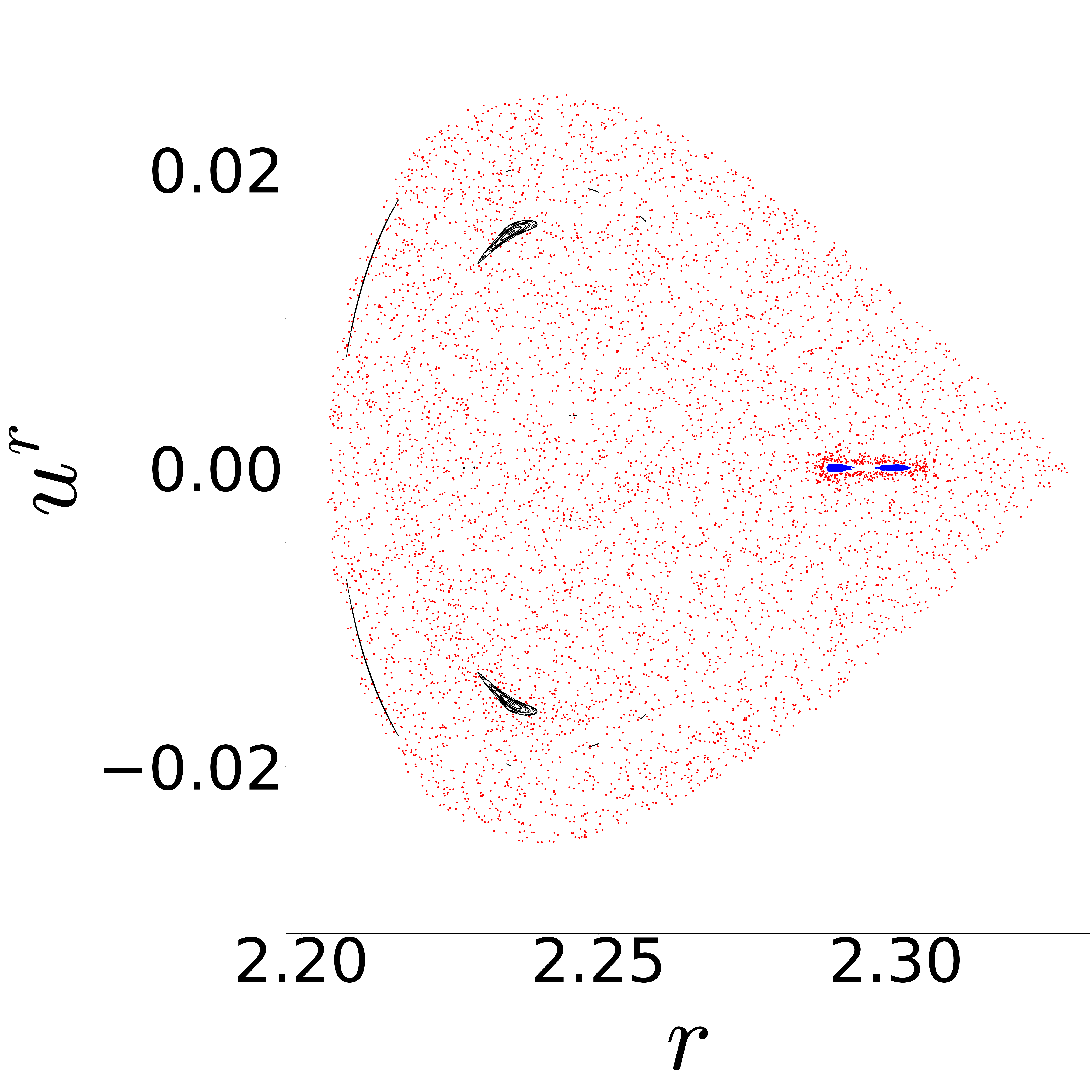}
\caption*{}
\end{subfigure}
\quad
\begin{subfigure}[h]{0.4\linewidth}
   \includegraphics[width=0.8\linewidth]{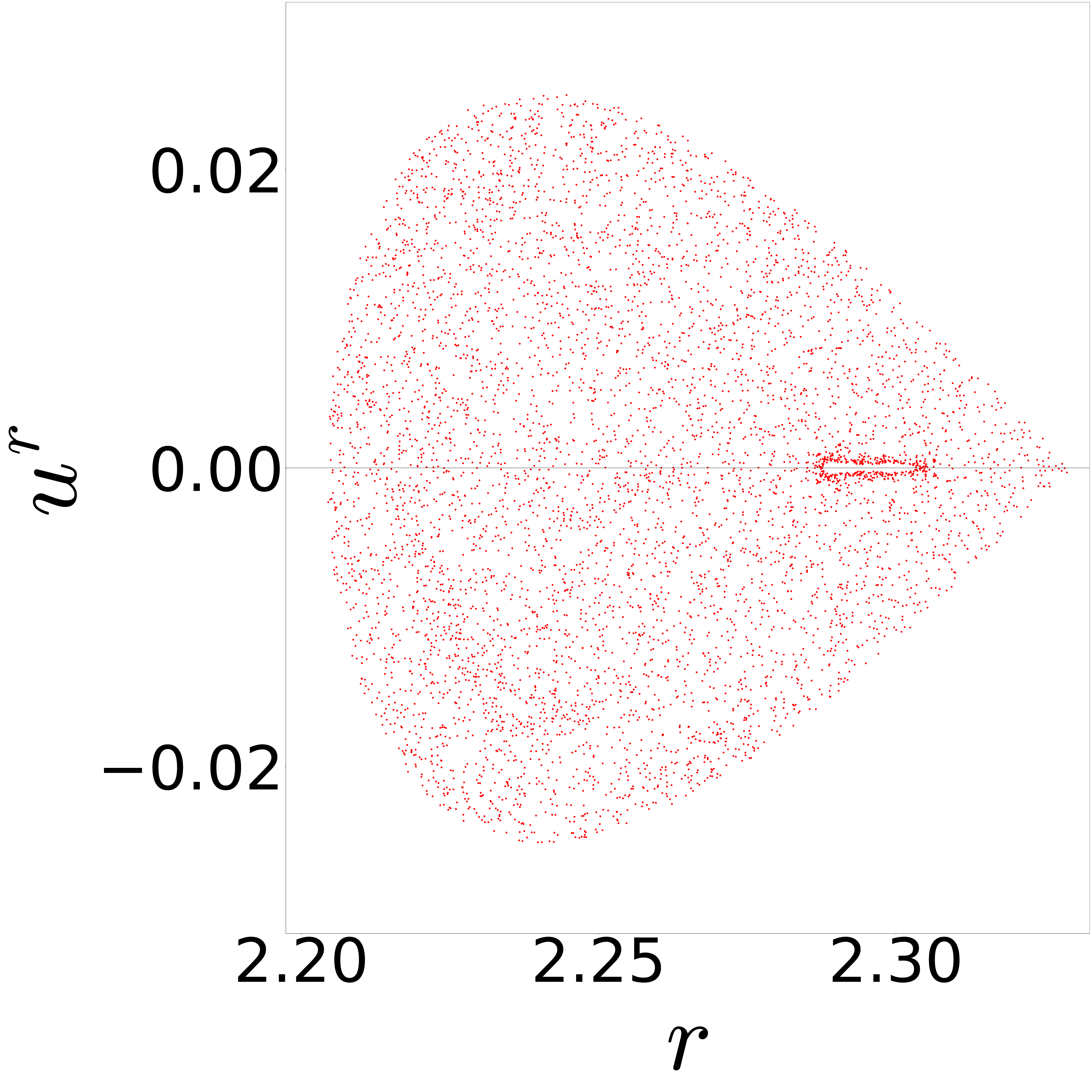} 
\caption*{}
\end{subfigure}%
\caption{Poincar\'e section in the $(r,u^r)$-plane for a light ray starting at $r_0=40M$.The impact parameters are $b=4.304332M$ and $\alpha=0.0054019M$. While the orbit of the incoming light ray is mainly chaotic, randomly filling the section, it also behaves regularly, spending time very close to a periodic orbit (dense cluster of points).}
\label{fig:stick1}
\end{figure}

We will now follow a light ray that is launched from $r_0=40M$ and with impact parameters $b=4.304332M, \alpha=0.0054019M$.  The photon  enters the pocket through the narrow throat and stays trapped for $106000M$ of the affine parameter, before escaping to infinity. Two Poincar\'e sections, one also depicting the previously discussed periodic orbits of the domain and one for the incoming orbit alone, are shown in Fig. \ref{fig:stick1}. 

While the motion is chaotic for the most part with points randomly distributed in the Poincar\'e section, the light ray spends a significant time period close to the boundary of a periodic orbit, as indicated by the density of points in that region.

\begin{figure}[h] 
\centering
  \begin{subfigure}[b]{0.32\linewidth}
    \centering
      \includegraphics[width=0.8\linewidth]{0.0054019_orbpoinc_compressed.pdf} 
    \caption*{} 
    \label{fig7:a} 
  \end{subfigure}
    \begin{subfigure}[b]{0.32\linewidth}
    \centering
      \includegraphics[width=0.8\linewidth]{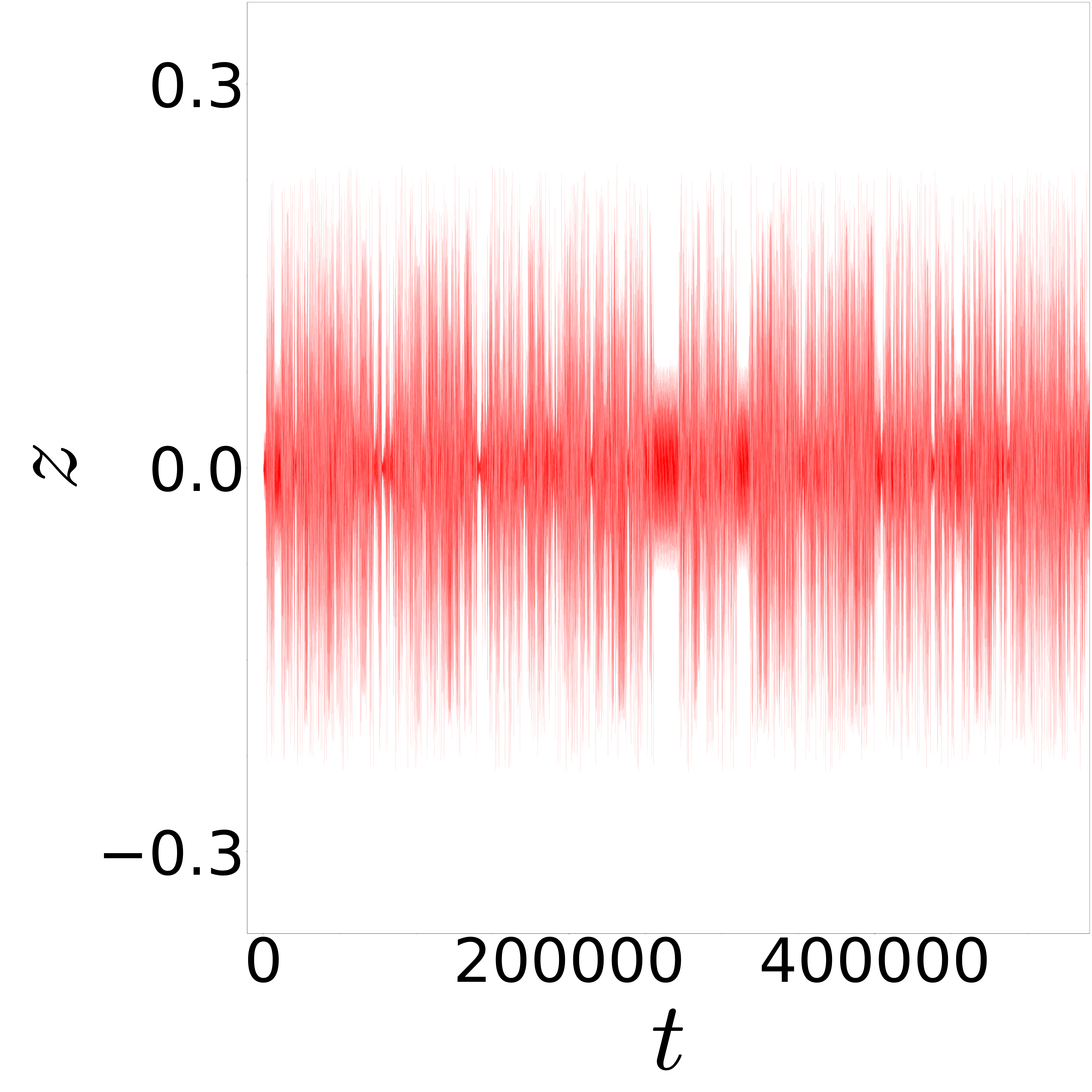}
    \caption*{} 
    \label{fig7:c} 
  \end{subfigure}
   \begin{subfigure}[b]{0.32\linewidth}
    \centering
     \includegraphics[width=0.8\linewidth]{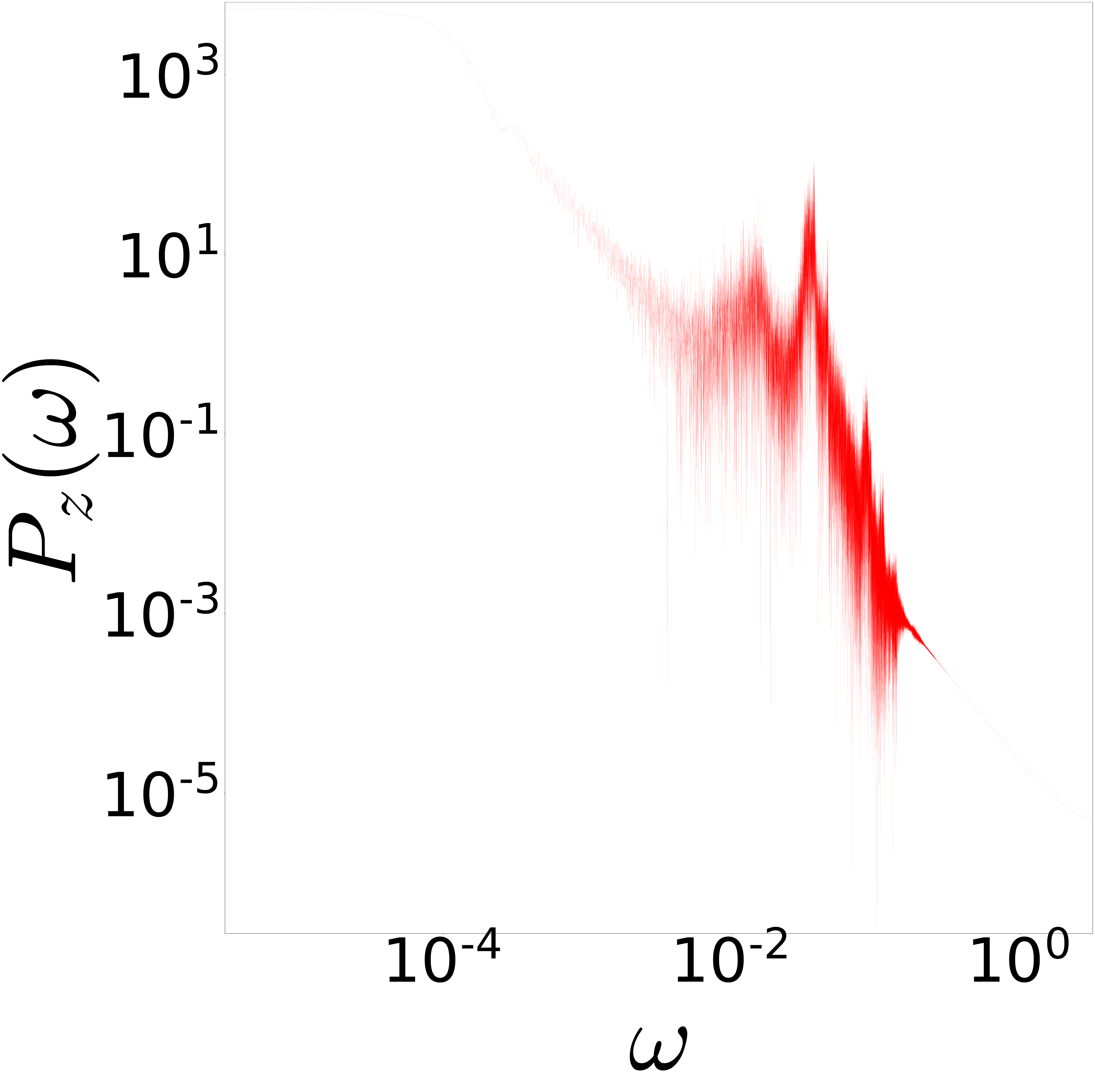}
    \caption*{} 
    \label{fig7:d} 
  \end{subfigure} 
  \begin{subfigure}[b]{0.32\linewidth}
    \centering
      \includegraphics[width=0.8\linewidth]{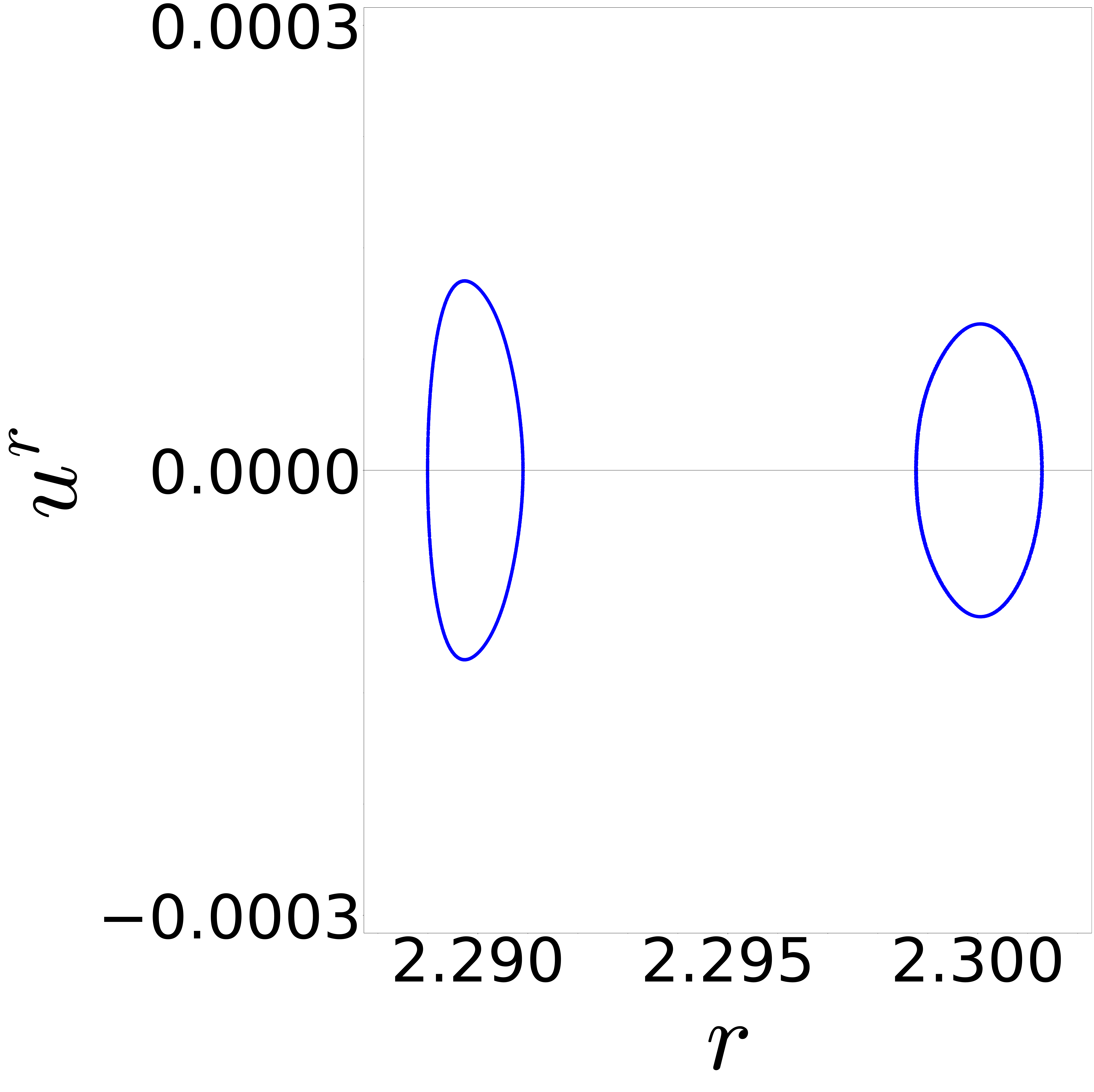}
    \caption*{} 
    \label{fig7:b} 
  \end{subfigure} 
  \begin{subfigure}[b]{0.32\linewidth}
    \centering
      \includegraphics[width=0.8\linewidth]{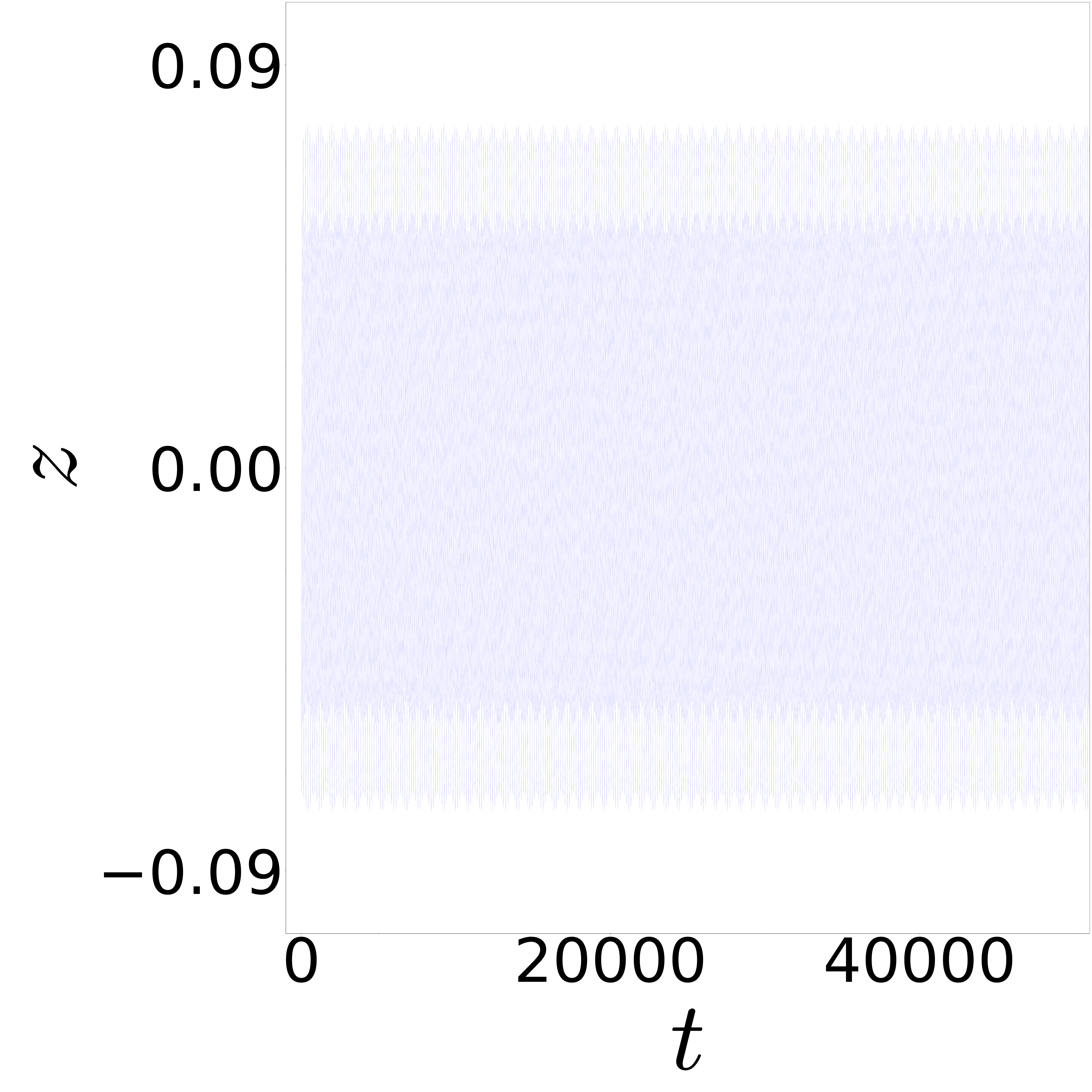}
    \caption*{} 
    \label{fig7:d} 
  \end{subfigure} 
  \begin{subfigure}[b]{0.32\linewidth}
    \centering
      \includegraphics[width=0.8\linewidth]{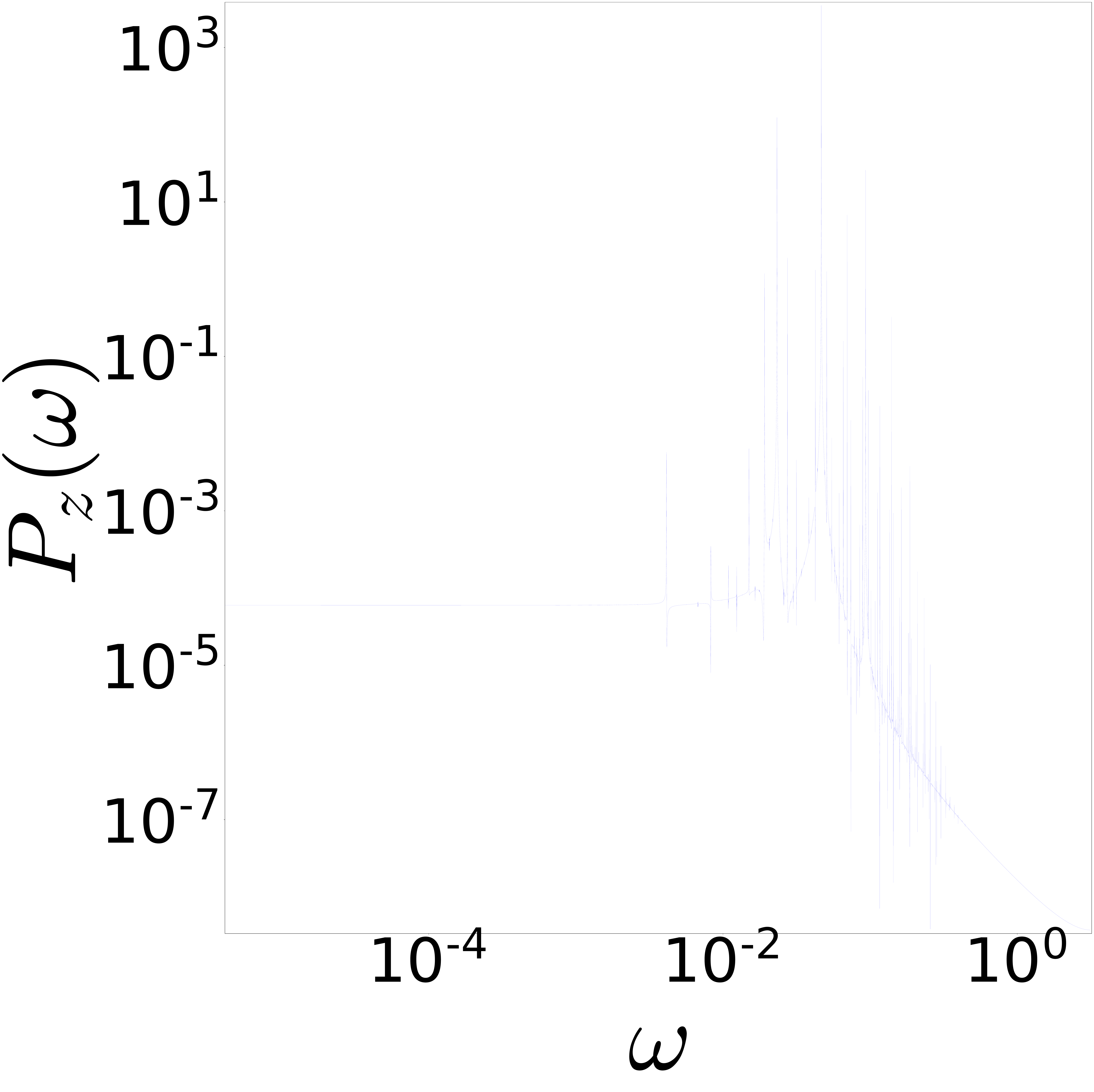}
    \caption*{} 
    \label{fig7:d} 
  \end{subfigure} 
  \caption{Comparison of a chaotic (red) and regular (blue) orbit. The impact parameter of the chaotic (red) orbit is $\alpha=0.0054019M$. We show from left to right, Poincar\'e sections in the $\theta=\pi/2$ plane, the time series of their z component and the respective power spectra. The power spectrum of the chaotic orbit obeys the so-called $1/f$ fluctuations meaning that the orbit is \textit{sticky.}}
  \label{fig:stickyvsperiodic}
  \end{figure}

In order to characterise the orbit's chaos, we take the time series of its $z$ position and compute the corresponding Fourier spectrum and compare it against that of a regular periodic orbit. This comparison is shown in Fig. \ref{fig:stickyvsperiodic}, where we have the plots for the chaotic orbit and the plots for the periodic orbit. In the figure, one can see that even though the motion is chaotic, there are two periods of time for which the mean value of the $z$ time series is constant, specifically at $ t\in [255000M,270000M]$ and $t\in [310000M,320000M]$. Additionally, the power spectrum of the time series obeys the so-called $1/f$ fluctuations \cite{Dutta1981,Weissman1988,Manneville1980,Kohyama1984}. These characteristics indicate that the orbit is \textit{sticky}. %
The Poincar\'e section in this case clearly indicates, by the density of points, to which island the orbit tends to stick, but this is not always the case given that additional islands of stability may exist that are not clearly visible in the plane we've chosen. In our example the light ray enters the allowed region and spends some time interval of "mixing" in the chaotic sea. When it arrives in the vicinity of a sticky set, in this case the blue periodic 2:2 tube orbit, it becomes non-chaotic sticking to the island's boundary for a significant time interval before it returns back to the chaotic sea and starts mixing once again. This process is repeated randomly \cite{Afraimovich1998}. 

\begin{figure}[h]
 \centering
\begin{subfigure}[h]{0.4\linewidth}
\includegraphics[width=\linewidth]{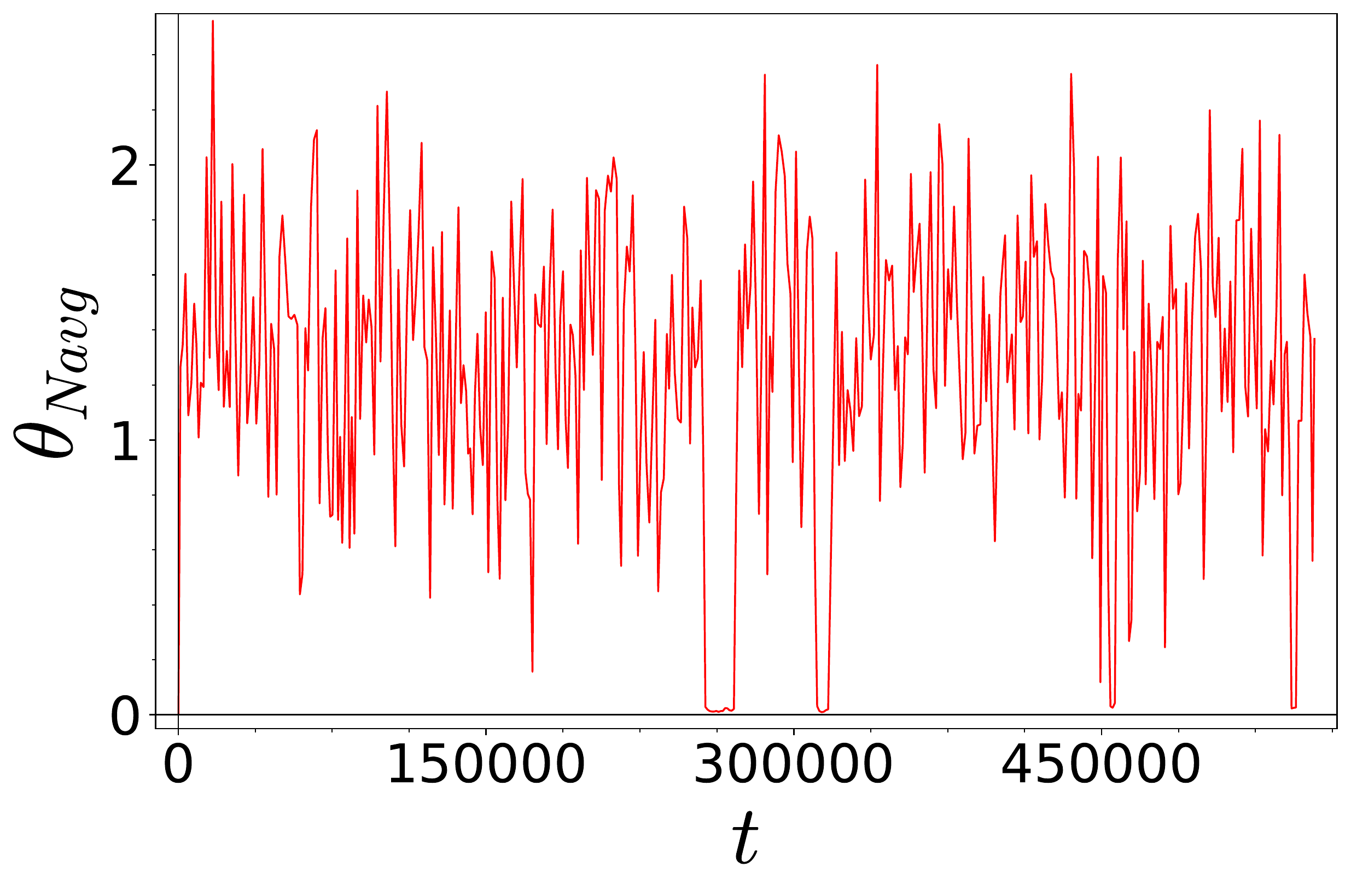}
\caption*{}
\end{subfigure}
\quad
\begin{subfigure}[h]{0.4\linewidth}
\includegraphics[width=\linewidth]{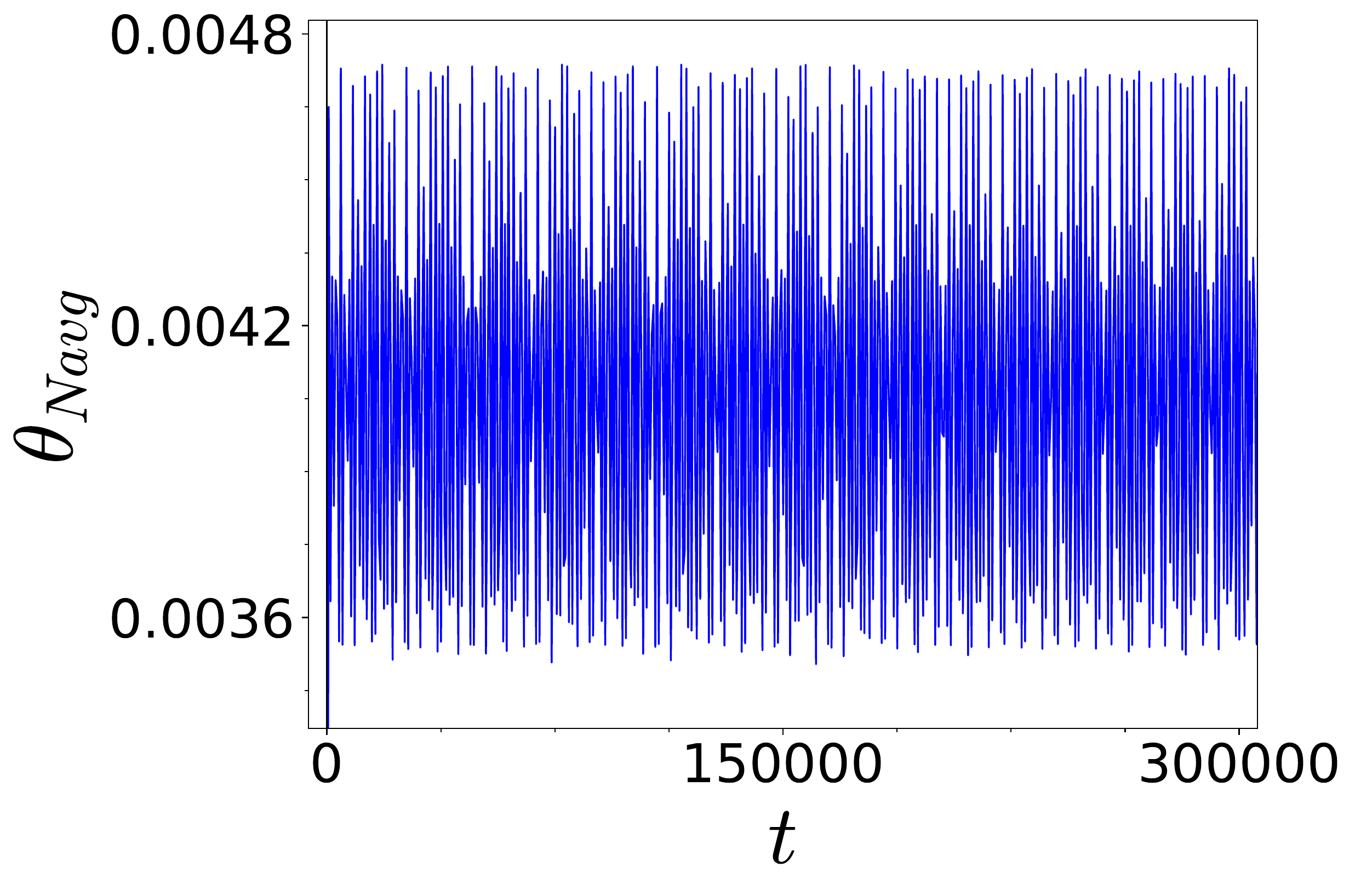}
\caption*{}
\end{subfigure}%
\caption{Time series of the average rotation angle $\theta_{Navg}$ for the chaotic (red) and regular (blue) orbits of Fig.~\ref{fig:stickyvsperiodic}.}%
\label{fig:sticky1_rot}
\end{figure}

Calculating the rotation number for the chaotic orbit, we can see that there are two cases where its average value is stable, in agreement with the time intervals for which the mean value of the z component time series is periodic. The rotation number for both the chaotic and the periodic orbit is shown in Fig. \ref{fig:sticky1_rot}.

\begin{figure}[h] 
\centering
  \begin{subfigure}[b]{0.32\textwidth}
    \centering
        \includegraphics[width=0.8\linewidth]{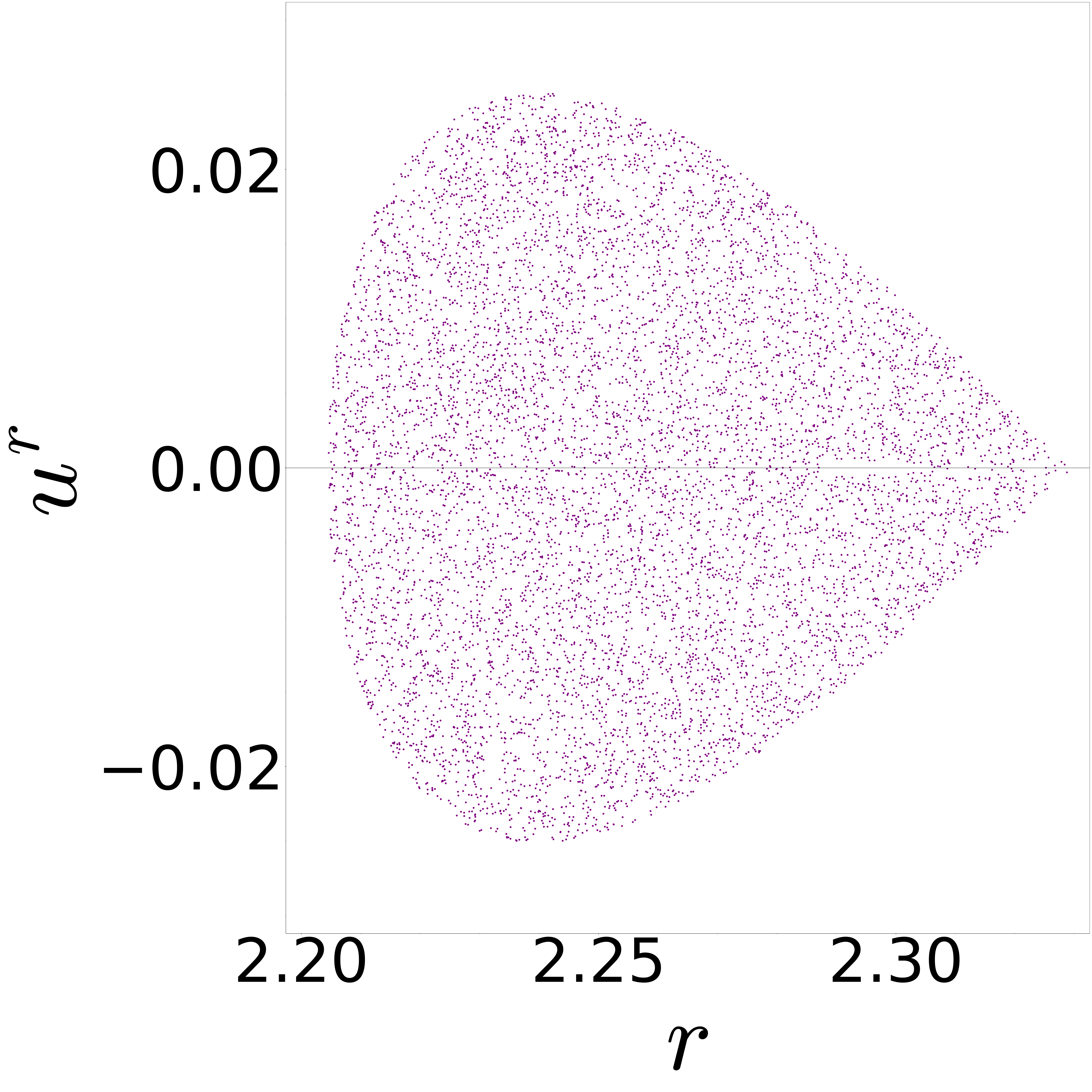} 
    \caption*{} 
    \label{fig7:a} 
  \end{subfigure}
    \begin{subfigure}[b]{0.32\textwidth}
    \centering
      \includegraphics[width=0.8\textwidth]{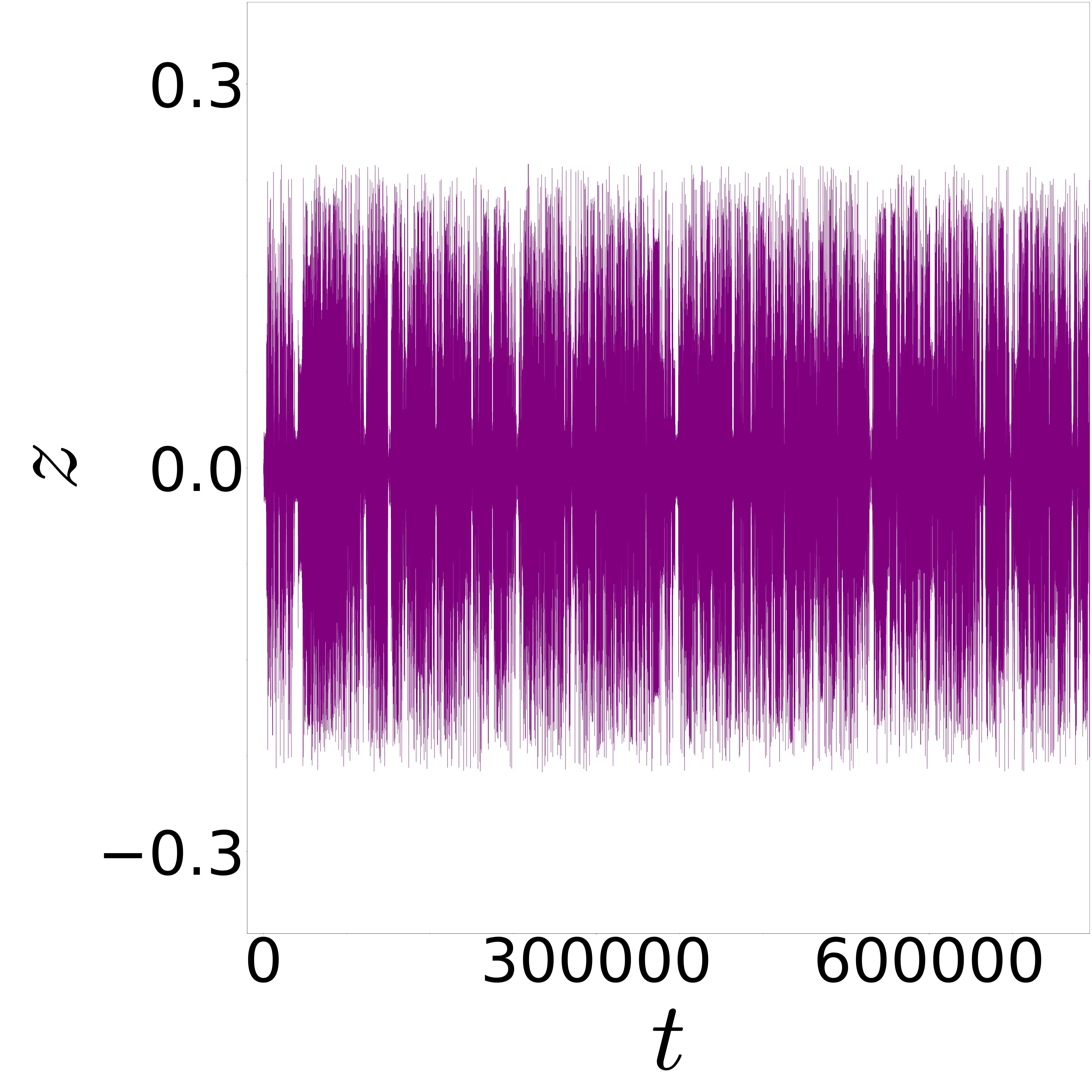}
    \caption*{} 
    \label{fig7:c} 
  \end{subfigure}
    \begin{subfigure}[b]{0.32\textwidth}
    \centering
       \includegraphics[width=0.8\textwidth]{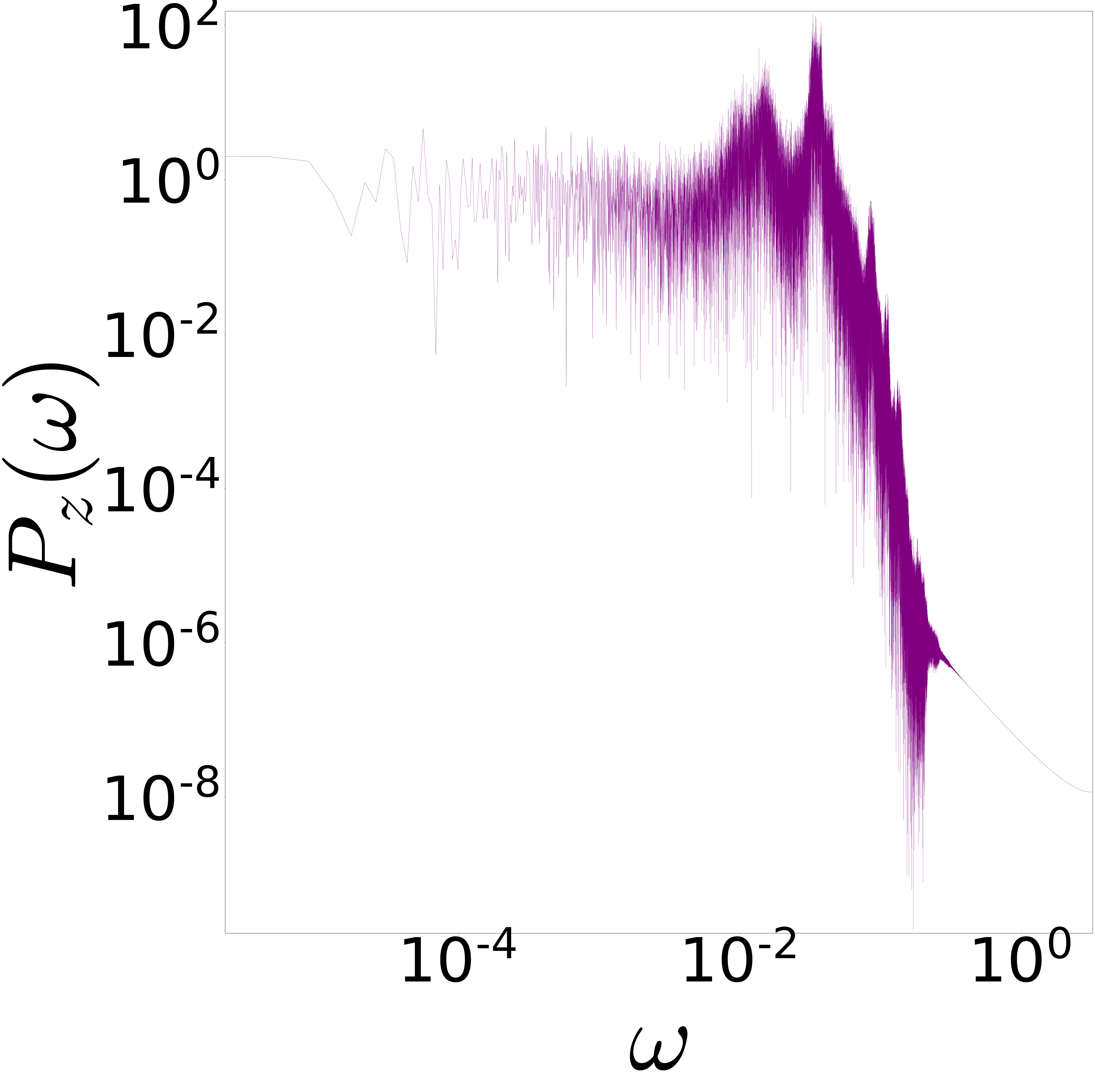}
    \caption*{} 
    \label{fig7:d} 
  \end{subfigure} 
    \caption{Example of a white-noise orbit (in purple) with impact parameter, $\alpha=0.00541M$. We show from left to right, the Poincar\'e section in the $\theta=\pi/2$ plane, the time series of the z component and the power spectrum. The power spectrum of the purple chaotic orbit does not obey the so-called $1/f$ fluctuations (compared to the red power spectrum in Fig \ref{fig:stickyvsperiodic}) meaning that the orbit is not \textit{sticky.}}
  \label{fig:whitevssticky}
 \end{figure}

Additionally to the sticky chaotic orbits there can exist chaotic orbits that are not sticky and therefore are qualitatively different. We present here an orbit that despite of being trapped for a long period of time in the pocket, does not exhibit sticky behaviour and is therefore of the \textit{white-noise} type. %
Fig. \ref{fig:whitevssticky} shows the Poincar\'e section, the time series of the $z$ component and the power spectrum of the orbit. As one can see, the white-noise type orbit does not exhibit any kind of stagnant motion in its time series and the power spectrum does not follow the 1/f power law. It is, therefore, not sticky. Furthermore one can see from the power spectrum how the \textit{white-noise} orbit gets its name, i.e., the flat part of the spectrum. Finally the power spectra of both white-noise and sticky orbits are distinctively different from the power spectrum of the periodic orbit.  

To summarise, so far we have found that inside the pocket that is formed in a range of parameters for the HT spacetime, photons can be trapped for long periods of time. In the allowed region of the phase space, there can be found islands of stability with periodic orbits that trap photons, while there also exist chaotic orbits that fill the allowed phase space in between the islands and can potentially escape to infinity. A class of these chaotic orbits can get temporarily trapped near stable islands of periodic orbits, and the trapping times can be long.  

\subsection{Cut-off Pocket}
\label{sec:cutoff}

\begin{figure}[h]
\begin{subfigure}[h]{0.45\linewidth}
\includegraphics[width=0.9\linewidth]{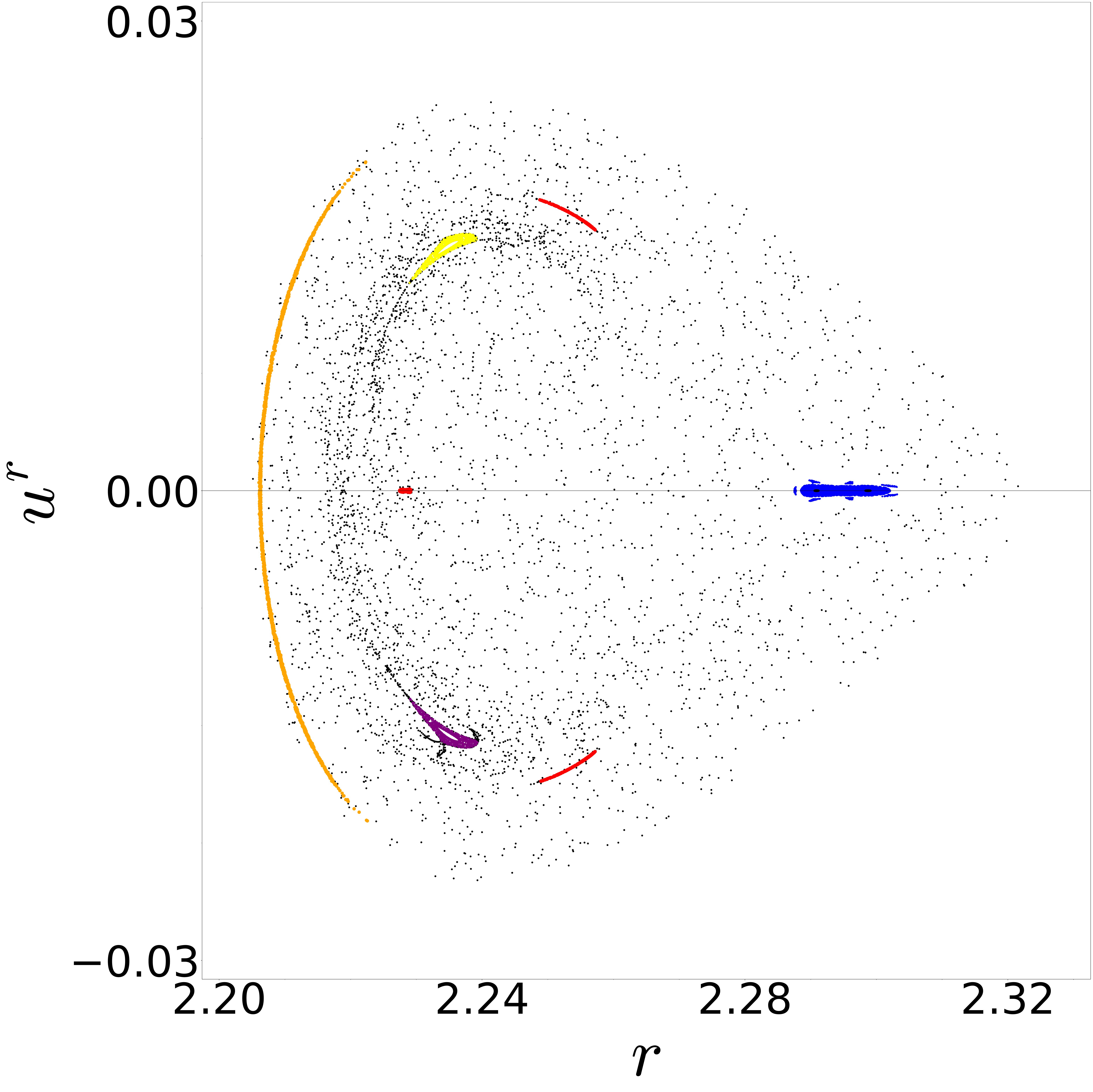}
\caption*{}
\end{subfigure}
\quad
\begin{subfigure}[h]{0.45\linewidth}
\includegraphics[width=0.9\linewidth]{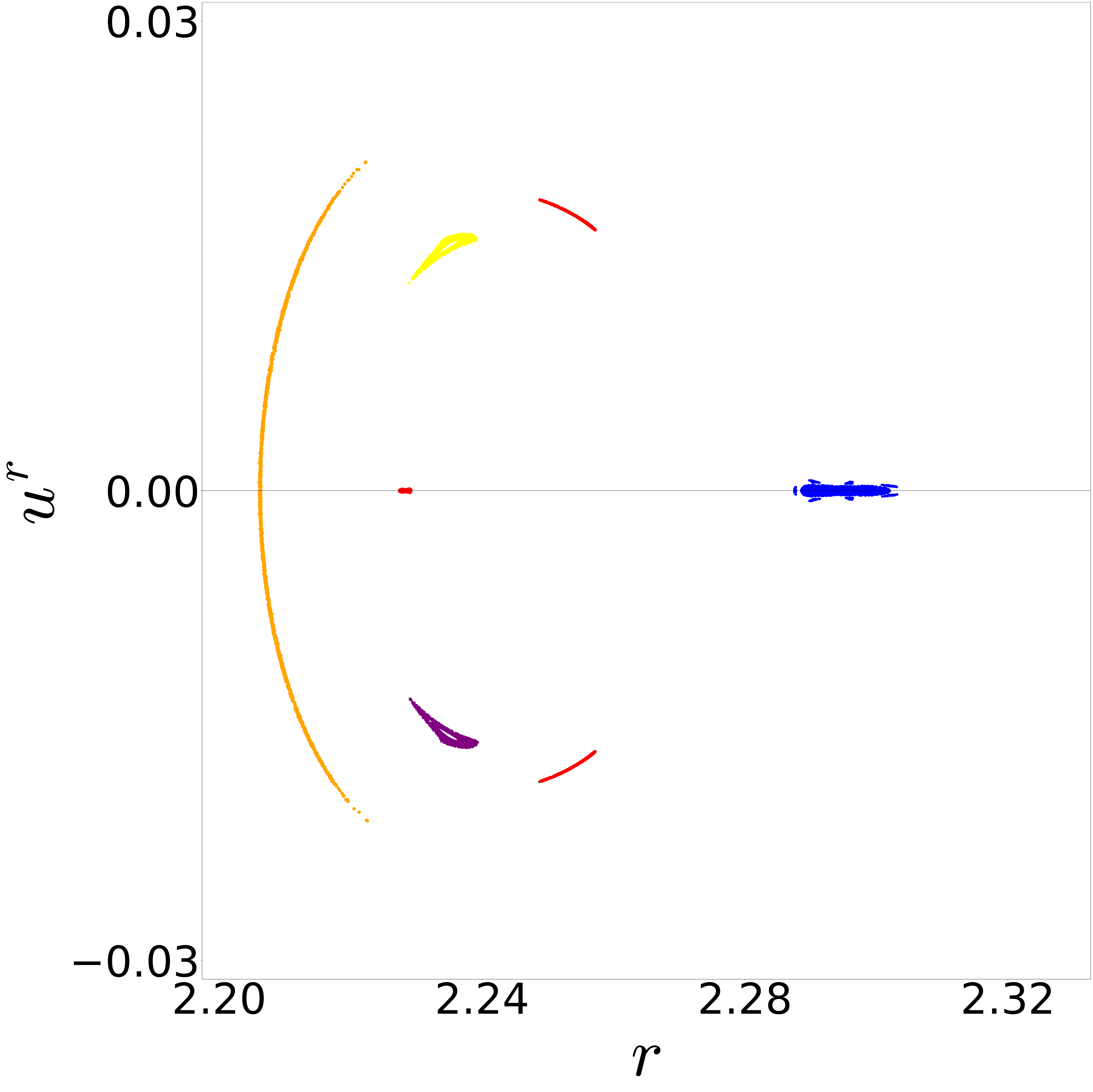}
\caption*{}
\end{subfigure}%
\caption{Poincar\'e section in the $(r,u^r)$-plane for the cut-off pocket. The impact parameter is $b=4.30434M$. Left: The domain is mostly chaotic for the whole available space, with small islands of stability surviving. Right: The islands of stability, colour coded for orbits belonging to different island groups.}
\label{fig:cutoff_all}
\end{figure}

The natural next step to take is to investigate the properties of the completely separated pocket where light rays can not enter from the exterior nor escape to it. This is a fully closed Hamiltonian system where stable bounded null orbits, possibly originating from the accreting matter, are possible to exist. While these cannot be directly observable in the qualitative characteristics of the object's shadow for example, they allow for energy to be trapped in the pocket that may have some other interesting phenomenology. Furthermore, in the case of ultra-compact objects, this sort of trapping could be related to the presence of trapped gravitational wave modes that could affect the late-time gravitational-wave ringdown, producing echoes, making it thus distinct from that of a usual black hole \cite{Cardoso:2014sna,shipley2019strongfield}. Furthermore, it is worth investigating in an effort to better understand the properties of the marginally open system.

\begin{figure}[h]
\centering
\includegraphics[height=0.40\textwidth]{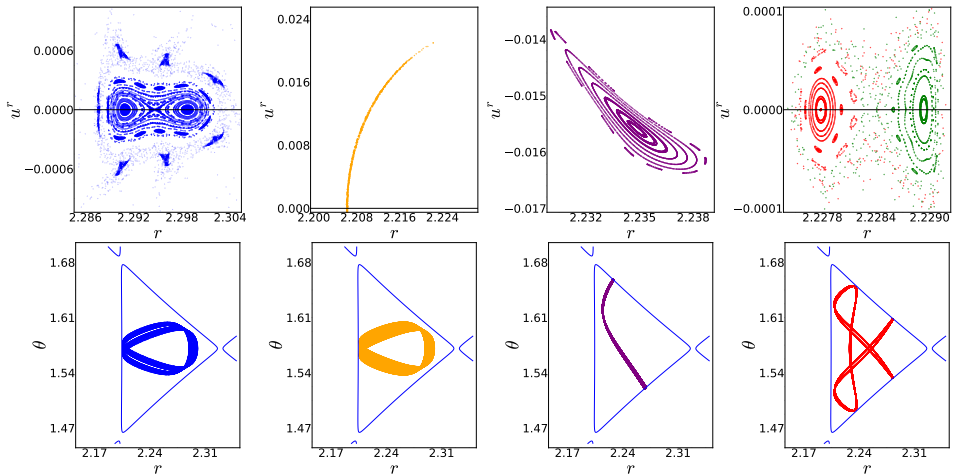}
\caption{Islands of stability and the corresponding orbits, for the cut-off pocket.}%
\label{fig:cutofforbits}
\end{figure}

We choose the usual values for the spin parameter and the quadrupole deviation and increase the impact parameter $b$ from the value we used for the open pocket, to $b = 4.30434M$ that gives a closed pocket. We numerically integrate several hundred null geodesics and show the Poincar\'e section in Fig. \ref{fig:cutoff_all} for this cut-off pocket. While sharing many qualitative characteristics with the case of the marginally open pocket, there are interesting differences that hint to the dynamical evolution of the system. 
Fig. \ref{fig:cutofforbits} shows some of the islands of stability observed, along with the corresponding orbit form. Here we have kept the same colour codes that we used for the islands presented for the open system. 
%
%
This system is still very close to the marginally open one, yet some interesting features appear. It seems that some of the islands observed in the open system exhibit a tendency to merge. It will therefore be interesting to explore how the cut-off pocket evolves as we further increase the impact parameter up to the point it almost disappears.

\subsection{Transition to Chaos}
\label{sec:transition}

It is now our goal to investigate how the dynamics change for the cut-off pocket, a necessarily closed Hamiltonian system, as we change its size. We remind that the shape and size of the pocket depends only on the impact parameter $b$, as it is the only parameter entering the effective potential for the photons. The value $b=4.304332M$, corresponds to the marginally open HT system that was studied in Section \ref{sec:marg}, and we set it as the \textit{threshold parameter} $b^{*}$. For $b\geq b^{*}$, the system is closed and its size continuously decreases up until it ultimately disappears for $b\approx 4.3063M$. Since in the marginally open HT system (Fig.\ref{fig:marg1}) the chaotic domain occupies almost the whole available space with small islands of stability surviving in the chaotic sea, we expect it to be the end state of the evolution of a system that transitions from order to chaos. We will therefore start from higher values of $b$ where the pocket is close to its disappearance (or appearance) and continuously decrease the impact parameter down to the value $b\approx b^{*}$.

\begin{figure}[b]
\centering
\includegraphics[height=0.28\textheight]{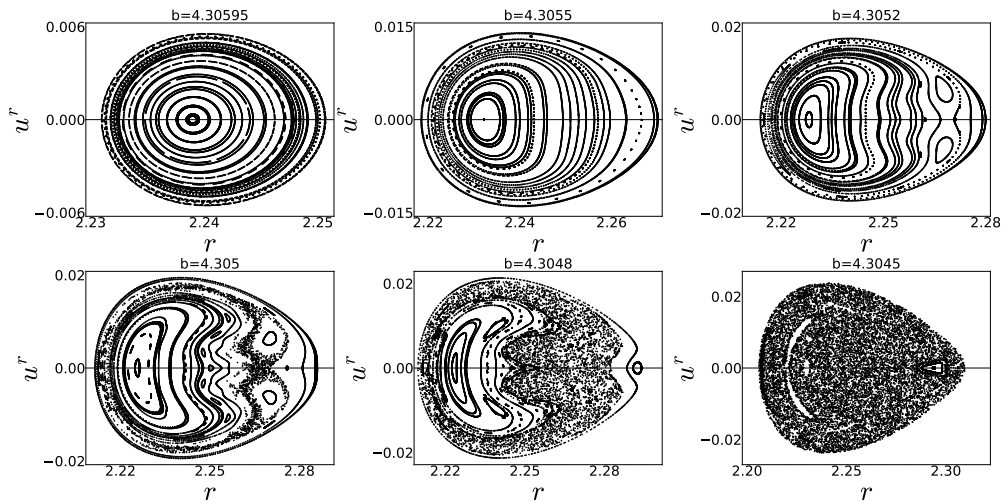}
\caption{Orbital structure of the phase space $(r,u^r)$ for several values of the impact parameter $b$ demonstrating the evolution of the islands of stability and the emergence of the chaotic regions.}
\label{fig:transpoinc}
\end{figure}

Fig. \ref{fig:transpoinc} presents Poincar\'e sections for different values of the impact parameter $b$ in the aforementioned range. For higher values of $b$ the motion is completely ordered. As $b$ decreases, a chaotic layer along with several bifurcated families and families of higher resonance appear. The number of chaotic orbits is highly increased and for $b\approx b^{*}$ they occupy almost the whole available space. It is now evident that the islands studied in the marginally open HT system are products of bifurcation. Comparing for example the blue pair of islands of the marginally open (Fig. \ref{fig:margblue}) and the cut-off (Fig. \ref{fig:cutofforbits}) pockets, it is visible that the islands are a product of bifurcation of the unstable point that exists between them. As $b$ decreases and the transition to the open system begins, the unstable point is destroyed, the distance between the islands is increased and the stable points that they surround will eventually become unstable and a new cycle of \textit{bifurcations and destruction} will begin. As this process is repeated, chaos dominates. This can probably be related to the existence of islands of stability of increasing multiplicity, that we observed for the marginally open system. 
It is also worth noting that the initial circular shape of the pocket gradually becomes nut-like with sharper corners. Again this behaviour is reminiscent of the closed H\'enon-Heiles system \cite{Henon-Heiles-1964AJ,Zotos2014,Zotos2017}. %

\section{The Shadow of a Hartle-Thorne spacetime}
\label{sec:5}

We turn now to the calculation of the shadow of a non-Kerr compact object described by the HT spacetime. Our goal is to find how the bifurcation of the light-ring and the formation of the pocket, affect the observed shadow. 

Our setup for the construction of a shadow is the following. We consider an observer at a large distance from the compact object. Assuming that a shadow is the \textit{dark silhouette of the compact object/BH against a spherical background illumination}, we divide all light rays into two categories, those that go to spatial infinity after being scattered by the compact object and those that reach the ``surface''/horizon and get absorbed. We consider light sources densely distributed all over the universe, except in the region between the BH/compact object and the observer. This way all the past-oriented rays of the latter category of the two we defined will meet a background source of light, while the former will not. Each light ray that connects the observer to a source will correspond to a bright spot on the observer's field of view, while those that do not, to darkness. The boundary that separates the two, is created by light rays that go neither to infinity nor are lost to the compact object. They are instead trapped within the spacetime \cite{perlick2021}.

This setup aims to provide the {\it mathematical} shape of the shadow and is not intended to give the actual appearance of a realistic shadow image observed by some telescope. This is the textbook way of studying the morphology of a shadow \cite{Bardeen:1972fi,Cunha2015PhRvL,Cunha2016IJMPD,Cunha2016PhRvD,Cunha2017PhysRevD,Cunha2018GReGr,Lima:2021las,Medeiros:2019cde,perlick2021,Shipley-Dolan2016,shipley2019strongfield,WangPhysRevD2018}. Some more details on the algorithm as well as the details on selecting where the object's ``surface'' is (i.e., $r_s$), are given in \ref{sec:App:0}.

Since the observer is at a large distance from the compact object, the two impact parameters as defined by Bardeen \cite{Bardeen:1972fi,1973ApJ...183..237C}, that are related to the angular size of the shadow on the observer's field of view, will be given by the definitions, 
\be b=p_{\phi}/p_t, \quad \textrm{and} \quad \alpha = p_{\theta}/p_t, \nn\ee
where we remind that the definition for $b$ that we will be using from now on gives the opposite sign for the impact parameter on the observer's image plane, as discussed earlier. The impact parameter $b$ is the apparent displacement of the image perpendicular to the projected axis of symmetry while $\alpha$ is the displacement parallel to the axis. These have dimensions of length so they can be related to angles in the observer's sky if we divide them by the distance $r_0$. We also note that our naming convention is slightly different from that of Bardeen.

In the following subsections we will see how the results of Section \ref{sec:4} are connected to the emergence of several interesting features of the shadow that a compact object described by the HT metric, casts. The fact that the effective potential of the HT metric can form pockets, leads to photons being trapped in these pockets for long periods of time (quasi-bound orbits). The chaotic motion in this pocket is expected to relate to chaotic patterns emerging in the observers image since a sensitive dependence on initial conditions between the pixel and the geodesic creating it, exists \cite{Cunha2016PhRvD}. The formation of pockets in the effective potential is directly related to the existence of stable light-rings. Multiple light-rings generate, in a sense, families of periodic orbits and invariant manifolds that define the shape of a compact objects shadow. A number of key features such as multiple disconnected shadows, regions of chaotic motion in the image plane and principal shadows with non-convex boundaries are related to their existence \cite{shipley2019strongfield}.

\subsection{HT shadow for $\chi=0.4$ and $\delta q = 0.1$}

\begin{figure}[h]
    \centering
    \includegraphics[width=0.4\textwidth]{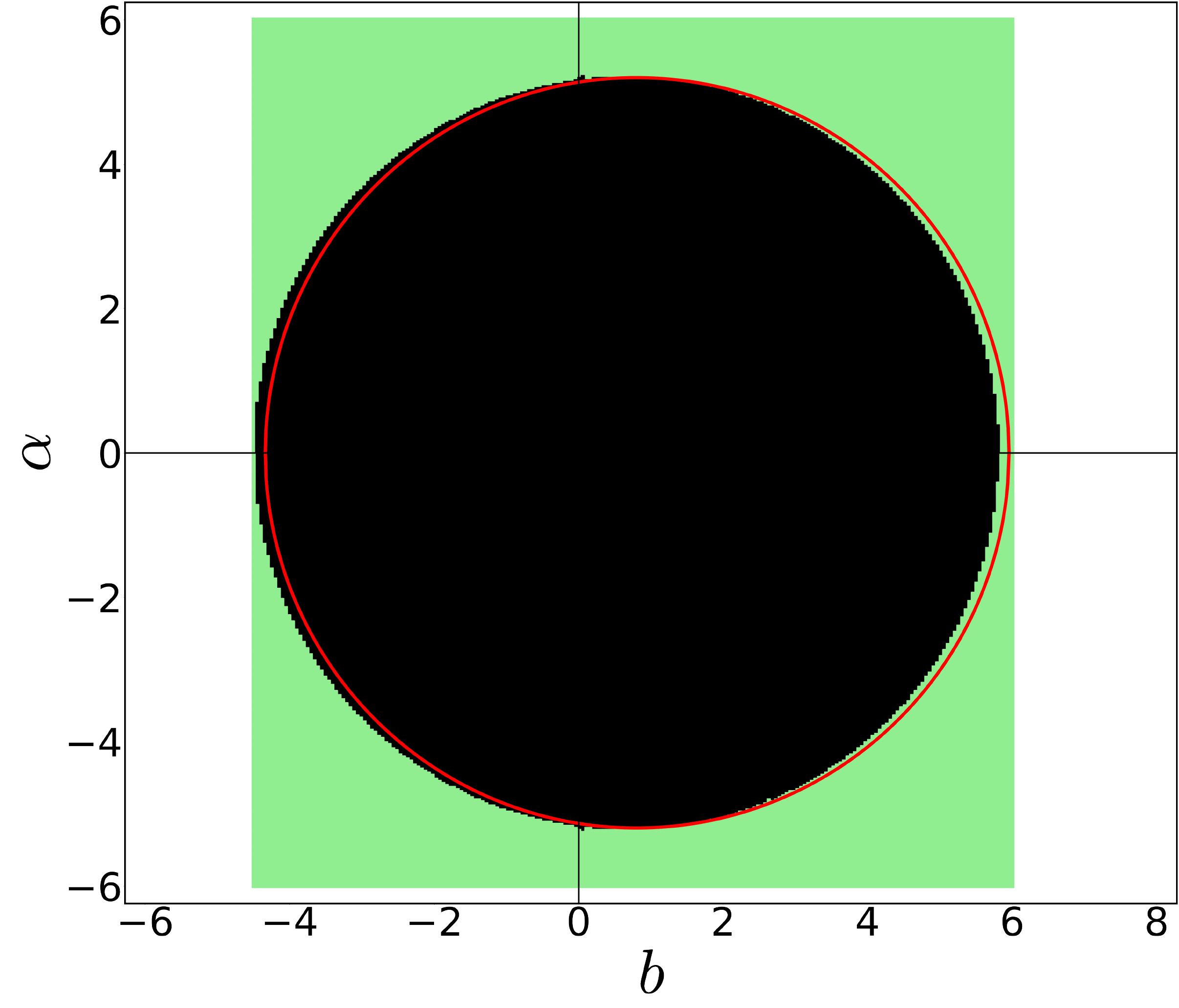}
    \caption{Shadow of a compact object described by the HT spacetime with a spin parameter of $\chi=0.4$ and quadrupole deviation from the Kerr metric, $\delta q =0.1$, viewed from $\theta=\pi/2$. The red shadow curve is that of a Kerr black hole with the same spin.}
    \label{fig:HT0401}
\end{figure}

The first HT shadow that we compute is that of a compact object with a small quadrupole moment deviation $\delta q=0.1$ and a spin parameter of $\chi=0.4$. We expect it to be close to that of a same spin $\chi=0.4$ Kerr black hole given that the deformation is relatively small. In this range, none of the phenomena studied in the previous section is present and only one co-rotating light-ring exists, on the equatorial plane $\theta = \pi/2$ (of course we also have a counter-rotating light-ring and the spheroidal orbits in between). We show the shadow in Fig. \ref{fig:HT0401} along with the shadow curve (red) of a Kerr black hole with the same spin parameter, analytically calculated from E. Gourgoulhon's available notebooks \cite{eric2018} for SageMath 9.2. Again, green pixels represent light rays that reach the observers screen from some source of light. The shadow is slightly transposed to the right along the horizontal axis but no qualitative difference is noticed in its shape, as expected.

\subsection{HT shadow for $\chi=0.4$ and $\delta q = 1$}
\label{}

\begin{figure}[b]
    \centering
    \includegraphics[width=0.42\textwidth]{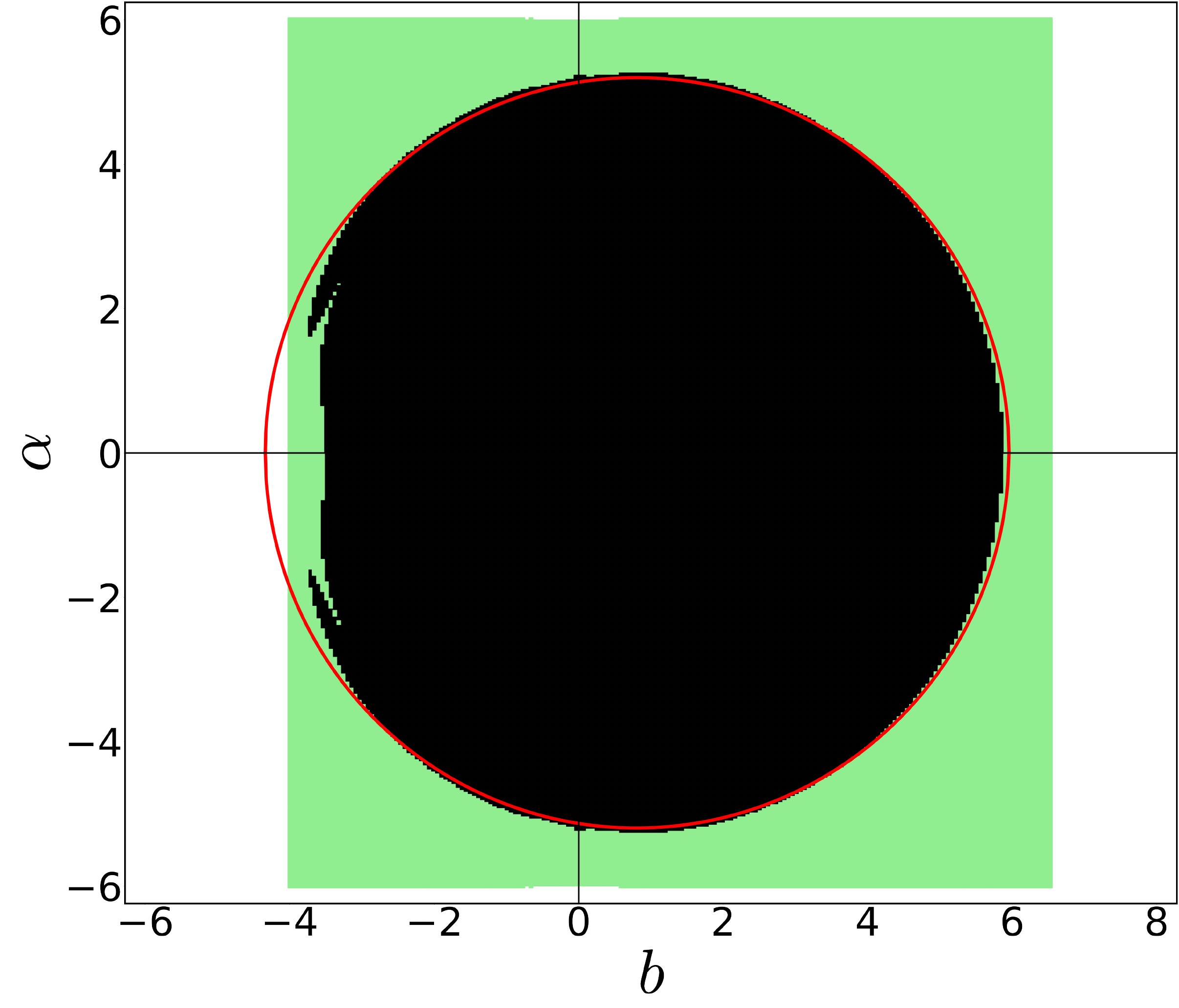}
     \includegraphics[width=0.37\textwidth]{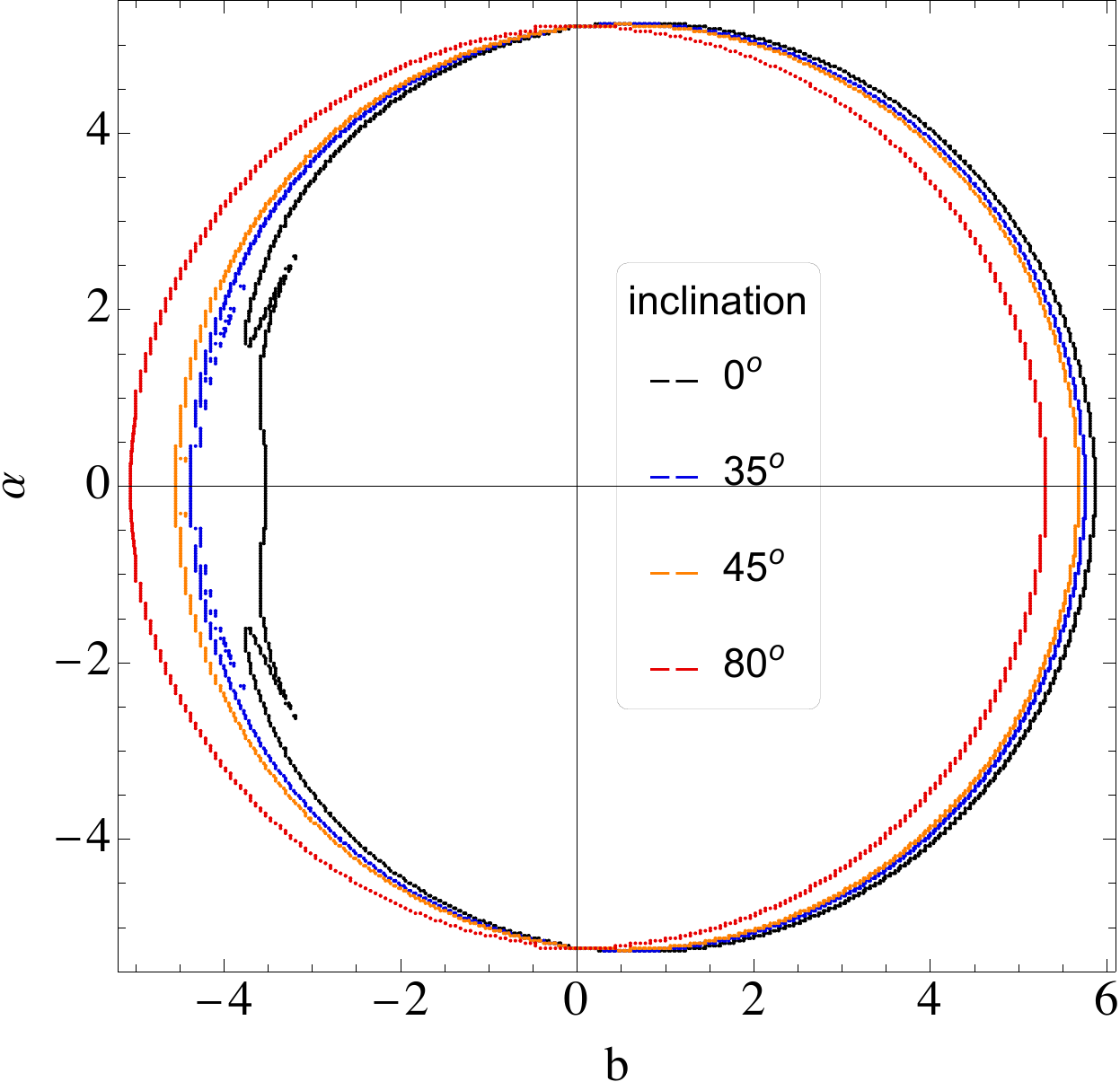}
    \caption{Shadow of a compact object described by the HT spacetime with a spin parameter of $\chi=0.4$ and quadrupole moment deviation, $\delta q = 1$. We assume $r_s=2.115866M$. Left: The shadow seen by an observer on the equatorial plane is compared against the shadow of a Kerr black hole with the same spin (red curve). Right: The different shadow curves correspond to the same object as in the left plot, but for different inclinations of the observer with respect to the equatorial plane.}
    \label{fig:HT041}
\end{figure}

We now compute the shadow of a compact object with a quadrupole moment deviation of $\delta q= 1$ and spin $\chi=0.4$. In this regime, we are past the critical spin parameter $\chi^*$. The equatorial light-ring has bifurcated to two non equatorial ones, symmetrical as to the equatorial plane $\theta=\pi/2$. We show the shadow in Fig. \ref{fig:HT041} together with a Kerr shadow for the same spin. Since the morphology of this shadow was expected to be richer, we set a grid of $213\times 213$ and increase the resolution to $90000$ pixels in total.

We are past the triple-light-ring regime, where only two non-equatorial rings exist (see Fig. \ref{fig:bifurc1}). These are shown to alter the shape of the shadow, forming ``eyebrow'' shapes for $b\approx -3.8M$. In this region the interior and the exterior regions of allowed motion are connected but there are light rays that do not fall into the object. These orbits circle the object, near the vicinity of the light-ring, and escape to infinity. This essentially means that the light ray reaches the throat of the separatrix, bounces off of it and scatters. One notices that there is a depression on the left of the figure that breaks the circular symmetry of the shape. If one were to observe such a shadow without enough resolution so as to resolve the eyebrows, the shape could be mistaken for that of a Kerr BH of higher spin (see Fig. \ref{fig:KerrShadow}). Finally, in Fig. \ref{fig:HT041} we also present the shadow for different inclinations for the observer. The inclination is defined as the angle the line of sight of the observer forms with the equatorial plane of the compact object. The different shadows show that at higher inclinations the shadow becomes more circular and looses the ``eyebrow'' features. This is something that is also observed for the Kerr shadow. The critical inclination for which the eyebrows are no longer visible is at around $35^o$.

\subsection{HT shadow for $\chi=0.327352$ and $\delta q =1 $}
\label{}

\begin{figure}[b]
\centering
\includegraphics[width=0.35\linewidth]{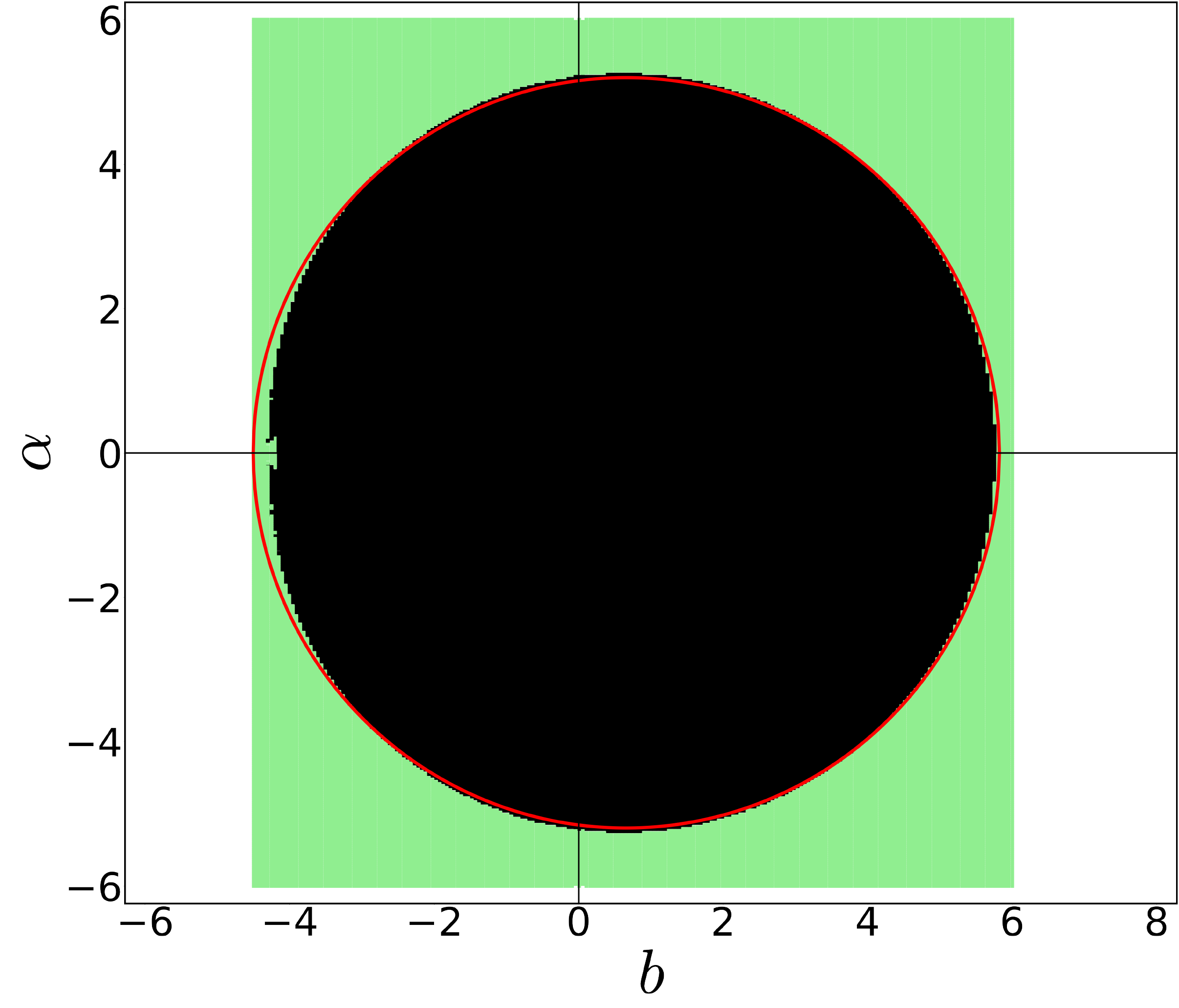}  \includegraphics[width=0.3\linewidth]{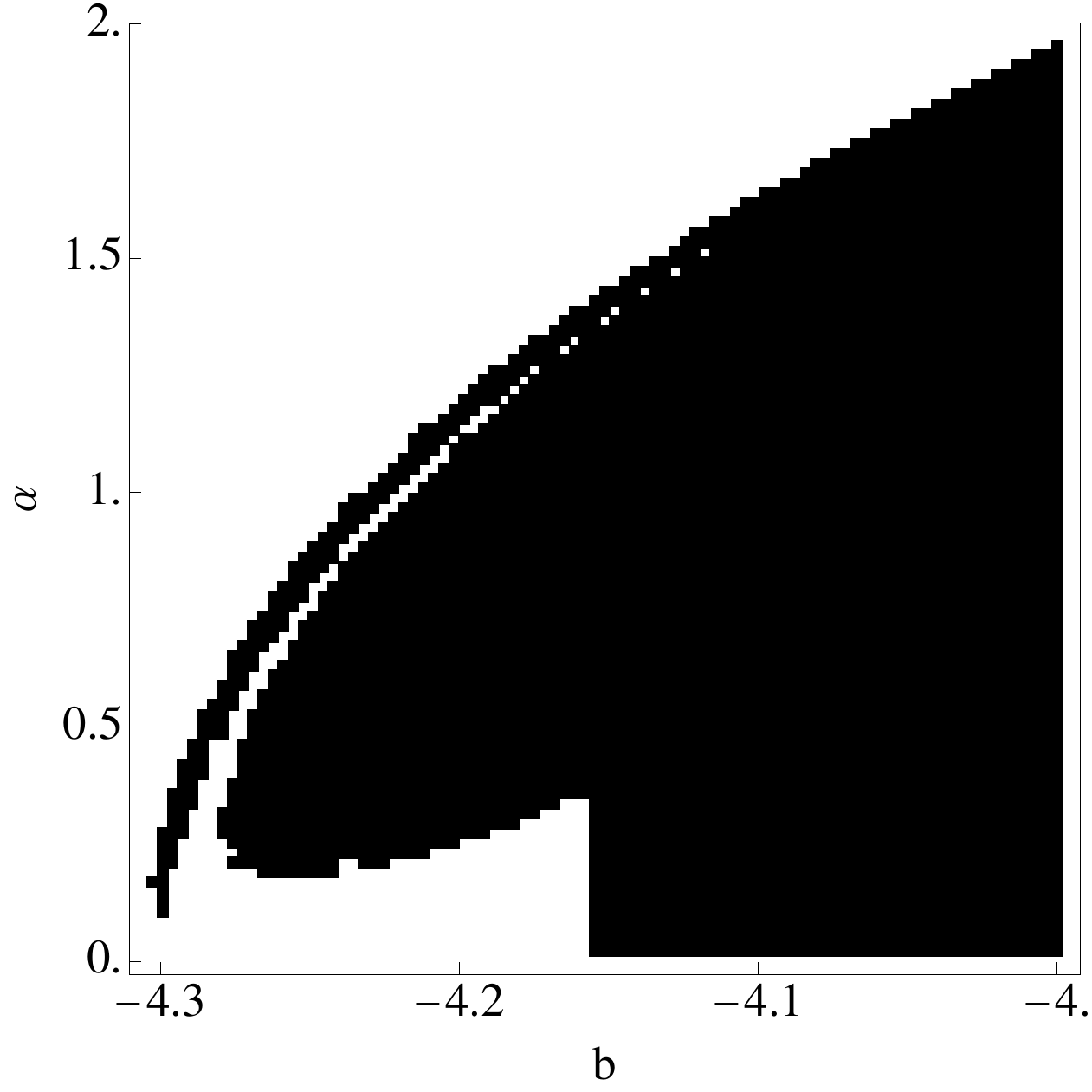}  \includegraphics[width=0.3\linewidth]{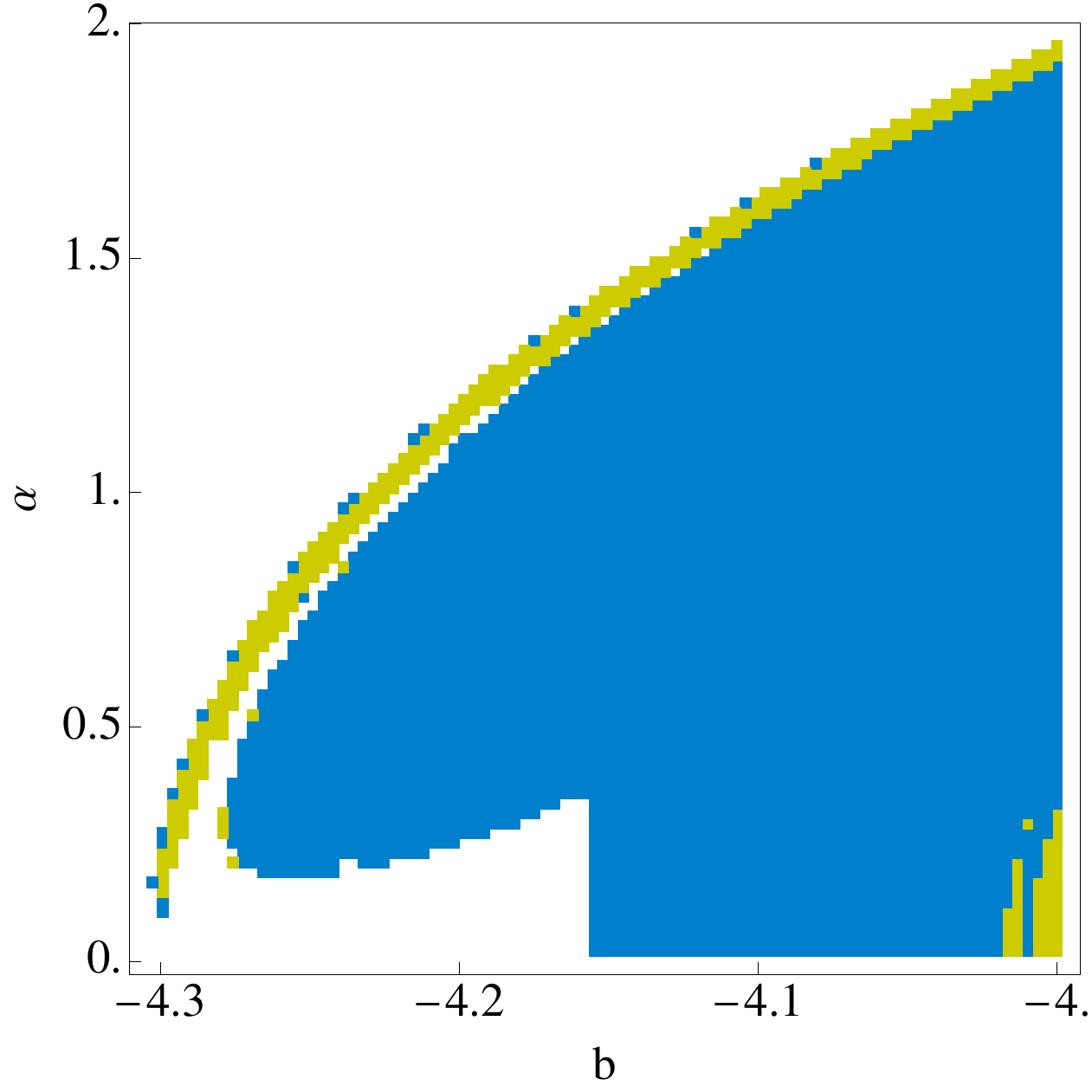} 
\caption{Left: Shadow of a compact object described by the HT spacetime with a spin parameter of $\chi=0.327352$ and quadrupole deviation from Kerr, $\delta q = 1$. We assume $r_s=2.0817767M$. In this case, there are three coexisting light-rings for the same impact parameter $b\approx -4.3M$. The red curve is that of a Kerr black hole with the same spin. Middle and Right: Close-up of the central left region of the left plot for $\alpha>0$. The shadow exhibits self-similarity related to the chaotic motion of light rays induced by the pocket. On the right, the colours indicate the throat through which the photons are lost.}
\label{fig:HT03271}
\end{figure}

The final shadow we compute is that of an object with $\delta q =1$ and $\chi = 0.327352$. We are now close to the critical spin parameter $\chi^*$ where the equatorial light-ring has bifurcated to two non-equatorial ones, but it is itself still present for the same impact parameter of $b\approx -4.3043M$ \cite{Glampedakis_Pappas2019}. In this regime, a pocket can form in the allowed motion region. The light rays can exhibit chaotic behaviour and get trapped for long time intervals. In Fig. (\ref{fig:HT03271}), we present the shadow of the object along with the shadow of a Kerr black hole with the same spin. The resolution is $90000$ pixels.

\begin{figure}[h] 
 \centering
  \begin{subfigure}[b]{0.34\linewidth}
    \centering
        \includegraphics[width=\linewidth]{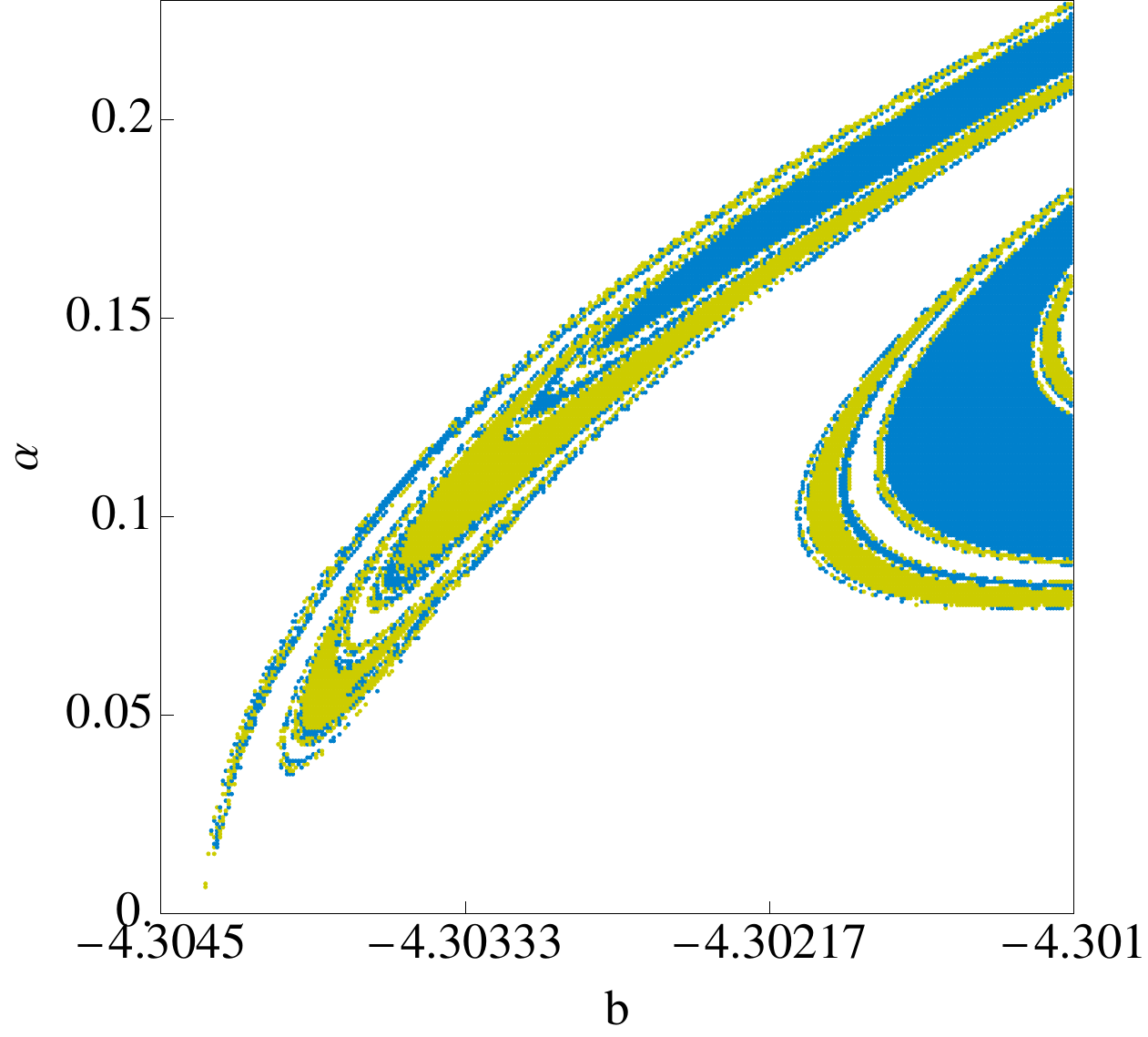} 
    \caption{} 
    \label{fig:5.11.a} 
  \end{subfigure}
  \quad
  \begin{subfigure}[b]{0.34\linewidth}
    \centering
    \includegraphics[width=\linewidth]{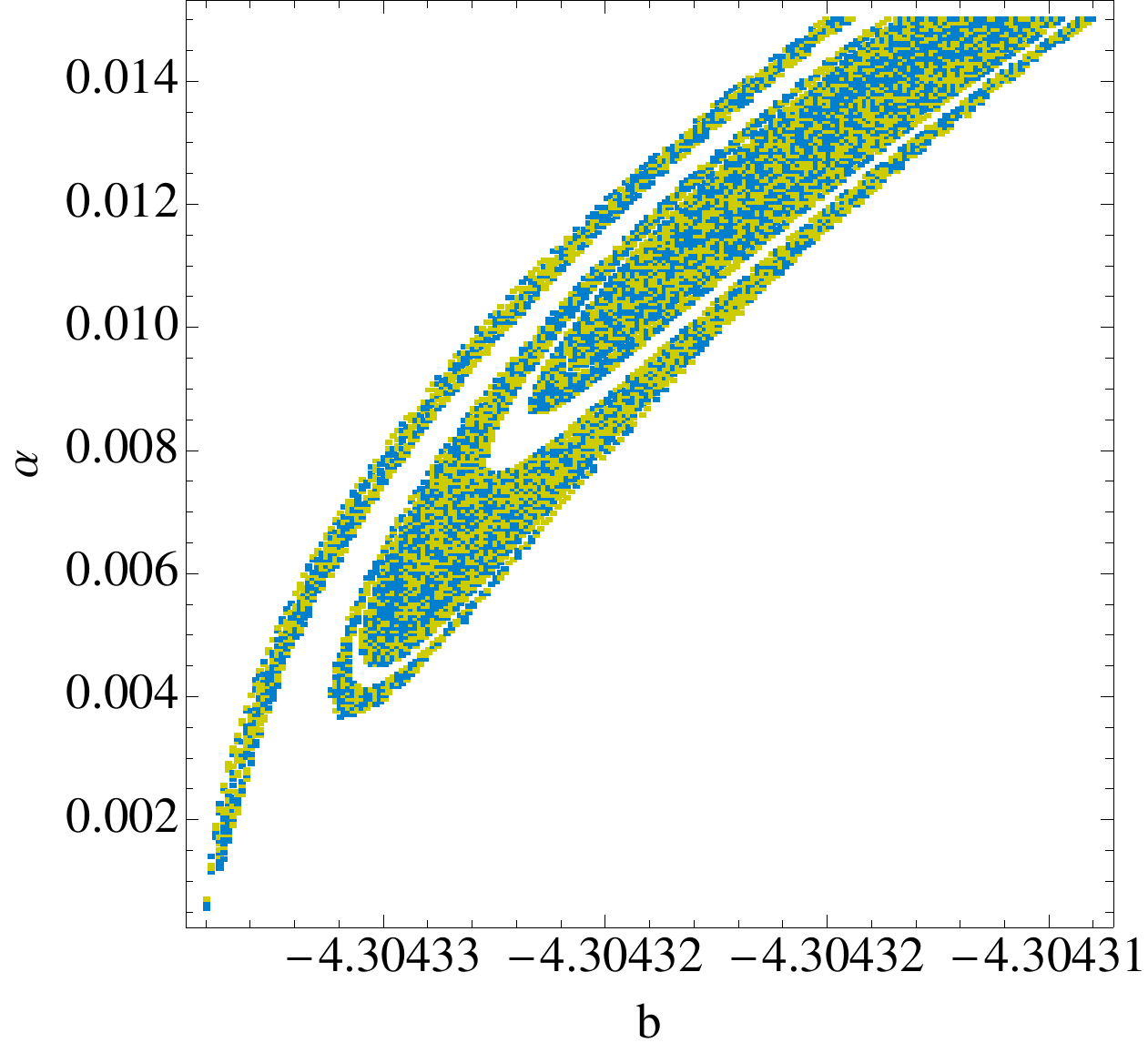} 
    \caption{} 
    \label{fig:5.11.b} 
  \end{subfigure} 
  \caption{Left: The self similar eyebrows show highly chaotic behaviour associated with the pocket. Right: A close up of the outer eyebrow exhibiting rich fractal structure. }%
  \label{fig:HTcloseup2and3}
  \end{figure}

In order to study the features of the shadow that are directly related to the pocket structure studied in Section \ref{sec:4}, the initially selected grid is not appropriate. The pocket shape of the separatrix is present for a small range of the impact parameter $b$ that is just visible in the large picture. Being in the region of chaotic motion where there is a great sensitivity in the initial conditions of the light ray, we have to zoom in in the region of interest and follow the same procedure for smaller and smaller intervals of the impact parameters to get as many qualitative features of the shadow as possible. In Fig.~(\ref{fig:HT03271}) we also show a close-up of the shadow in the regime where the pocket is present. A secondary eyebrow is visible, that is associated to the chaotic motion of light rays. In the rightmost plot %
we colour code the initial conditions as: (i) light-green for light rays that plunge into the object from the upper throat of the pocket; (ii) light-blue for light rays that plunge into the object from the lower throat of the pocket. %
We set a $300 \times 300$ grid and repeat this process, zooming in on the secondary eyebrow of Fig. \ref{fig:HT03271}. Zooming in further, there appears to be a sequence of thinner eyebrows on top of eyebrows (see Fig. \ref{fig:HTcloseup2and3}), exhibiting a self-similar hierarchy, a behaviour associated to the presence of the pocket. We also see that the throat through which the photons are lost has some chaotic structure in the shadow.

\subsection{Eyebrow Formation and Self-Similarity}
\label{}

As we have mentioned, the eyebrow features of the shadow and the fractal structure observed are associated to the presence of the off-equatorial light-rings and the pocket as well as the related chaotic motion of light rays. We will try to explain how the eyebrows form and how they are correlated, following the same principles as Yumoto et. al \cite{Yumoto_2012} did for the Majumdar-Papapetrou di-hole. 

We start with the case of the second shadow computed for the parameters $\chi=0.4$ and $\delta q =1$ (Fig.~\ref{fig:HT041}). In this regime, (see Fig. \ref{fig:bifurc1}) %
the object has two non-equatorial light-rings that correspond to light rays reflecting between the walls of each throat that connects the allowed motion region with the object. Focusing on one of the throats, since the spacetime is axisymmetric, we can set two critical impact parameters $b_+$ and $b_-$ for light rays that respectively reflect in the upper and lower wall of the throat just before escaping to infinity. The impact parameters $b_- < b < b_+$ pass through the throat and fall onto the object to form the eyebrow shape of the shadow. This can be seen more clearly in the relevant discussion in the \ref{sec:App:1}. %

In the case of the shadow computed for $\chi=0.327352$ and $\delta q =1$ (see Figs.~\ref{fig:HT03271}-\ref{fig:HTcloseup2and3}), a pocket is formed allowing the trapping of light rays for a long time. The existence of this pocket is directly related to the presence of multiple stable light-rings. Cunha et.al \cite{Cunha2016PhRvD}, showed that photons can approach and approximately resonate with each light-ring or any combination thereof. This creates a hierarchy of ''resonances'' that leads to the formation of self-similar fractal eyebrows that are infinitely many and thinner.

\section{Higher order HT, light-ring bifurcation, pockets, and chaos}
\label{sec:App:2}

  \begin{figure}[h] 
  \centering
  \begin{subfigure}[b]{1.\linewidth}
    \centering
    \includegraphics[width=0.98\linewidth]{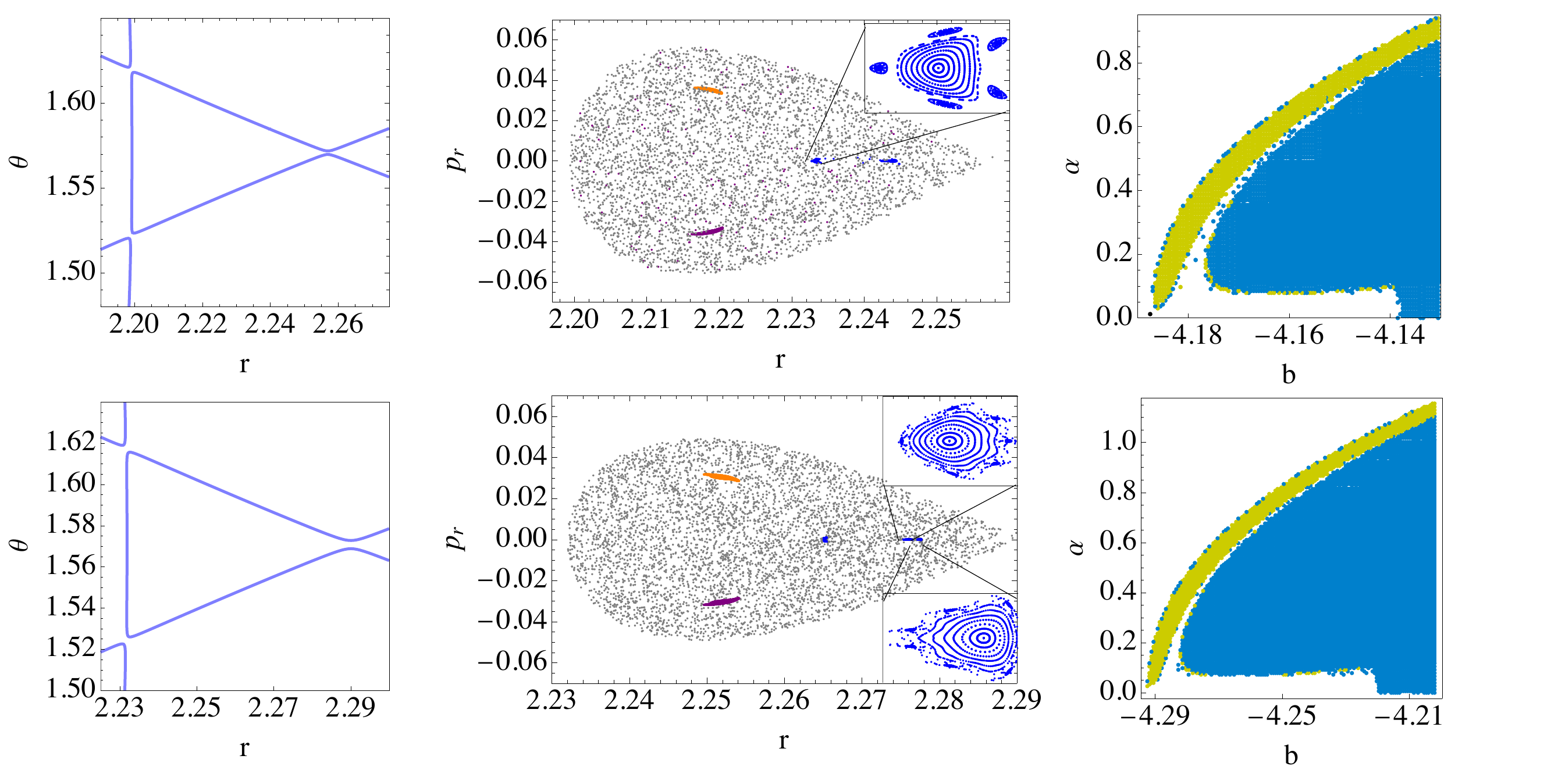} 
    \caption*{} 
    \label{fig:5.11.a} 
  \end{subfigure}
  \caption{Pocket, Poincar\'e section (in the $(r,p_r)$-plane), and fractal shadow for different values of the free parameters $(\chi,\delta q, \delta s_3)$ in the $\mathcal{O}(\Omega^3)$ expansion of the HT metric. The top row corresponds to $(0.384862,0.5,0.1)$, while the bottom corresponds to $(0.3263094,1,0.5)$. The impact parameters $b$, for the marginally open throats to exist, are $4.187521M$ and $4.292566M$ respectively. In the Poincar\'e sections we have magnified one of the regions with periodic islands that are highlighted with colour.}%
  \label{fig:robustness}
  \end{figure}
A question one may ask is how robust these results are to the order of the HT solution that we have chosen. To explore this we have also looked into the presence of the pocket in higher order expansions of the HT solutions. For example, the $\mathcal{O}(\Omega^3)$ HT has been produced in \cite{Benhar_etal2005,Glampedakis2018PhRvD} and expressed in terms of the deformation parameters, $\delta q$, that we have seen, and $\delta s_3$, which is the spin octupole-deviation parameter from the Kerr spin octupole, $S_3=-\chi^3 M^4(1-\delta s_3)$. The bifurcation of the light-ring and the feature of pockets can be found in the case of the $\mathcal{O}(\Omega^3)$ as well, just as it is found in the $\mathcal{O}(\Omega^2)$ HT spacetime, and in Fig. \ref{fig:robustness} we show such pocket examples for different combinations of the parameters $(\chi,\delta q, \delta s_3)$. The existence of the pockets is enough to indicate the existence of chaotic behaviour, as we have found for the $\mathcal{O}(\Omega^2)$ HT photon geodesics. It has been demonstrated in the literature that such pockets tend to create chaotic behaviour for both null and timelike geodesics (see for example \cite{Letelier2001,Gueron:2001ex,Cunha2016PhRvD,Shipley-Dolan2016,WangPhysRevD2018}) by introducing stable and unstable points \cite{Cunha2017PhysRevLett}, therefore we expect this to be a general feature. Even so, we have calculated Poincar\'e sections and shadows in the aforementioned cases as well, which can be seen in Fig.~\ref{fig:robustness}. The figure shows that what we have seen for the $\mathcal{O}(\Omega^2)$ extends to the $\mathcal{O}(\Omega^3)$ HT too.

These results and their robustness with respect to the HT solution's order, were in retrospect not surprising, since in the literature a connection has been made between this sort of chaotic behaviour and the introduction of prolate deformations to compact objects \cite{Letelier2001,WenBiaoHan2010,WangPhysRevD2018}. 
Furthermore, the eyebrow feature of the shadow, related to the presence of the off-equatorial light-rings, is an even more general feature, since the off-equatorial light-rings also exist for other spacetimes, apart from the $\mathcal{O}(\Omega^3)$ HT spacetime, such as the Johannsen-Psaltis spacetime \cite{Johannsen:2011dh} with prolate deformation \cite{Glampedakis_Pappas2019} or other spacetimes with prolate deformations, such as the case of the di-hole \cite{Shipley-Dolan2016} or the Manko-Novikov spacetime \cite{WangPhysRevD2018}. We are thus confident that the features observed here are a relatively general feature of spacetimes with prolate deformations. The eyebrow feature will be further explored in forthcoming work.

\section{Conclusions}

The generic end result of a collapse process in GR is a Kerr rotating black hole, where all perturbations present initially are radiated away after some time. This canonical picture of astrophysical black hole formation is being put to the test in recent years with both the development of gravitational wave astronomy and the observations made by the LIGO/VIRGO detectors, as well as the direct observation in the electromagnetic spectrum of the vicinity of black holes by the EHT. This allows for the thorough testing of the nature of black holes, a systematic investigation of their properties, and the exploration of possible deviations from GR.

For this reason theoretical models for compact objects deviating from Kerr black holes have been developed, that can behave as so called black hole mimickers and could in principle take the place of Kerr as black hole candidates. In addition, in order to test generic deviations at the level of the metric from the Kerr metric, several artificial spacetime metrics have been developed that parametrically deviate from Kerr. One such seasoned alternative metric that one could use is the HT slowly rotating spacetime, that both has the ability to parametrically deviate from Kerr while also being a natural construction of a rotating spacetime around some compact object. Furthermore, the HT spacetime has been shown to have some intriguing properties in a range of the parameter space that is relevant for black hole mimickers \cite{Glampedakis2018PhRvD}. For those parameters, HT could have some very interesting phenomenology, it can have off-equatorial light-rings and it can have a light-ring triplet that forms a pocket of allowed motion \cite{Glampedakis_Pappas2019}. 

The pocket shape of the separatrix, allows for the trapping of photons for large enough time intervals. Fixing the free parameters of the metric to appropriate values, leaves the separatrix shape dependant only on the impact parameter $b$ of the photon.  As $b$ increases and the pocket's escapes become narrower, the trapping time increases and chaos arises. At the same time, the system essentially transits from an open Hamiltonian with three escapes, to a closed one. The throats of the pocket that connect the region to the compact object, are the first ones to close, while in this case a narrow escape to spatial infinity continues to exist, and the system is \textit{marginally open}. %
Multiple stable bounded null orbits exist in this case along with unstable and chaotic ones. Photons that are launched from infinity and enter the pocket through the narrow throat can be completely chaotic, demonstrate white-noise, or exhibit time intervals where the motion is stagnant in the boundaries of stable periodic orbits. Therefore, while the motion of the latter \textit{sticky} photons is mainly chaotic, there are time intervals where it behaves as periodic. %
The respective Fourier spectra follow the $1/f$ power law in the low frequency regime, a tell-tale of \textit{Stickiness}. Further increase of $b$ separates the pocket from all escapes and a \textit{cut-off pocket} forms, which is a closed Hamiltonian system. The motion is completely ordered for higher values of $b$, right before the cut-off pocket disappears. This means that as $b$ decreases the system transits from order to chaos. 
Behaviour similar to the one found here, associated with the existence of a pocket has been found in the literature for other systems as well, such as the Majumdar-Papapetrou di-hole \cite{Shipley-Dolan2016,shipley2019strongfield}, or other \cite{Letelier2001,Gueron:2001ex,Cunha2016PhRvD,WangPhysRevD2018}, and therefore is a generic feature of having such pockets.

The aforementioned behaviour of the photon orbits affects significantly the shadow of the compact object. 
For a large enough quadrupole deviation, $\delta q =1$, and an appropriate spin, $\chi=0.4$, the shadow of a HT compact object exhibits a breakdown of circularity due to the equatorial light-ring's bifurcation to two non-equatorial ones. Photons that bounce off the throats of the pocket and escape to infinity, confer eyebrow features to the shadow's shape, while the shadow's deviation from circularity resembles that of rapidly rotating Kerr BHs. In the regime where the three light-rings coexist (around $\chi=0.327352$) the shadow of the object is circular and nearly coincides with the shadow of a Kerr BH with the same spin parameter. The circle breaks for a small range of the impact parameter $b$ due to the existence of the pocket and the three co-rotating light-rings, and small eyebrows are formed. The eyebrows in this case, appear to be infinitely many and thinner, exhibiting a self-similar hierarchy, a typical property of fractals. 
This occurs due to photons sticking to different combinations of the existing stable bounded null orbits in the pocket and has been found for other systems as well \cite{Cunha2016PhRvD,Shipley-Dolan2016,WangPhysRevD2018}. The observed fractal structure will be further explored in future work, while the light-ring bifurcation process is also worth exploring from the dynamical systems perspective.

These results, related to the shadow of such non-Kerr objects, are of great interest to the EHT and its future incarnations \cite{EHT2019ApJ.I,EHT2019ApJ.V,EHT2019ApJ.VI}, when enough resolution will be available to hopefully be able to observe and resolve higher order light-rings or the fractal features found in this work as well as in other cases of UCOs \cite{Cunha2016PhRvD}. Such features will be smoking gun indications of the non-Kerr nature of such an object. Furthermore, it is worth looking for more spacetimes that have these properties and perform a systematic study. 

Regarding the shadow observations using interferometers, further work needs to be done in modelling the astrophysical environment and the sources of the emitted photons in order to see how flux is accumulated and how that affects the actual observed image. Another interesting question is whether the pocket feature with the trapping of photons could lead to any sort of instability. Finally, the features explored here should also affect the quasi-normal modes of the system and this is something also worth exploring. 

We conclude by noting that the results presented here are robust to higher order expansions of the metric \cite{Benhar_etal2005,Yagi_2014,Yagi_2015,Glampedakis2018PhRvD}. The pocket feature and the resulting chaotic behaviour is not a result of poor approximation but seems to be a rather general feature of a spacetime with a prolate deformation, as it has been found in the literature in other spacetime cases as well \cite{Letelier2001,Gueron:2001ex,Cunha2016PhRvD,Shipley-Dolan2016,WangPhysRevD2018}. We can see that it is present also in the $\mathcal{O}$($\Omega^3$) HT spacetime, for different combinations of the free parameters $\chi,\delta q,$ and $\delta s_3$ 
(see Fig.\ref{fig:robustness}). Therefore, the phenomenology related to the pocket and the off-equatorial light-rings is expected to persist in general.

\ack

All the computations were done using programs written in Python language in the open-source software SageMath 9.2, using the SageManifolds extension.\footnote{ \url{https://sagemanifolds.obspm.fr/}} The main tool we used is the available built-in geodesics integrator that implements SciPy and invokes the LSODA algorithm. In the majority of the computations, the integration of several thousand geodesics was needed, a rather slow procedure given the complexity of the HT metric. Since a rather basic PC and not a workstation was used for most of the computations, we implemented the multi-processing Python module in order to exploit every available resource. This dropped the overall data processing time to $1/10$-th of an equivalent serial program. Notebooks with the codes used and demo calculations are available on github.\footnote{\url{https://github.com/KstrsKostas/GRsmcode}}

The authors would like to thank the developers of SageMath and SageManifolds and in particular Eric Gourgoulhon, for the great job they have done developing it and creating documentation and examples which we thoroughly exploited. Many computations have been done independently using Mathematica as it is hard to teach an old dog new tricks. The Mathematica licence was provided by the Aristotle University of Thessaloniki. Some results presented in this work have been produced using the Aristotle University of Thessaloniki (AUTh) High Performance Computing Infrastructure and Resources.

\appendix

\section{Ray-tracing algorithm}
\label{sec:App:0}

With the HT spacetime being non-integrable, an analytical calculation of the shadow is not possible. Therefore we use numerical methods to integrate the geodesic equations. The method that we use is the so called \textit{backwards ray-tracing} \cite{shipley2019strongfield}, and we implement it using SageMath 9.2. In brief, the algorithm works as follows: (i) We set a fixed starting point for our light rays far enough so that the spacetime there can be considered practically flat. Furthermore we assume that the observer is on the equatorial plane, therefore our initial conditions are $\left\{r_0=100M,\theta_0=\pi/2\right\}$;   
(ii) We create a large grid of resolution $N \times M$, where $N$ and $M$ is the number of different values of our impact parameters $b$ and $\alpha$ respectively; (iii) We numerically integrate the geodesic equations for every pair of $\left\{b, \alpha\right\}$ while using conditions in order to determine whether the light ray escapes or plunges into the compact object; (iv) We finally scatter plot the shadow using each point as a pixel. \footnote{On a technical note, the HT geodesics required significantly more time to be numerically integrated than the much simpler case of Kerr. In order to counter this issue, we used Python's multiprocessing module which highly increased the overall computation speed.}

In the case of BH's, pathologies such as curvature singularities are hidden behind horizons and are therefore inaccessible to distant observers. They are, in that sense, harmless \cite{Cardoso:2019rvt}. Since UCO's are horizonless (i.e., have a surface) and we use the HT metric to describe their exterior spacetime, we assume that possible irregular regions will not exist outside the surface and therefore will not interfere with the numerical integration of the geodesics. The two main issues encountered in this case are the Killing Horizon, that is an $(r,\theta)$ surface defined by $\mathcal{D}=0$, and a singularity caused by $g_{rr}=0$, as we discussed briefly in Section \ref{sec:2}. %
Therefore, to avoid these problems we assume that any photon that reaches a minimum radius of $r_s$, that depends on the parameters of the compact object, is lost since it will either hit the surface at that point or will eventually hit the surface of the compact object which is a little further in. \footnote{In general the pathologies that we need to avoid do not interfere with the allowed region of motion for the photons, given by the zero velocity separatrix. There is only one instance that this happens for some of the models, for a small range in $b$, where the $g_{rr}=0$ surface slightly crosses the separatrix and we use the coordinate radius where this happens to select $r_s$. More details on this in the \ref{sec:App:1}}

\begin{figure}[h]
    \centering
    \includegraphics[width=0.35\textwidth]{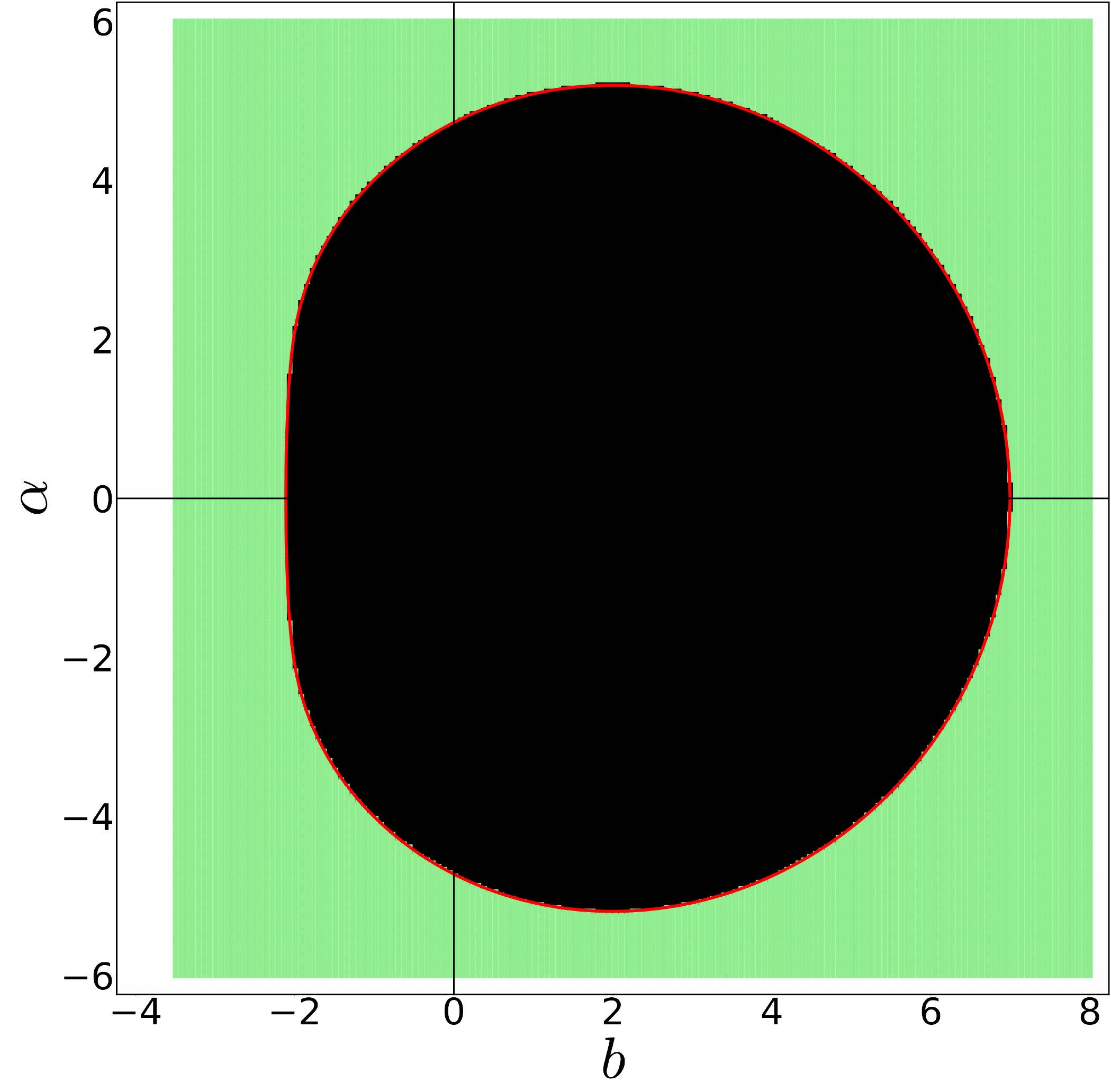}
    \caption{Shadow of a spinning Kerr black hole with a spin parameter of $\chi=0.998$, viewed from $\theta=\pi/2$, computed with the backwards ray-tracing method. The red curve represents the analytical solution of the shadow.}
    \label{fig:KerrShadow}
\end{figure}

To test our algorithm, we started by computing the shadow for a Kerr BH with a spin parameter of $\chi=0.998$. In Fig. \ref{fig:KerrShadow} we present here the shadow we computed and compare it to the known result for the same spin. We created a grid of $150\times 150$ points for positive values of the impact parameter $\alpha$, since the shadow is symmetric with respect to to the horizontal axis, and drew a shadow consisting of 45000 light rays. The green pixels correspond to light rays that escape to infinity while the black ones are lost to the BH. We can see that the results of the code are quite accurate and can therefore proceed to calculate the HT shadows.

\section{Anatomy of the HT shadow}
\label{sec:App:1}

In this Appendix we address the issue of the radius $r_s$ beyond which we assume the photons are lost (to the surface of the compact object), and show how it affects the shadow, and that it does not affect the interesting features related to the off-equatorial light-rings. To illustrate the various possibilities for the orbits, we assume the compact object model with $\chi=0.4$ and $\delta q=1$, for which we had selected the surface where the photons are lost to be at $r_s=2.115866M$. 

To see how the surface at $r_s$ and the separatrix interact to form the shadow, we have calculated various orbits for a sequence of values of $b$ near the left edge of the shadow for which we have selected a few values of $\alpha$. The values of $b$ are selected so as to move from just outside the shadow towards the interior, crossing the shadow's outline at different points, while the values of $\alpha$ are chosen near the crossing points or at other locations of interest near the shadow. 

We remind here, that since we will be discussing the shadow, the definition for the impact parameter that we will be using is the same as we used in Section \ref{sec:5}, i.e., $b=p_{\phi}/p_t$, which means that co-rotating orbits have negative $b$.    
%
\begin{table}[h]
\caption{Points on the shadow image for which we calculate the various photon trajectories (see Fig. \ref{fig:surface}).}
\label{tab:points}
\begin{center}
\begin{indented}
\item[]\begin{tabular}{c c c c c }
    \br
    $b$ & $\alpha_{1}$& $\alpha_{2}$& $\alpha_{3}$& $\alpha_{4}$ \\
    \mr
    -3.716981 & $1.86792$& $1.89623$& $1.58491$& $0.1$ \\
    -3.603774 & $ 1.83962$ & $ 2.29245$ & $1.10377$ & $0.1 $ \\ 
   -3.490566 & $ 1.81132$ & $ 2.60377$ & $ 2.00943$ & $0.1 $  \\
   -3.377358 & $2.26415$ & $2.15094$ & $2.83019$ & $ 0.1$  \\
    \br
\end{tabular}
\end{indented}
\end{center}
\end{table}
%
\begin{figure}[h]
\centering
\begin{subfigure}[h]{0.243\linewidth}\centering
\includegraphics[height=1\textwidth]{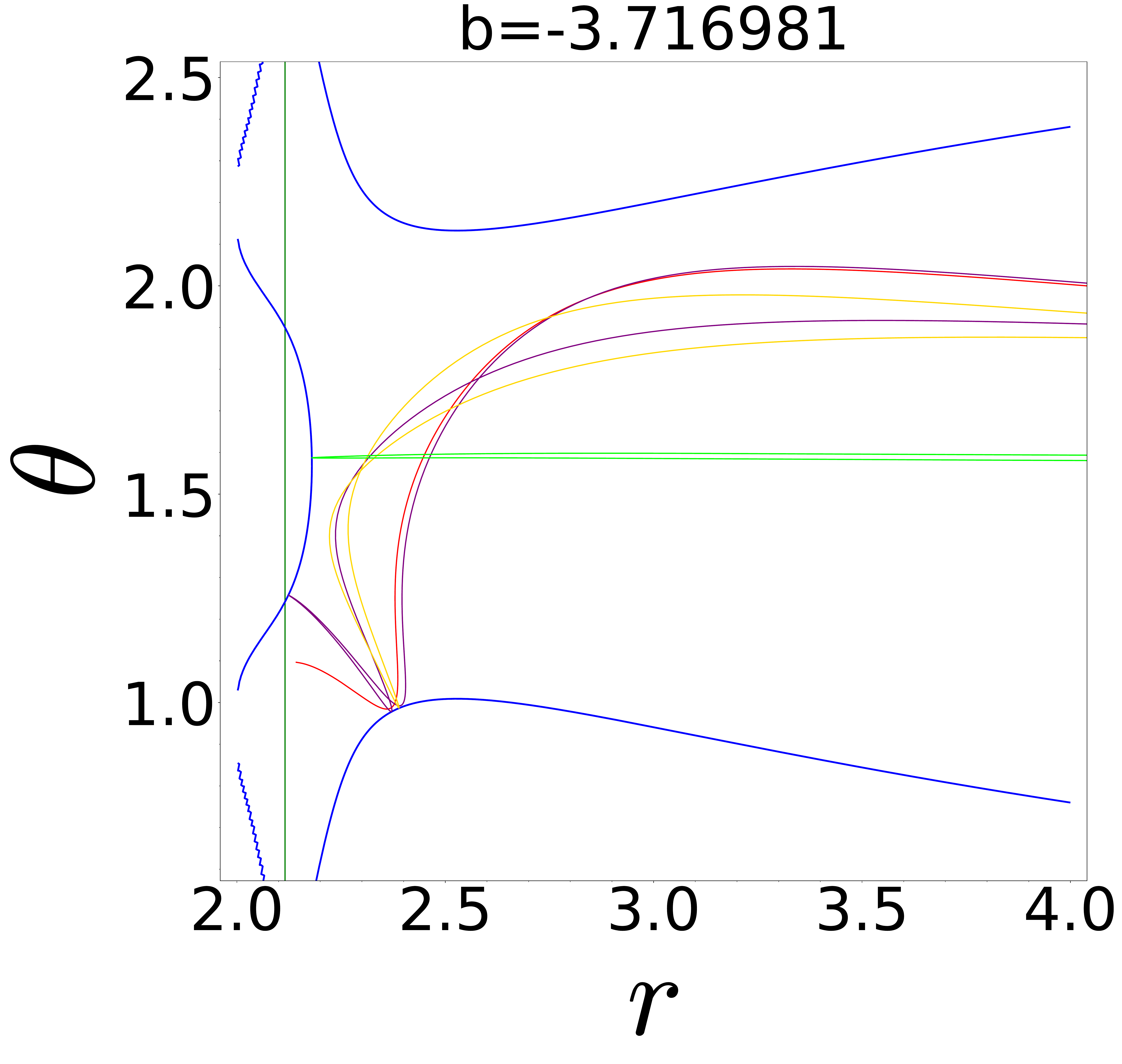}
\caption*{}
\end{subfigure}
\begin{subfigure}[h]{0.243\linewidth}\centering
\includegraphics[height=1\textwidth]{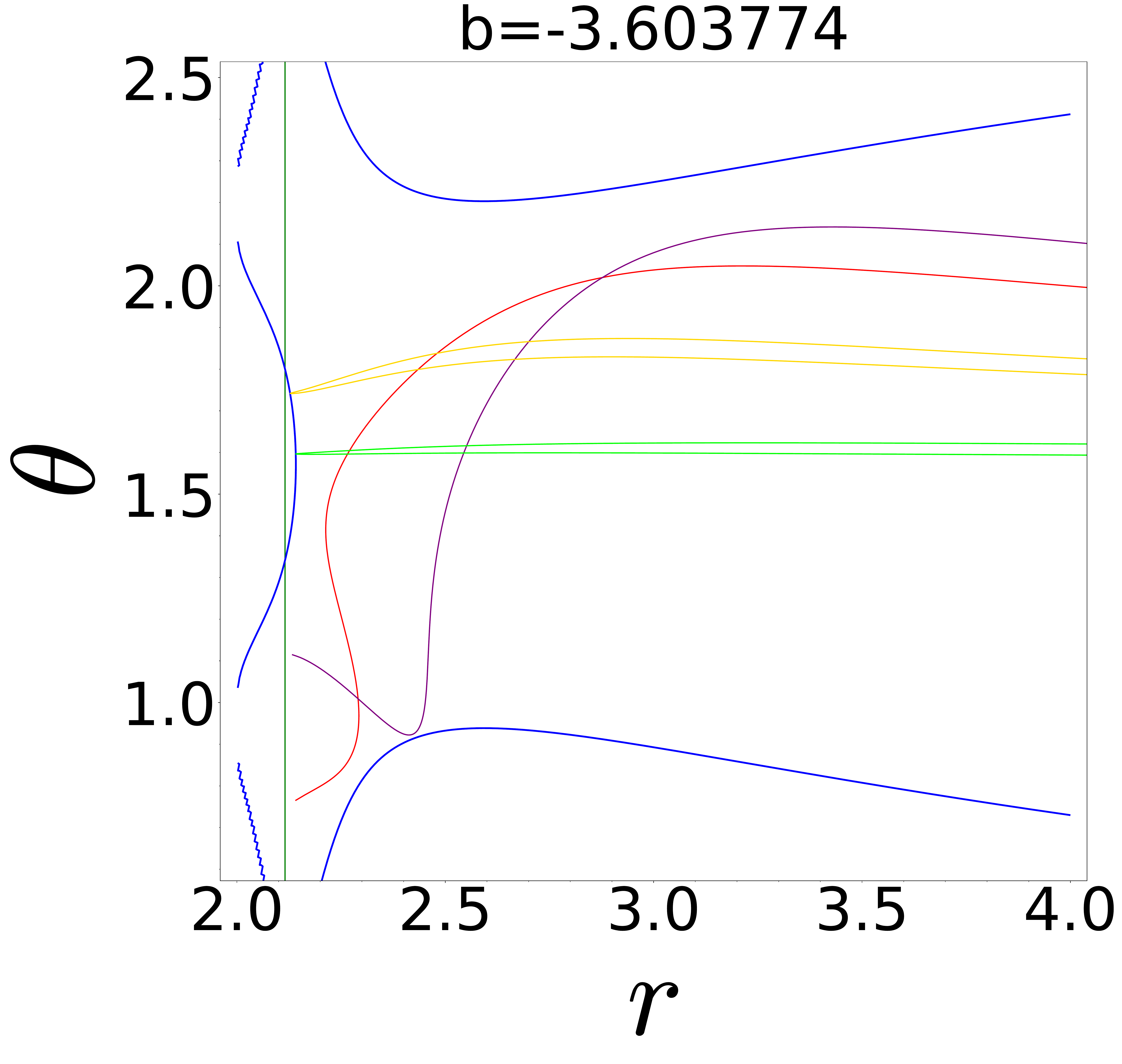}
\caption*{}
\end{subfigure}
\begin{subfigure}[h]{0.243\linewidth}\centering
\includegraphics[height=1\textwidth]{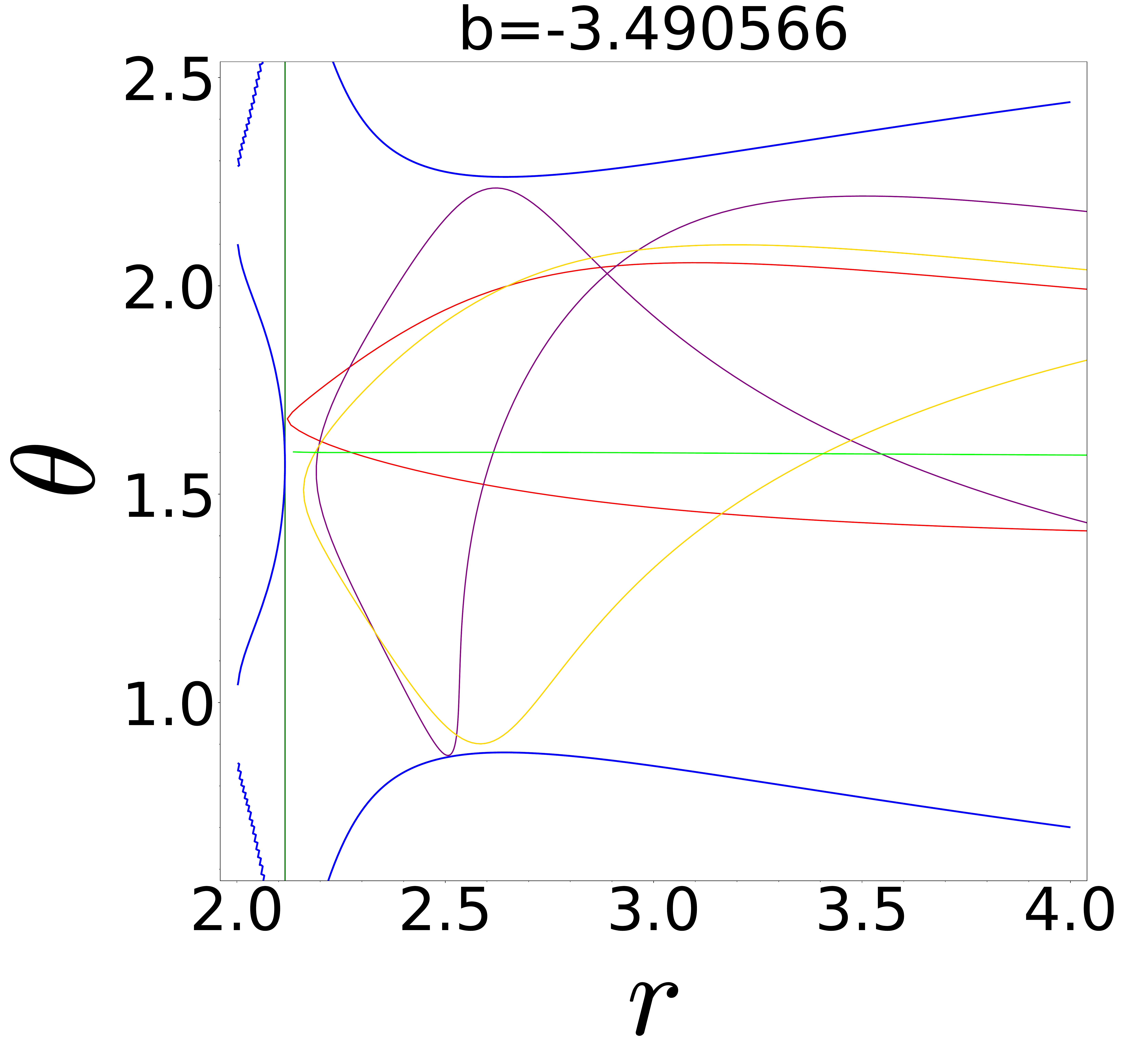}
\caption*{}
\end{subfigure}
\begin{subfigure}[h]{0.243\linewidth}\centering
\includegraphics[height=1\textwidth]{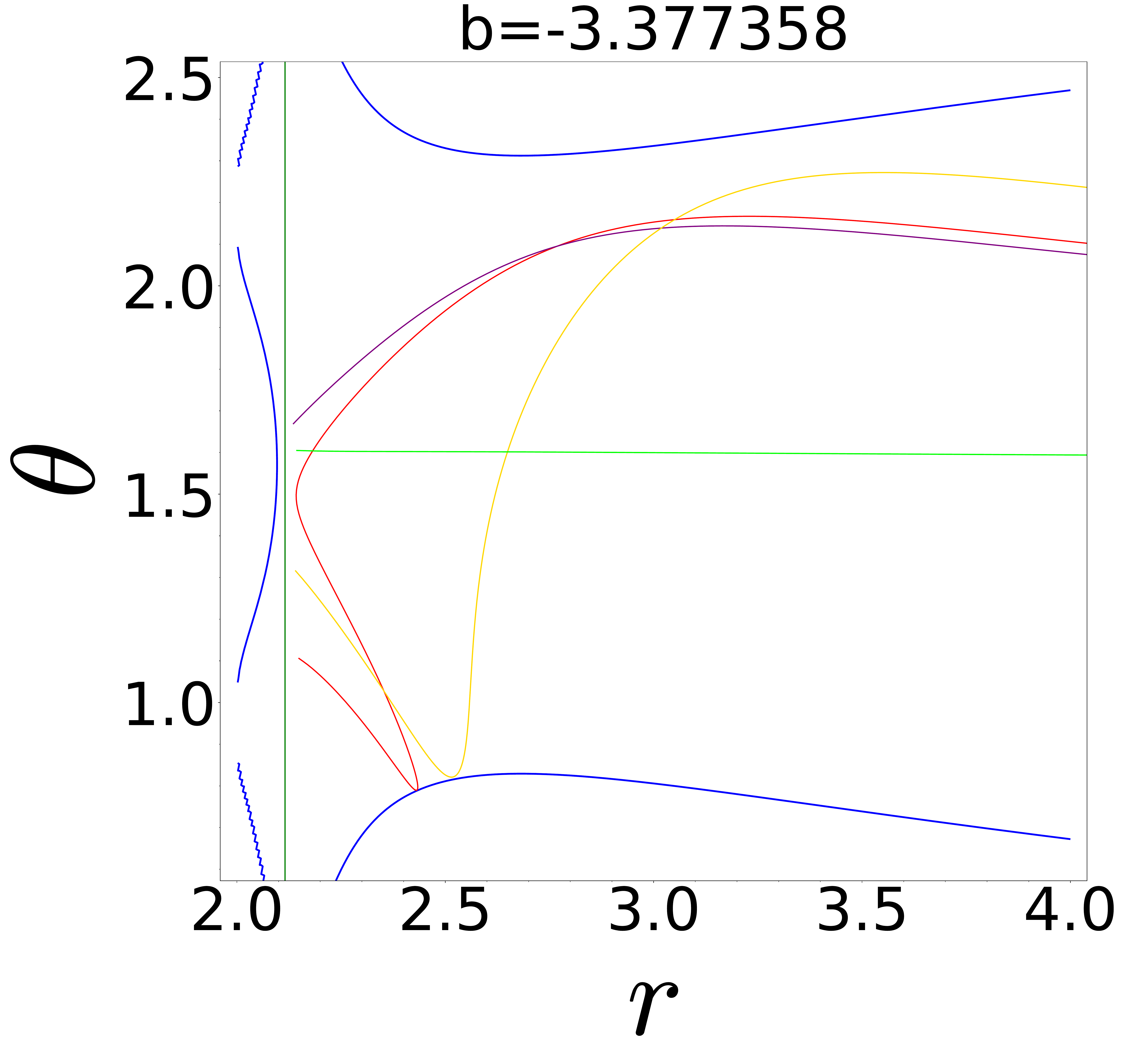}
\caption*{}
\end{subfigure}%
\\
\begin{subfigure}[h]{0.25\linewidth}\centering
\includegraphics[height=1\textwidth]{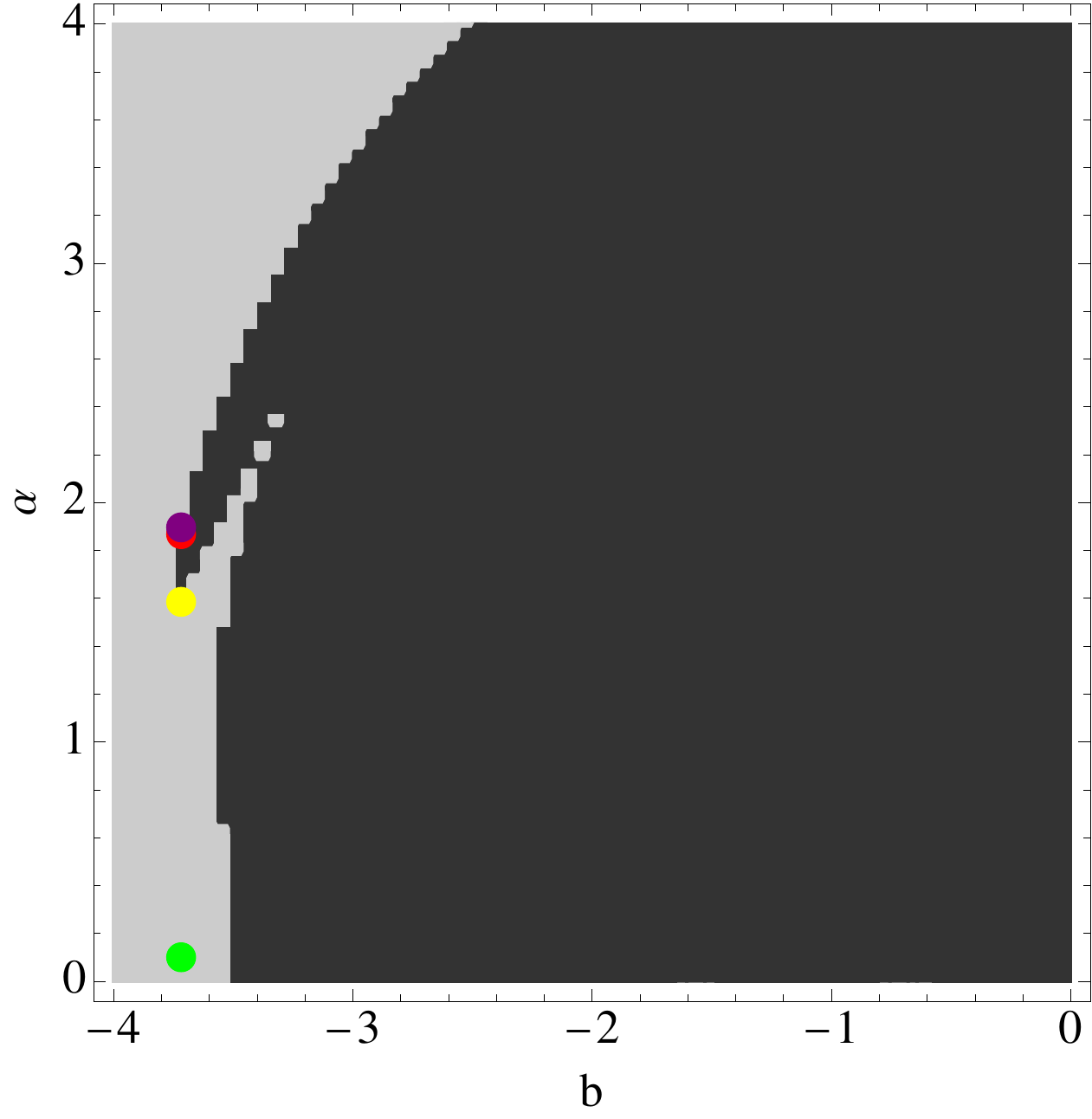}
\caption*{}
\end{subfigure}%
\begin{subfigure}[h]{0.25\linewidth}\centering
\includegraphics[height=1\textwidth]{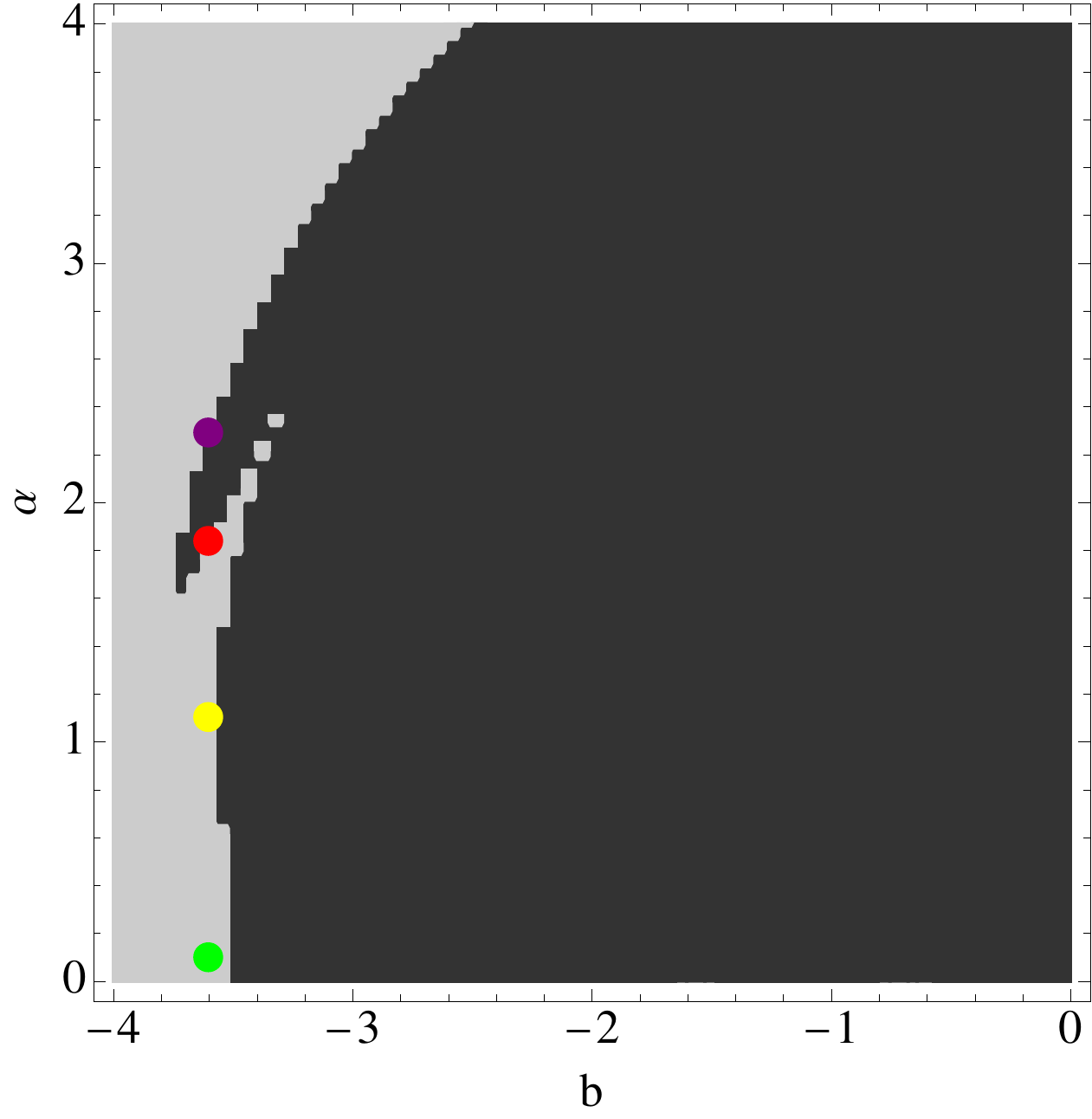}
\caption*{}
\end{subfigure}%
\begin{subfigure}[h]{0.25\linewidth}\centering
\includegraphics[height=1\textwidth]{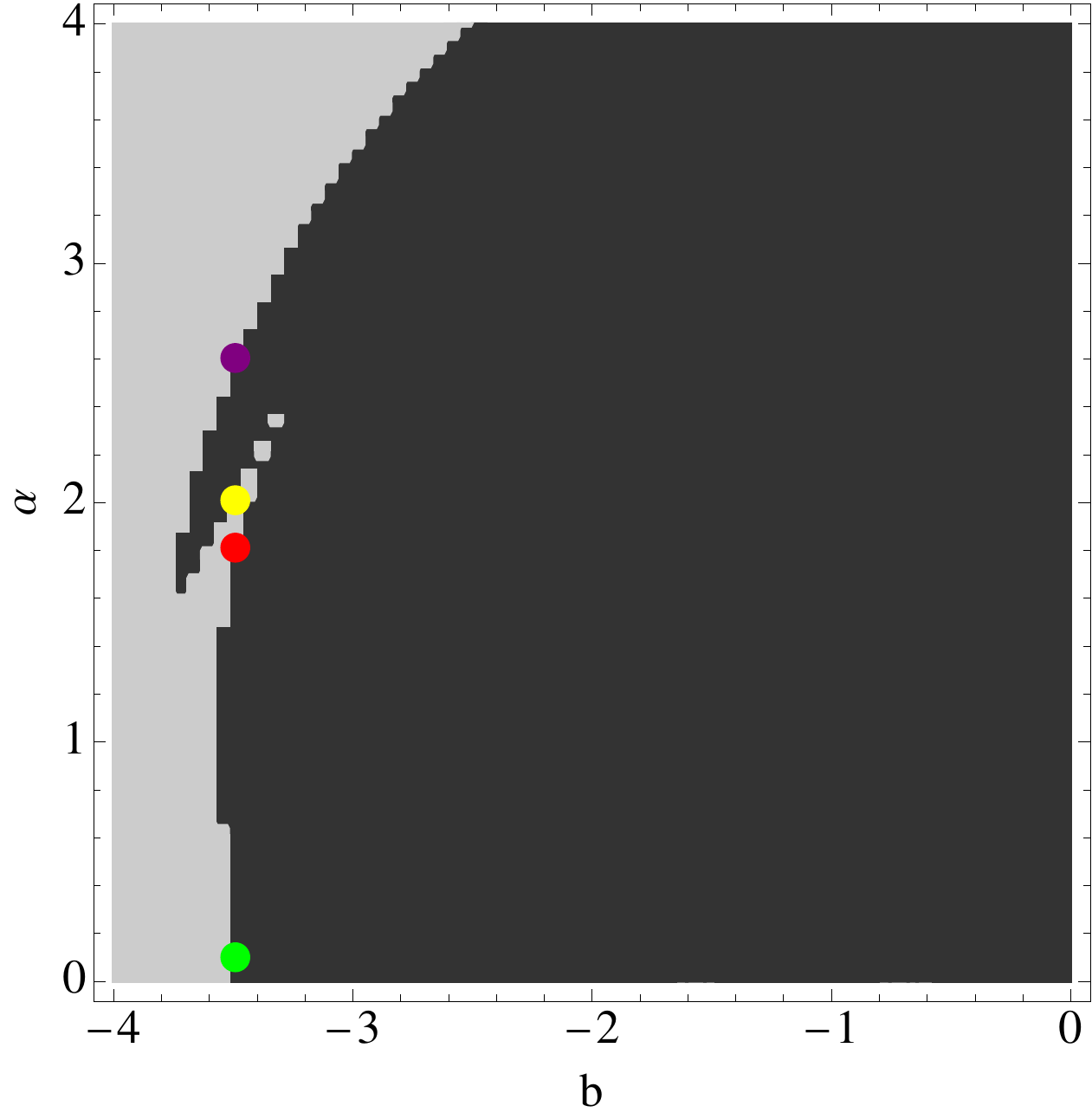}
\caption*{}
\end{subfigure}%
\begin{subfigure}[h]{0.25\linewidth}\centering
\includegraphics[height=1\textwidth]{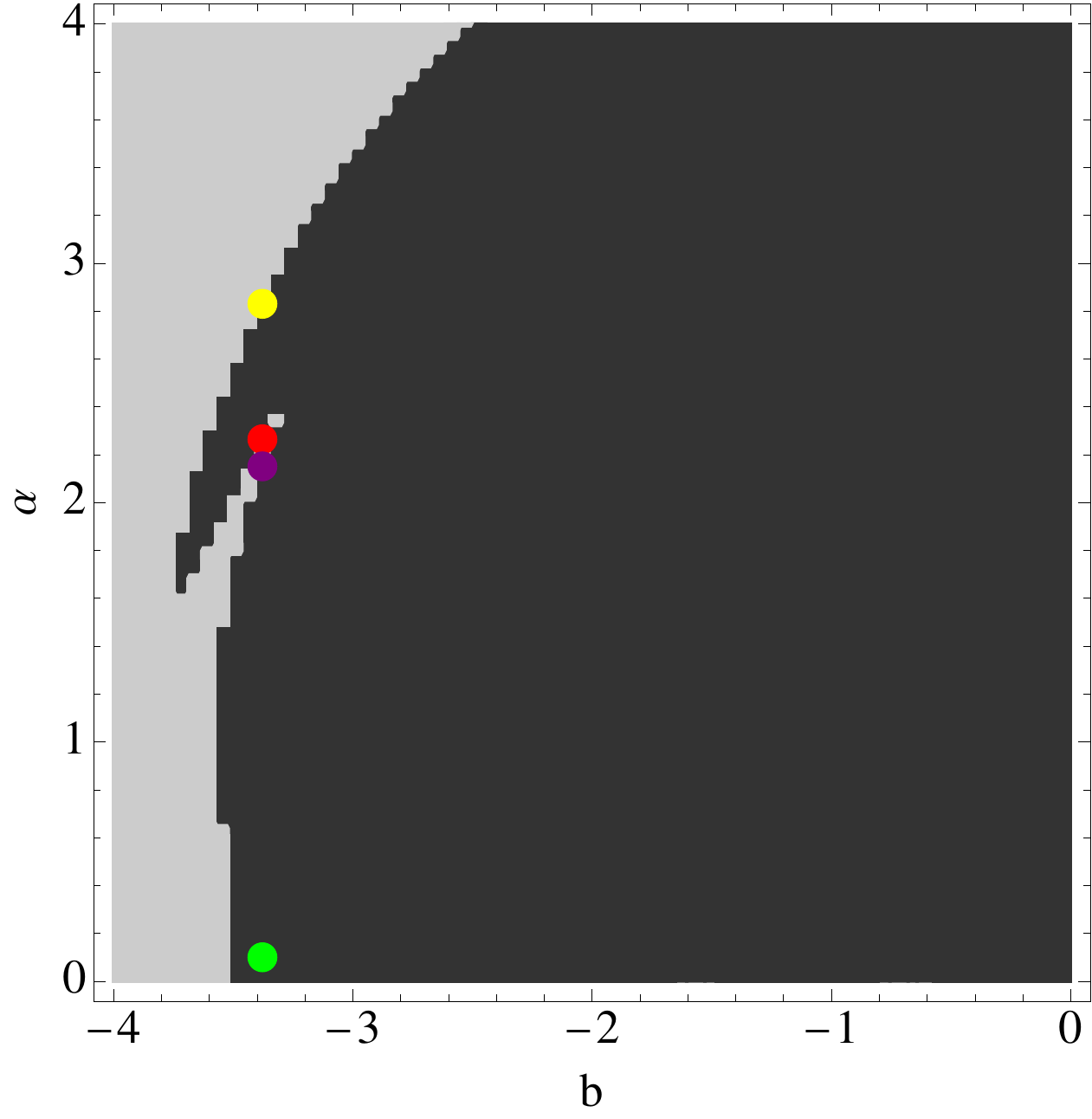}
\caption*{}
\end{subfigure}%
\caption{On the top row we plot 4 different orbits for 4 values of $\alpha$ (4 different colours), each time for a given choice of $b$ and show on the bottom row the points on the shadow image that these orbits correspond to (same colour coding). The blue line in the top row plots is the separatrix, while the constant $r$ line is the surface $r=r_s$. The compact object is the one with $\chi=0.4$, $\delta q=1$, $r_s\simeq 2.115866M$, as in Fig. \ref{fig:HT041}. }
\label{fig:surface}
\end{figure}
We take four cross sections of the shadow at different values of $b$, and for each cross section we choose four values of $\alpha$. The selected points are shown in Tab. \ref{tab:points}. For these points (the combinations of $b$, and $\alpha$ constitute initial conditions for photon trajectories) we calculate the corresponding orbits, which we plot in Fig. \ref{fig:surface} together with the shadow, on which we have placed these same points.

For the first column of plots on the left in Fig. \ref{fig:surface}, we can see that of the 4 trajectories selected, only the red one enters one of the throats and is lost (the integration terminating at $r_s$), while the yellow, purple and green are scattered back to infinity. These orbits correspond to points on the edge and inside the shadow, with the red one being inside the eyebrow formation. At the second column we see 2 out of 4 orbits being lost through the throat (red and purple) and the other 2 (yellow and green) being reflected by the separatrix near the equatorial plane. At the third column we see 3 orbits being reflected before getting to $r_s$ (yellow, red, and purple), while the green orbit is lost on $r_s$. Finally, on the last column on the right, we see that all the orbits are lost on $r_s$ and the corresponding points are all parts of the shadow.    

From the trajectories and the corresponding points on the shadow, we can see that choosing a surface $r_s$ does not interfere with the formation of the eyebrow feature of the shadow. This is particularly clear from the first two columns where the throat determines the range in which orbits are lost and therefore determines the thickness of the eyebrow. The third column also demonstrates that further in in the bright region inside the eyebrow, even though the surface at $r_s$ is almost completely outside the separatrix, there are trajectories that can still be reflected before getting to $r_s$ and escape. Therefore, even though the choice of $r_s$ affects the shadow (especially the orbits with small values of $\alpha$) it is not playing a crucial part in the formation of the eyebrows.

\section{Another example of sticky orbits}
\label{sec:App:3}

\begin{figure}[h]
\centering
\begin{subfigure}[h]{0.32\linewidth}
\includegraphics[width=\linewidth]{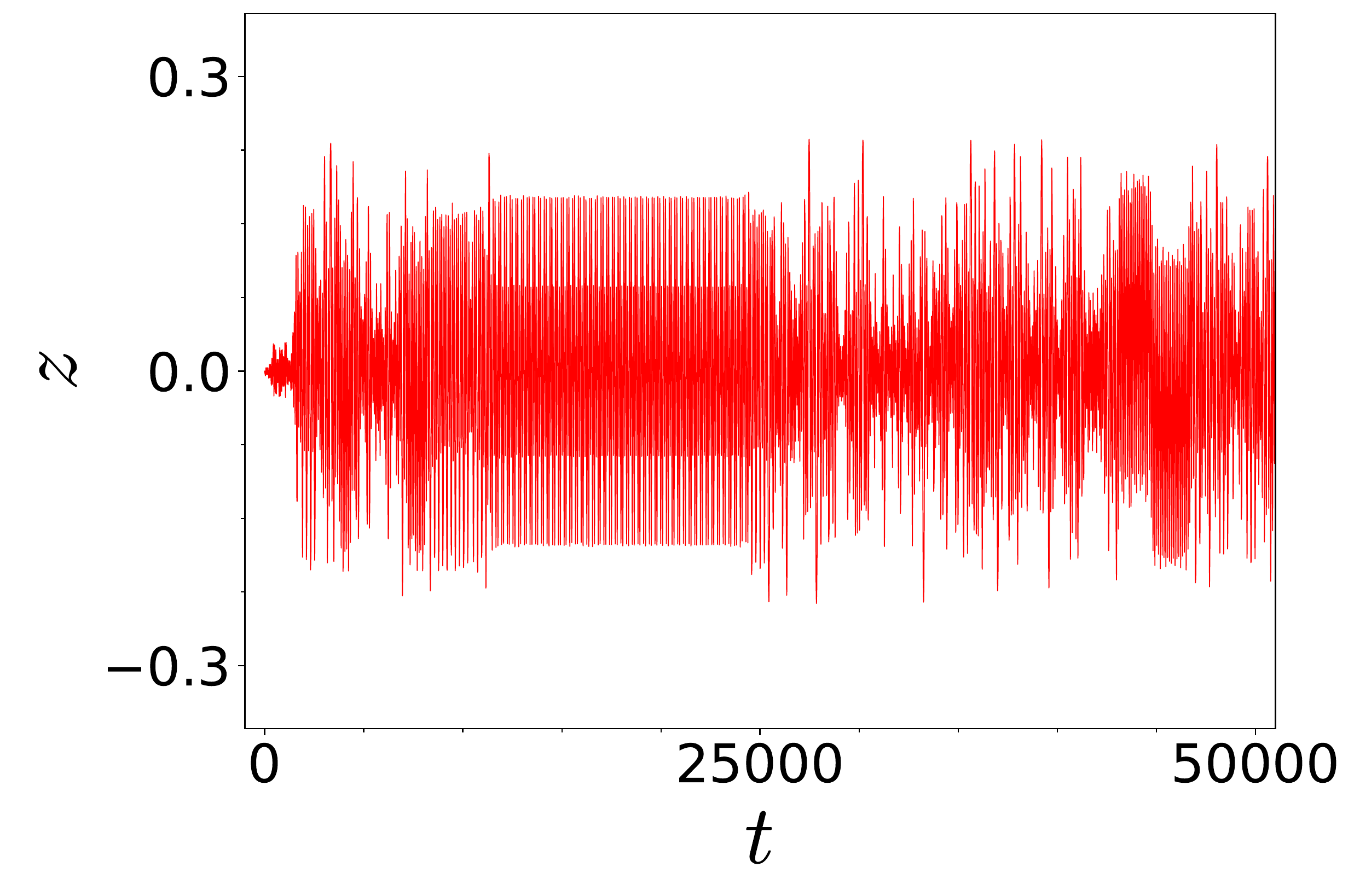}
\caption*{}
\end{subfigure}
\begin{subfigure}[h]{0.32\linewidth}
\includegraphics[width=\linewidth]{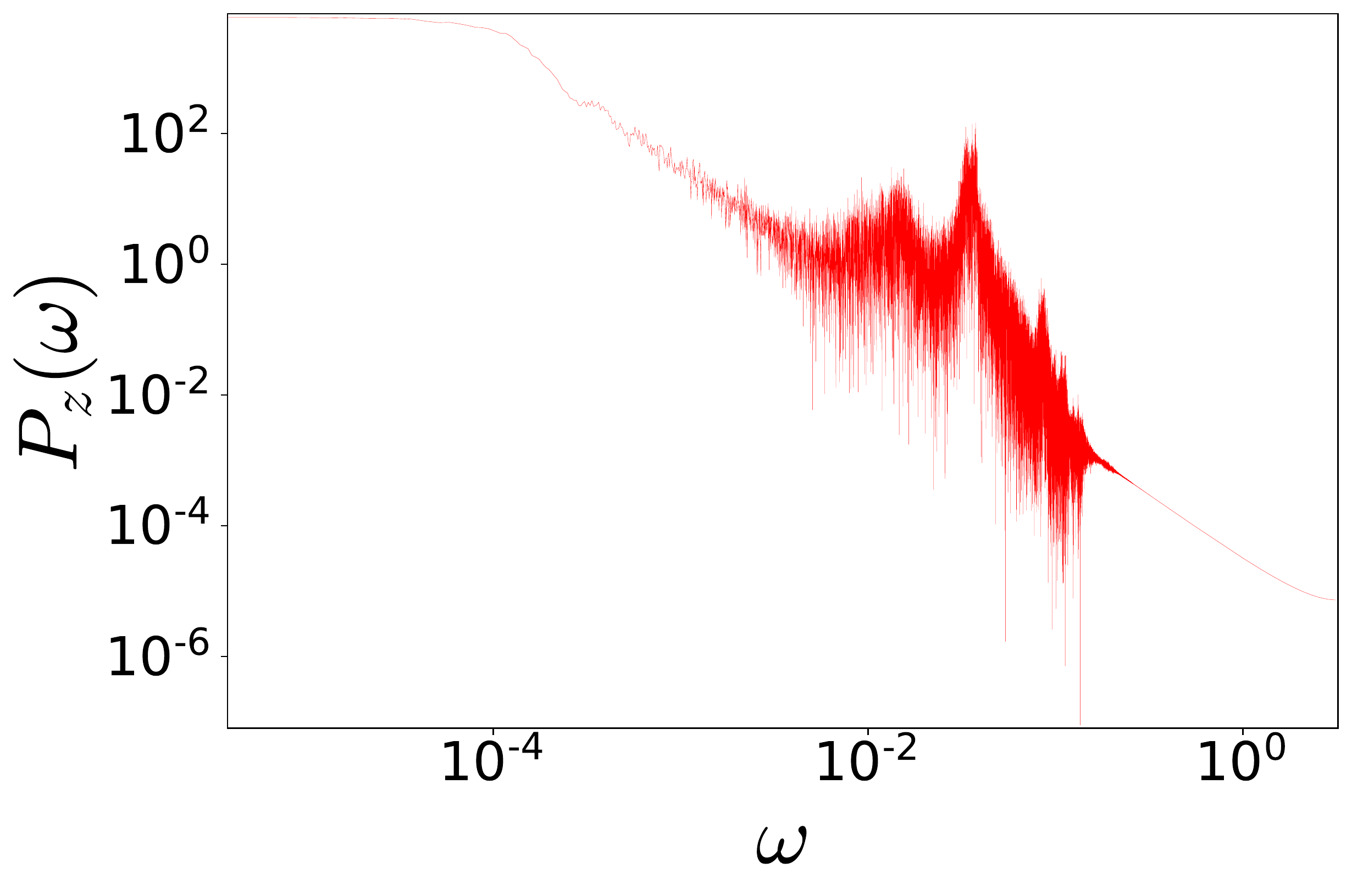}
\caption*{}
\end{subfigure}
\begin{subfigure}[h]{0.32\linewidth}
\includegraphics[width=\linewidth]{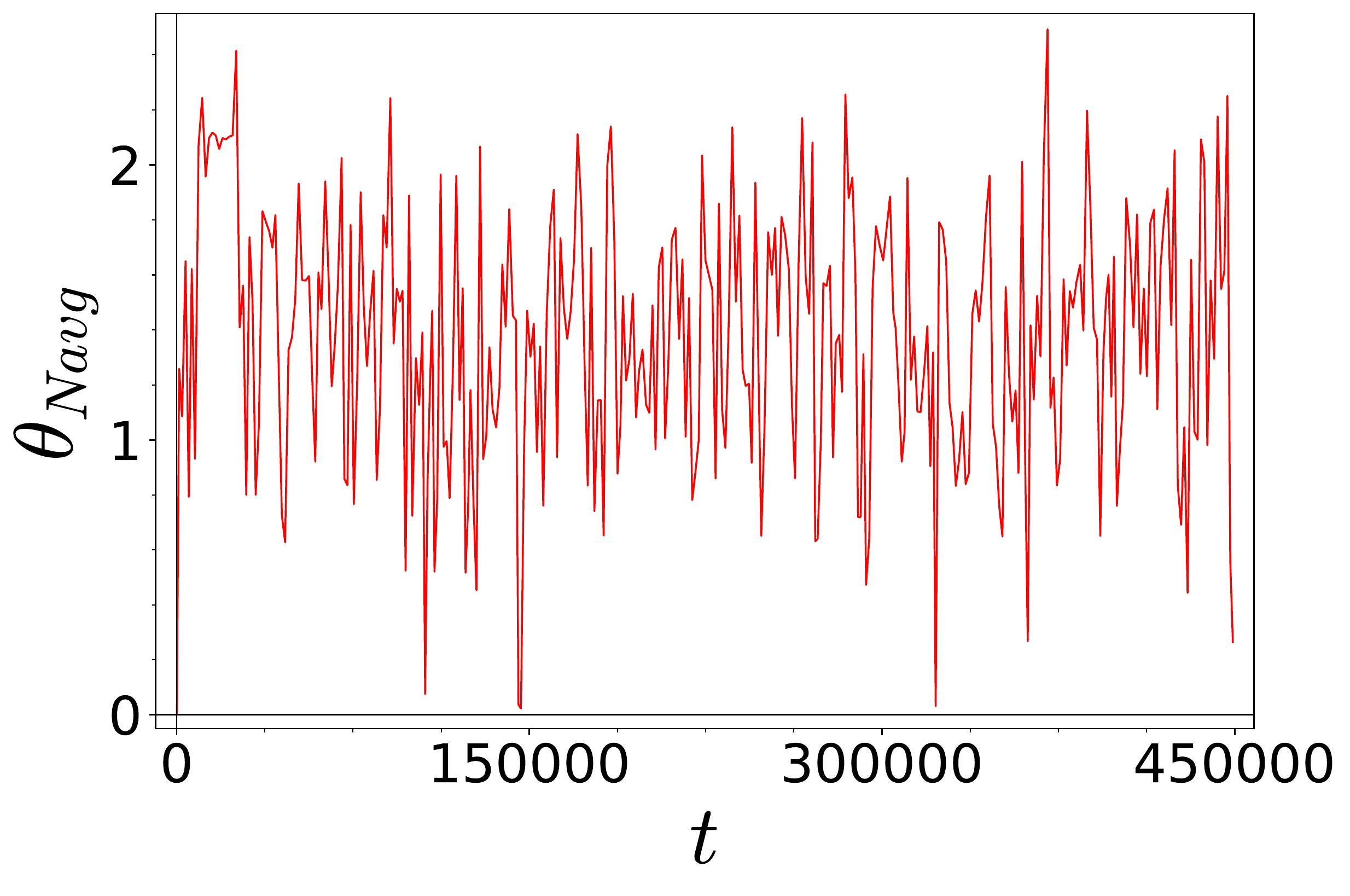}
\caption*{}
\end{subfigure}%
\caption{Left plot: The time series of the $z$ component of a chaotic orbit with impact parameter $\alpha=0.005402M$. Middle plot: The corresponding power spectrum. It obeys the so-called $1/f$ fluctuations meaning that the orbit is \textit{sticky}. Right plot: The average rotation angle. The rotation angle is stable when the orbit exhibits regular behaviour.}
\label{fig:sticky2_1}
\end{figure}

In the main text we provided one example of sticky orbits in order to give an idea of their properties. 
To further illustrate the properties of sticky orbits we also provide here an additional example with an orbit starting at $r_0=40M$ and with $\alpha=0.005402M$, i.e., an orbit with an initial deviation from the previously computed incoming orbit of the order $10^{-7}$ for the impact parameter $\alpha$. The time series of the $z$ component shows periods of time where the mean value is constant and looks similar to the previous case. However this orbit is a little different from the previous one. This chaotic orbit sticks to one of the triple islands that are located away from $u^r=0$. Its power spectrum also obeys the $1/f$ fluctuations validating that the orbit is sticky as can be seen in Fig.~\ref{fig:sticky2_1}.%
%
%
Zooming in to the domain of stickiness for the time series of $z$, we can see that the orbit initially is chaotic and at around $t\simeq10000M$ it sticks to an island of periodic orbits where is stays until the time of $t\simeq 25000M$, at which point it returns to the chaotic sea. As we can see, the orbit does not stay in the chaotic sea for long, it revisits some periodic islands in a seemingly random manner. 
This is a typical behaviour of sticky orbits. %
To further check the sticky part of the orbit we calculate the average rotation angle $\theta_{N\textrm{avg}}$. %
In contrast to the previous example (Fig. \ref{fig:sticky1_rot}), the angle now is not close to zero. This is due to the position of the islands the orbit sticks to, with respect to the location of the invariant point we have chosen. As the orbit oscillates between the islands it is seen to have some average angle with respect to the invariant point.


\section*{References}
\bibliographystyle{plain}
\bibliography{biblio.bib}
\end{document}